\RequirePackage{lineno}
\documentclass[%preprint,
twocolumn,
nobibnotes,
eqsecnum,
amsmath,amssymb,
amsmath,amssymb, aps,
]{revtex4}

\usepackage{epsfig}
\usepackage{amsfonts}
\usepackage{amsmath}
\usepackage{amssymb}
\usepackage{bm} 
\usepackage[framemethod=PStricks]{mdframed}
\usepackage{lipsum}
\usepackage{float}
\usepackage{color}
\usepackage{pdflscape}
\usepackage{graphicx}
\usepackage[counterclockwise]{rotating}

\newcommand{\be}{\begin{equation}}
\newcommand{\ee}{\end{equation}} 
\newcommand{\bea}{\begin{eqnarray}} 
\newcommand{\eea}{\end{eqnarray}}

\DeclareMathOperator{\Ei}{Ei}

\begin{document}

\title{Vortex Identification from Local Properties of the Vorticity Field}
\author{J.H. Elsas$^{1,2}$ and L. Moriconi$^2$}
\affiliation{$^1$Department of Mechanical Engineering, The Johns Hopkins University, \\
Baltimore, MD 21218, USA}
\affiliation{$^2$Instituto de F\'\i sica, Universidade Federal do Rio de Janeiro,
C.P. 68528, 21945-970, Rio de Janeiro, RJ, Brazil}

\begin{abstract}

A number of systematic procedures for the identification of vortices/coherent structures have been developed as a way to
address their possible kinematical and dynamical roles in structural formulations of turbulence. It has been broadly acknowledged, 
however, that vortex detection algorithms, usually based on linear-algebraic properties of the velocity gradient tensor,
{\black{can be}} plagued with severe shortcomings and {\black{may become, in practical terms,}} dependent on the choice of 
subjective threshold parameters in their implementations. In two-dimensions, a large class of standard vortex identification 
prescriptions turn out to be equivalent to the ``swirling strength criterion" ($\lambda_{ci}$-criterion), which is critically 
revisited in this work. We classify the instances 
where the {\black{accuracy of the $\lambda_{ci}$-criterion is affected by nonlinear superposition effects}} and propose an alternative 
vortex detection scheme based on the local curvature properties of the vorticity graph $(x,y,\omega)$ -- the ``vorticity curvature 
criterion" ($\lambda_\omega$-criterion) --  which 
improves over the results obtained with the $\lambda_{ci}$-criterion in controlled Monte-Carlo tests. A particularly 
problematic issue, given its importance in wall-bounded flows, {\black{is the eventual inadequacy}} of the $\lambda_{ci}$-criterion for 
many-vortex configurations in the presence of {\black{strong background shear}}. We show that the $\lambda_\omega$-criterion is 
able to cope with these cases as well, if a subtraction of the mean velocity field background is performed, in the spirit of the 
Reynolds decomposition procedure. A realistic comparative study for vortex identification is then carried out for a direct numerical 
simulation (DNS) of a turbulent channel flow, including a three-dimensional extension of the $\lambda_\omega$-criterion. In contrast 
to the $\lambda_{ci}$-criterion, the $\lambda_\omega$-criterion indicates in a consistent way the existence of small scale isotropic 
turbulent fluctuations in the logarithmic layer, in consonance with long-standing assumptions commonly taken in turbulent boundary 
layer phenomenology.

\end{abstract}

%\pacs{47.27.-i, 47.27.De, 47.27.N-}
\maketitle

%\begin{linenumbers}
\modulolinenumbers[5]

\section{Introduction}\label{sec:introduction}
    The twofold question on whether long-lived vorticity-carrying structures - coherent structures for short - can survive up to higher Reynolds numbers and 
  play an important dynamical role in turbulence, with particular attention to the problems of isotropic and wall-bounded flows,
	has been for a long time a matter of great interest in the fluid dynamics community \cite{rob,adrian1,maru1,wall,jim,tardu, dennis}.
	
    From a modeling perspective, the vorticity field $\vec \omega$ of incompressible flows (our focus in this work) 
  can be considered to be a more fundamental observable than the velocity field $\vec v$, once the latter can be derived 
  from the former through
  \begin{linenomath*} \be
  v_i = - \epsilon_{ijk} \partial^{-2} \partial_j \omega_k \ , \ \label{biot-savart}
  \ee \end{linenomath*}
  where, above, $\partial^{-2}$ stands for the {\black{inverse Laplacian}} operator. Of course, Eq. (\ref{biot-savart}) 
  is nothing more than the Biot-Savart law in the fluid dynamical context. 
  
    One aims, in the so called ``structural formulation of turbulence", to achieve an expressive reduction in the number of 
    degrees of freedom from the introduction of kinematical or dynamical models of coherent structures, the 
  spatial support of strongly correlated vorticity lines. These special vorticity domains are then taken to be the sources of 
  the turbulent velocity field, straightforwardly recovered with the help of Eq. (\ref{biot-savart}). It is interesting to point 
	out that while structural modeling is still a very open problem, one finds, within the framework of wavelet compression 
	techniques strong support for pursuing this direction of research \cite{farge,farge1,farge2}.
	 
    Among the several types of turbulent flows, the turbulent boundary layer (TBL) is a particularly rich stage for
  the production and interaction of coherent structures \cite{tardu}, like streamwise and
  hairpin vortices (often bunched in packets), the latter remarkably anticipated several decades ago by Theodorsen
  \cite{theo} and Townsend \cite{town2}. Due to the variable sizes of these structures, which are
  directly related to their distances from the wall, as depicted in the attached eddy hypothesis \cite{town2, maru_adrian}, 
	the TBL turns out to be a dynamical system characterized by strong multiscale couplings.
  
    The pioneering structural approach of Perry and Chong \cite{perry_chong} has underlined in many alternative ways
  subsequent investigations of the TBL along the years \cite{sree,perry_maru,perry_henbest,mori,maru_science,mathis}, devoted 
	to the study of boundary layer phenomena like viscous drag, the existence of enhanced intermittent velocity fluctuations near the
  wall region and the crossover between turbulent kinetic energy production and dissipation, all of these being points of
  potential applied relevance. Nevertheless its appealing physical picture, the structural approach has been unable, so far, to
  address in a predictive way a relevant phenomenological framework like the law of the wall. {\black{An even more ambitious aim
  for the structural program would be to provide a foundation for the broadly used Reynolds-averaged phenomenological models (like the k-epsilon model) \cite{schli,pope}. In these approaches, one has to resort to {\it{ad hoc}} closure assumptions which relate the Reynolds stress tensor to the mean properties of the flow. This mathematical object could, as a matter of principle, be derived
  from the statistical modeling of the energetically most important vortical structures. 
  
  While at the present state of knowledge, the aforementioned ideas are still essentially speculative, we show in this work that the structural approach, as based on an accurately validated vortex identification procedure, can offer an interesting insight into the physics of wall bounded flows, if one restricts attention on issues of turbulent isotropization.}}
        
    A major problem in the structural formulation of turbulence -- paradoxically as it may sound -- is the ambiguous meaning
  of the coherent structure concept itself, {\black{as long ago emphasized in seminal papers by Hussain \cite{hussain_cs1,hussain_cs2}.}} An operational answer to this question is to define a coherent structure as the compact flow configuration that is obtained, from numerical or experimental data, through the application of some postulated identification algorithm. 
  
  Galilean invariant vortex identification methods usually rely on the information encoded in velocity gradients, which tag regions of the flow characterized by ``swirling motions" in locally co-moving reference frames.{\black{ An interesting physical picture underlying the usefulness of velocity gradients in the identification of coherent structures has to do with the empirical fact that they are correlated with zones of quasi-uniform momentum \cite{mein_adrian}. Therefore, velocity gradients are enhanced around the boundaries of such zones, and provide, in this way, ``shear envelopes", which are ultimately the reason for the phenomenon of coherent structure persistence, as observed in the dynamics of hairpin vortices \cite{zhou_adrian_bala} }}
  
    Most of the discussions on structural aspects of turbulence adopt {\black{Eulerian}} vortex detection methods like 
   the Q-criterion \cite{okubo,weiss,hunt}, the $\Delta$-criterion \cite{perry_chong_cant} and its closely related swirling
  strength criterion ($\lambda_{ci}$-criterion) \cite{zhou,chaka}, or the $\lambda_2$-criterion \cite{jeong_hussain}.
  In all of these criteria, a scalar field, derived from the velocity gradient tensor, is used as a ``marker" to indicate if a
  given point in the flow belongs or not to a vortex. Vortices, are therefore, identified as the connected regions mapped by
  such scalar fields.
	
{\black{
Other classes of vortex identification methods
shift from the definition of ``scalar markers", 
to representative flow configurations, either by
selecting the most energetic ones instantaneously or
by retrieving flow patterns by means of statistical
averaging procedures. For the sake of completeness,
we list below a brief description of five of these
approaches.

(i) In the proper orthogonal decomposition, one tries 
to extract the relevant flow modes that are, on the average, 
more energetic, by solving associated eigenvalue problems
\cite{holmes}.

(ii) A computer-science inspired approach uses
artmap neural networks as a classification tool,
in which a self refining algorithm is used to
identify relevant structures \cite{ferre-gine}.

(iii) Wavelet denoising theory can provide a decomposition
of the velocity field on a complete set of orthogonal
spatially localized modes, in which the
more energetic ones turn out to be associated with
coherent structures \cite{farge}.

(iv) ``Lagrangian coherent structures" can be defined
from the investigation, along pathlines, of
the local dynamical system of fluid element motions
\cite{haller, beron}

(v) Conditionally averaged flow configurations,
representing coherent structures, can be obtained
from a subset of flow realizations that
satisfy certain prescribed statistical signatures,
a procedure which is closely related to
the method of linear stochastic estimation
\cite{adrian_eddies, adrian_review}.}}
  
    Even though there are studies which have pointed out the {\black{pros and cons}} of the available vortex identification
  methods \cite{chaka,eddies,adrian_chris_liu,chaka2,kolar,chen}, systematic investigations of their limitations are still in order. {\black{Commonly 
  noted problems}} are related to shape distortions of retrieved vortices and the subjective definition of threshold parameters, {\black{sometimes
  necessary to increase the efficiency of the identification algorithms}}. As we will emphasize in the following, a less
  obvious (but not less important) difficulty is associated with the effects produced on vortex identification by a shearing
  environment, as in free shear turbulence, turbulent boundary layers or channel flows. 
  
    The velocity gradient-based vortex identification strategies so far addressed in the literature are essentially equivalent,
  in two-dimensions, to the $\lambda_{ci}$-criterion. This is a key point in our discussion, which relies on a careful study
  of how the $\lambda_{ci}$-criterion performs for a variety of controlled ``synthetic" two-dimensional flow configurations. {\black{It turns out that there are serious challenges}} with the use of the $\lambda_{ci}$-criterion, which have motivated us to
  introduce an alternative vortex identification prescription, referred to as the vorticity curvature criterion
  ($\lambda_\omega$-criterion), a vortex identification method entirely based on local properties of the vorticity field.
    
    Our results are centered on the analysis of two-dimensional coherent structures, which are important actors, for instance, in the {\black{quasigeostrophic approximation for the
		dynamics of the atmosphere and the ocean (low Rossby number regime, planetary length scales)}} \cite{vallis}, in purely two-dimensional turbulent systems \cite{boffetta}, and also in 
		the properties of streamwise/wall normal plane sections of turbulent boundary layer flows \cite{adrian_mein_tom,camussi,wu,dennis_nickels},
		which reveal the existence of spanwise vortex tubes.
		We introduce and study the problem of vortex identification for large ensembles of synthetic two-dimensional vortex systems and 
		subsequently investigate, by means of a turbulent channel flow direct numerical simulation (DNS), the statistical features of boundary 
		layer vortices from the point of view of both the $\lambda_{ci}$ and the $\lambda_\omega$ criteria. 
  
    This work is organized as follows. To make the paper as self-contained as possible, we provide, in Sec. \ref{sec:swst}, a
  detailed definition of the $\lambda_{ci}$-criterion, and classify, from the analysis of simple two-dimensional vortex
  configurations, its main issues. In order to overcome the observed difficulties with the $\lambda_{ci}$-criterion, 
	an essentially threshold-free vortex identification method, the $\lambda_\omega$-criterion, is proposed and discussed in Sec.
  \ref{sec:vcurv}, which is found to considerably improve vortex detection for most of the problematic cases. 
    
  Monte-Carlo simulations of synthetic vortex systems are introduced in Sec. \ref{sec:montecarlo}, as a way to evaluate how
  the $\lambda_{ci}$-criterion and the $\lambda_\omega$-criterion automated algorithms perform for a large number of samples. 
	We find, at this point, poor results for both vortex identification methods for the case of vortices in the presence of a strong 
	background shear. To cope with that, we devise a background shear subtraction procedure, meaningful for statistically stationary 
	flows, which points out the better, and reasonably good, performance of the $\lambda_\omega$-criterion when compared to the one of 
	the $\lambda_{ci}$-criterion.
  
    We, then, move to the analysis of a more realistic scenario in Sec. \ref{sec:dns}, provided by the numerical simulation of a
  turbulent channel flow. Having in mind all the issues discussed in the previous sections, it turns out that while the
  $\lambda_{ci}$-criterion fails to indicate isotropization of small scale turbulent fluctuations in the TBL logarithmic layer, the
  $\lambda_\omega$-criterion can do so very succesfully, which is a remarkable phenomenological result within the 
	context of the structural formulation. We also discuss, in Sec. \ref{sec:3dvc}, the extension of the $\lambda_\omega$-criterion 
	to the case of fully three-dimensional flows, including some preliminary visualizations for the turbulent channel structures 
	obtained in this way. Finally, in Sec. \ref{sec:conclusions}, we summarize our findings and point out directions of further research.  
 
\section{Swirling-Strength Issues}\label{sec:swst}    
    The $\lambda_{ci}$-criterion for vortex identification relies on the analysis of the instantaneous topology of the
  velocity vector field  \cite{zhou}. In two dimensions (our main interest in this paper), one wants to single out points
  of the flow that can be classified either as sources or sinks of streamlines.
  In more concrete terms, set as $(x_1,x_2)=(0,0)$ the position of an arbitrary point in the flow, which has instantaneous
  vanishing velocity in the co-moving reference frame. Taking the velocity field to be ``frozen", we can write down the
  linearized equation of motion for a particle that {\black{follows}} the frozen streamlines of the flow in a neighborhood of the
  origin as
    \begin{linenomath*} \be 
    \dot x_i = A_{ij} x_j  \ , \
    \ee  \end{linenomath*} 
		where $A_{ij} = \partial_j v_i|_{x=0}$ is the $i,j$ matrix element of the velocity gradient tensor {\black{${\bf{A}}$}}.
		It is not difficult to show that spiraling orbits around the origin (the focus of motion) are necessarily associated with
  complex eigenvalues of {\black{${\bf{A}}$}}. The eigenvalue equation reads
  \begin{linenomath*} \bea
    &&\det({\partial_j v_i - \lambda \delta_{ij})} = \nonumber \\
		&&=\lambda^2 -\lambda \partial_i v_i 
           + \det ( \partial_j v_i)  = 0 \ . \ \label{det}
 \eea \end{linenomath*}
    {\black{The ``swirling strength" field is the scalar quantity defined as the imaginary part, taken as positive,}} 
    of the complex eigenvalue $\lambda \equiv \lambda_{cr} +i \lambda_{ci}$. The $\lambda_{ci}$-criterion, thus, 
    postulates that vortex domains are regions of the flow which have non-zero swirling strength. For incompressible two-dimensional flows, things are a bit simpler, once Eq. (\ref{det}) tells us that these regions are the loci of points where the velocity gradient determinant is positive.

\begin{figure}[ht]
      \includegraphics[width=1.05\linewidth]{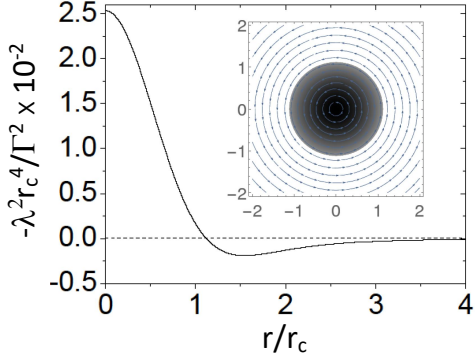}
      \vspace{-1.3cm}
      
    \caption{The dimensionless velocity gradient determinant for the Lamb-Oseen vortex as a function of $r/r_c$. Inset:
		density plot of the swirling strength field and the vortex streamlines (coordinates are given in units of $r_c$).}
    \label{fig:swst1}
  \end{figure}

    To exemplify the analysis, we illustrate how the $\lambda_{ci}$-criterion works for the prototypical Lamb-Oseen vortex
  \cite{batchelor}, which is in fact an important building block in structural studies \cite{mori,jeff,herpin_a,herpin}. 
  {\black{Let $\epsilon_{ij}$ be the two-dimensional Levi-Civita symbol.}} 
  The Lamb-Oseen vortex is defined by the divergence free velocity field, {\black{with components}}
  \begin{linenomath*} \be 
    v_i = \epsilon_{ij} x_j F(r) \ , \  \label{lo}
  \ee  \end{linenomath*} 
  where 
  \begin{linenomath*} \be 
    F(r) = \frac{\Gamma}{2 \pi r^2} \left ( 1 - e^{-\frac{r^2}{r_c^2}} \right ) \ . \ \label{fr}
  \ee \end{linenomath*} 
    Above, $r_c$ and $\Gamma$ denote the vortex core radius and its asymptotic circulation, respectively. The velocity
  gradient determinant can be easily derived as
  \begin{linenomath*} \bea
    &&\det (\partial_j v_i) =  F [ F + r F']  = \left ( \frac{\Gamma}{2\pi r^2} \right )^2 \nonumber \\
		&&\times \left [1-2e^{-\frac{r^2}{r_c^2}} +e^{-\frac{2r^2}{r_c^2}} \left (3 - \frac{2 r^2}{r_c^2} \right ) \right ]
	 \label{detLambOseen}
  \eea \end{linenomath*}
  and it is shown in Fig. \ref{fig:swst1} as a function of $r/r_c$. The interesting point here is that the velocity gradient
 determinant is positive only within a finite distance $r \leq \bar r$ from the origin, so that the Lamb-Oseen vortex is
 identified as the disk on the density plot given in the inset of Fig. \ref{fig:swst1}.
   From Eq. (\ref{detLambOseen}), we find that $\bar r$, and the vorticity
  flux across the disk, $\bar \Gamma$, are related to the corresponding vortex parameters as
    \begin{linenomath*} \be 
      r_c = \alpha \bar r  {\hbox{   and   }} \Gamma = \beta \bar \Gamma  \ , \ \label{alpha-beta}
    \ee \end{linenomath*} 
    where, in terms of the Lambert W function \cite{abramowitz},
    \begin{linenomath*} \be 
      \alpha \equiv \frac{1}{\sqrt{-\frac{1}{2} -W(-\frac{1}{2\sqrt{e}}})} \simeq 0.89  \label{alpha} 
    \ee \end{linenomath*} 
    and
    \begin{linenomath*} \be 
      \beta = \frac{1}{1-e^{-\alpha^2}}  \simeq 1.4 \label{beta} \ . \
    \ee \end{linenomath*} 
    It is common to assume, as a first approximation, that the connected regions highlighted by the $\lambda_{ci}$-criterion
  have, even in many-vortex two-dimensional systems, circular shapes, so that the relations given in (\ref{alpha-beta}) can be
  used to recover, in an automated fashion, the radius and the circulation parameters of the identified vortices.
  These same parameters can be obtained, alternatively, but with greater computational cost and comparable accuracy, 
  from fittings, in the spotted regions, of the recorded velocity fields to the Lamb-Oseen pattern, Eqs. (\ref{lo}) and (\ref{fr}) \cite{herpin_a,herpin}.
  
    Serious difficulties can arise in the implementation of the $\lambda_{ci}$-criterion when two or more vortices get close enough 
		to each other, or if they are in the presence of a shearing background. However, there are no comprehensive works in the literature 
		which attempt to define the conditions for the accurate use of this vortex identification method. Therefore, we put forward 
		below, as a necessary stage for improvement over the $\lambda_{ci}$-criterion, an informal (and not exhaustive) classification of 
		its important problematic issues for the case of two-vortex systems. To render our discussion free of ambiguities, whenever we refer 
		to strict two dimensional vortices throughout the paper, we mean precisely Lamb-Oseen vortices.
  
  \vspace{1.5cm}
    \leftline{(i) {\it{Vortex Shape Distortion and Coalescence}}}
  \vspace{0.3cm}
    
\begin{figure}[ht]
      \begin{minipage}{0.45\linewidth}
        \includegraphics[width=\linewidth]{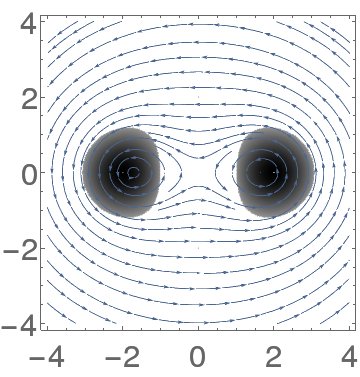}
          \put (-100,85){\makebox[0.05\linewidth][r]{(a)}}
      \end{minipage}
      \quad
      \begin{minipage}{0.45\linewidth}
        \includegraphics[width=\linewidth]{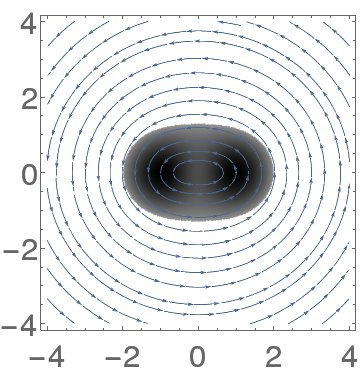}
          \put (-100,85){\makebox[0.05\linewidth][r]{(b)}}
      \end{minipage}
      
      \begin{minipage}{0.45\linewidth}
          \includegraphics[width=\linewidth]{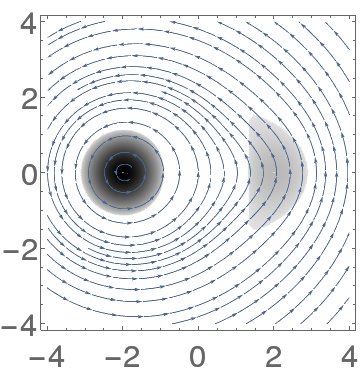}
        \put (-100,85){\makebox[0.05\linewidth][r]{(c)}}
      \end{minipage}
      \quad
      \begin{minipage}{0.45\linewidth}
          \includegraphics[width=\linewidth]{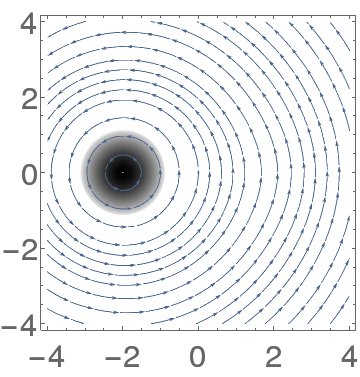}
          \put (-100,85){\makebox[0.05\linewidth][r]{(d)}}
      \end{minipage}
      
      \caption{In all of the four depicted cases, vortex pairs have the same core radius. Coordinates are
      given in units of $r_c$. Let $\Gamma_L$ and $\Gamma_R$ be the circulations of the left and right vortices, 
      respectively. (a) Shape distortions of two near vortices with $\Gamma_L=\Gamma_R$; (b) vortex coalescence for a
      configuration with vortex centers separated by $2r_c$ and $\Gamma_L =\Gamma_R$; (c) configuration with vortex centers
      separated by $4r_c$ and $\Gamma_L = 5\Gamma_R$; (d) the same separation as in (c), but with $\Gamma_L = 10\Gamma_R$. 
      The right vortex escapes detection by the $\lambda_{ci}$-criterion.}
     \label{fig:swst2}
    \end{figure}

    As it is shown in Fig. \ref{fig:swst2}a, the shapes of two vortices get distorted as they approach
  each other, up to the point where they coalesce into a single vortex structure, as in Fig. \ref{fig:swst2}b,
	due to the fact that the streamlines with opposite flow directions can mutually cancel in the region between them. 
	Nevertheless the fact that there are two local swirling strength peaks in the merged region, it is not an obvious 
	task how to disentangle them in practical automated analyses.
  
    {\black{In order to solve}} the vortex merging problem we could define a threshold parameter $T$ and select regions of the flow
  which have $\lambda_{ci} > T$. This can actually break the coalesced structures back to two vortices again, but as a side
  effect other vortices in the system would be erased from detection. It is also likely that many other coalesced vortices in
  the flow would not be split in this way. Once there is not a clear prescription on how to define $T$, its choice is essentially
  subjective, and the threshold solution is far from being a well-established procedure. It should be clear, however, that
	there should be some room, in principle, for the implementation of iterative thresholding algorithms like the ones used in 
	denoising theory \cite{farge}.
	
  \vspace{0.3cm}
  \leftline{(ii) {\it{Ghost Vortices}}}
  \vspace{0.3cm}
  
    Considering two vortices with the same radius, for instance, if one of them has larger circulation, shape distortion is,
  as expected, more pronounced for the vortex with smaller circulation. Instead of coalescence, however, the weaker vortex can
  disappear completely from the flow, if it happens to be close enough to the strong one. These situations are depicted in
  Figs. \ref{fig:swst2}c and \ref{fig:swst2}d.
    
  {\black{
  \vspace{0.3cm}  
  \leftline{(iii) {\it{Background Shear Effects}}}}}
  \vspace{0.3cm}
  
      The most dramatic issues on the identification of vortices by means of the $\lambda_{ci}$-criterion are probably the ones
  associated to background shear effects, which for evident phenomenological reasons, are especially important in wall-bounded
  flows.
  
    \vspace{0.3cm}
	   \begin{figure}[ht]
      \includegraphics[width=1.05\linewidth]{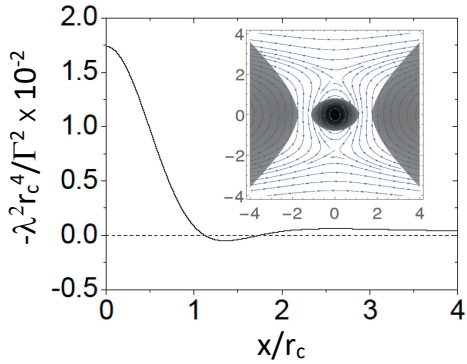}
      \vspace{-1.3cm}
      
    \caption{The dimensionless velocity gradient determinant along the $y=0$ axis, for a vortex of positive circulation $\Gamma$ and radius $r_c$
		  in the presence of a horizontal background shear of negative vorticity $\bar \omega = -0.05  \Gamma/r_c^2$. {\black{The $x$ coordinate is given in units of $r_c$}}.
			Inset: density plot of the swirling strength field for this flow configuration.}
    \label{flaps}
  \end{figure}
	
    \begin{figure}[ht]
      \begin{minipage}{0.45\linewidth}
        \includegraphics[width=\linewidth]{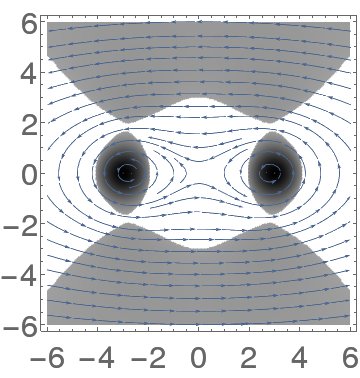}
          \put (-100,90){\makebox[0.05\linewidth][r]{(a)}}
      \end{minipage}
      \quad
      \begin{minipage}{0.45\linewidth}
        \includegraphics[width=\linewidth]{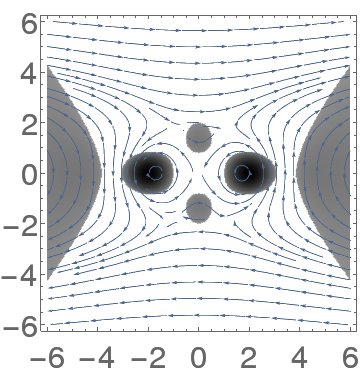}
          \put (-100,90){\makebox[0.05\linewidth][r]{(b)}}
      \end{minipage}
      
      \caption{The background shear is horizontal and both vortices have positive circulation $\Gamma$ and
			radius $r_c$. Coordinates are given in units of $r_c$. The background vorticity is $|\bar \omega| = 0.05
      \Gamma/r_c^2$. (a) two vortices in a background shear of positive vorticity; (b) two vortices in a
      background shear of negative vorticity.}
      
      \label{fig:swst3}
    \end{figure}
    
        Take, as an illustrative example, the constant background shear with vorticity $\bar \omega$, described by the velocity
  field {\black{with components}} $(v_x,v_y) = (-\bar \omega y,0)$, which can be superimposed to the velocity field produced by a vortex or a couple of
  vortices. Of course, the presence of background shear modifies the velocity gradient determinant. Analogously to Fig.
  \ref{fig:swst1}, the velocity gradient determinant is plotted in Fig. \ref{flaps} for $y=0$, as a function of $x/r_c$. 
  Differently from the free vortex case, the velocity gradient determinant becomes positive again at some distance
  from the origin, a fact that is related to the existence of two disconnected and spurious unbounded regions - henceforth
  referred to as ``flaps" -  which surround the real vortex, as shown in the inset of Fig. \ref{flaps}. Depending on the intensity and 
	relative sign of the background vorticity, the vortex can disappear and only the flaps remain, or the flaps can coalesce
  with the vortex, forming a large, unbounded, structure.
  
    In the test situation where we have two Lamb-Oseen vortices with identical circulations in the presence of a constant
  background shear, the flaps still show up, as can be seen in Fig. \ref{fig:swst3}a and \ref{fig:swst3}b. Furthermore, it
  turns out that if the background vorticity is opposite to the ones of the two vortices, then, besides the flaps, two spurious
  vortices appear. More complex patterns arise if additional vortices are superimposed to the background shear flow, 
  once flaps and spurious vortices can also mutually interact. 
  
  \vspace{0.3cm}
  {\black{\leftline{(iv) {\it{Spurious Vortices}}}}}
  \vspace{0.3cm}
    
    Spurious vortices can be misleadingly identified by the $\lambda_{ci}$-criterion in
    many-vortex configurations. These regions have, in general, relatively small area and circulation, making them, 
    even if sometimes numerous, mostly non influential to the overall properties of flow, with the exception of counting 
    statistics. Disregarding other aspects of Fig. \ref{fig:swst3}b, the two vertically aligned and disconnected spots 
    shown there are examples of spurious vortices generated from the approximation of two real vortices, further enlarged 
    by the presence of background shear, identified in the picture as the two darker disconnected compact regions.
    \vspace{0.3cm}
    
     The four general instances discussed above clearly indicate that the analysis of coherent structures through the use of
  the $\lambda_{ci}$-criterion, even though meaningful in cases where the vortex density and the vorticity of the background 
  shear are small enough, can lead to inaccurate results, mainly in the investigation of turbulent flows, characterized by 
  strong multiscale intermittent fluctuations of vorticity and strain.
    
    {\black{In the next section, we put forward an alternative 
  vortex identification method, which has the local vorticity field as its main ingredient, and is devised to mitigate the 
  aforementioned deficiencies of the $\lambda_{ci}$-criterion.}}
  
\section{Vorticity Curvature Criterion}\label{sec:vcurv}

    {\black{As a key point in understanding the behavior}} of the  $\lambda_{ci}$-criterion in {\black{two-dimensional}} many-vortex systems, it is
  useful to point out the connection between this criterion and the differential-geometric properties of the stream function
  $\psi =\psi(\vec r)$. Note that in a dimensionless system of fluid dynamical units, the {\black{Gaussian}} curvature 
	$K$ \cite{kuhnel} of the stream function graph $(x,y,\psi)$ can be written as
    \begin{linenomath*} \bea
     &&K = \frac{\partial_1^2 \psi\ \partial_2^2\psi-(\partial_1\partial_2\psi)^2}{1+(\partial_1\psi)^2+(\partial_2\psi)^2}\\
     &&= \frac{\partial_1 v_1\ \partial_2 v_2 - \partial_1 v_2\ \partial_2 v_1}{(1+\vec v^2)^2} 
       = \frac{\det(\partial_j v_i)}{(1+\vec v^2)^2} \ . \ \label{Klambda}
    \eea \end{linenomath*}  
    It is clear, thus, from the comparison between (\ref{det}) and (\ref{Klambda}), that in incompressible two-dimensional
  flows a point belongs to a vortex, according to the $\lambda_{ci}$-criterion, if and only if its stream function graph has
  positive {\black{Gaussian}} curvature, like a dome. 
  
    For a typical vortex, which has {\black{two-dimensional vorticity $\omega(\vec r)$ (a pseudoescalar field)}}
    that decays faster than $1/r$, the streamfunction is
  asymptotically logarithmic, since 
    \begin{linenomath*} \bea
      &&\psi (\vec r)  = - \partial^{-2} \omega(\vec r) \nonumber \\
      &&= \frac{1}{2 \pi} \int d^2 \vec r' \log \left ( \frac{|\vec r - \vec r'|}{a} \right ) \omega (\vec r') \ , \
    \eea \end{linenomath*}  
    where $a$ is some (unimportant) arbitrary length scale in the flow. The Lamb-Oseen vortex, in particular, is associated 
  to the stream function
    \begin{linenomath*} \be 
      \psi   = \frac{\Gamma}{4 \pi} \left[\log(r^2/r_c^2)-\Ei(-r^2/r_c^2)\right] \ , \ \label{eq:psiOmega}
    \ee  \end{linenomath*}  
    where $\Ei (\cdot) $ refers to the Exponential-Integral function \cite{abramowitz}, which is dominated, far
  from the origin, by  the slowly varying logarithmic contribution in Eq. (\ref{eq:psiOmega}).
    
    The asymptotic logarithmic profile of the vortex stream function implies that there are strong non-linear superposition
  effects that affect the curvature  of the stream function graph associated to individual vortices in many-vortex systems.
  This is the main reason for all of the issues with the implementation of $\lambda_{ci}$-criterion, as discussed in the
  previous section. {\black{To understand this point in a more detailed way, consider a set of $N$ two-dimensional vortices,
  placed at positions $\vec r_i$, which are associated to respective streamfunctions $\psi_i(\vec r - \vec r_i)$, 
  where $i=1,2,...,N$. The streamfunction at a general position $\vec r$ of the flow, is given, therefore, as
  \begin{linenomath*} \be 
  \psi(\vec r) = \sum_{i=1}^N \psi_i(\vec r -\vec r_i) \ . \
  \ee  \end{linenomath*} 
  Since the individual streamfunction fields $\psi_i$ have spatial slow logarithmic variations, the above superimposed 
  streamfunction, $\psi(\vec r)$, can be considerably perturbed by the presence of other vortices in the system.}}
  
    The ideal set up to deal with vortex identification, thus, would be to base the analysis on the properties of spatially
  bounded fluid dynamical observables like the vorticity field carried by coherent structures. In two dimensions, the most
  immediate attempt along these lines would be to work with vorticity level curves, but this is a limited approach, since
  spurious vortices would proliferate and the subjective choice of thresholds would be unavoidable.
    
    If we insist on vorticity as a fundamental element in a local vortex identification scheme, an interesting heuristic
  proposal is simply to replace the stream function as it is used in the $\lambda_{ci}$-criterion by the vorticity field. Now,
  to find vortices we would look for positive curvature regions of the vorticity graph. This prescription is promising, but
  the inspection of simple cases suggests that some refinement is still in order. 
  
    Consider, for example, four identical vortices which are placed at the vertices of a square. It is not difficult to show
  that the {\black{Gaussian}} curvature of the vorticity graph is positive at the center of the square, even though there is no vortex
  there. Without loss of generality, if we take the real vortices to be ``bumps" of the vorticity graph (i.e., if they have
  positive vorticity), then the spurious vortex at the center is a bowl, with idiosyncratic positive vorticity. 
  
    In more mathematical terms, we just mean that while $\omega \partial^2 \omega$ is negative at the square vertices, it
  changes its sign at the center. This fact is the hint to establish a meaningful vortex identification prescription, the 
  $\lambda_\omega$-criterion, which relies on the local {\black{Gaussian}} curvature properties of the vorticity graph. To introduce it
  in detail, we first introduce some notation. {\black{Having in mind our two-dimensional context, define, from the vorticity field $\omega (\vec r)$,
  the {\it{pseudo-velocity field}}, with cartesian components}}
    \begin{linenomath*} \be 
    \tilde v_i (\vec r) \equiv \epsilon_{ij} \partial_j \omega (\vec r) \label{pseudov}
    \ee  \end{linenomath*} 
    and {\black{the {\it{pseudo-vorticity}} field}}
    \begin{linenomath*} \be 
    \tilde \omega (\vec r)  \equiv - \partial^2 \omega (\vec r) \ . \ \label{pseudo_omega}
    \ee  \end{linenomath*} 
    {\black{The streamlines associated to the pseudo-velocity field for the case of a single Lamb-Oseen vortex are qualitatively the same as the ones derived for the physical velocity field, so that they still represent a swirling motion. The main advantage in the use of above definitions is that while they do not spoil the physical meaning of what we consider to be a standard vortex, they are mathematical functions with more interesting local properties, like a fast gaussian decay as the radial distance from the vortex center increases.}}
    
    We can also write down the determinant of the pseudo-velocity gradient tensor as
    \begin{linenomath*} \be 
    \det ( \partial_j \tilde v_i) \equiv - \tilde \lambda^2 \ . \
    \ee \end{linenomath*} 
    {\black{Taking the imaginary part of $\tilde \lambda$ as positive}}, consider the scalar field   
    \begin{linenomath*} \be 
    \lambda_\omega \equiv \Theta(-\omega \partial^2 \omega) {\rm Im}\ \tilde\lambda 
    = \Theta(\omega \tilde \omega) {\rm Im}\ \tilde\lambda  \ , \ \label{lomega}
    \ee \end{linenomath*} 
    where $\Theta(\omega \tilde \omega)$ is the Heaviside filtering function that is expected to vanish for spurious vortices,
  like the one discussed in the preceding four-vortex example. Vortices are then identified by the $\lambda_\omega$-criterion
  as the connected regions of the flow where $\lambda_\omega \neq 0$.
  
    Comparing the $\lambda_\omega$-criterion to the $\lambda_{ci}$-criterion, we note that the essential advantage of the
  former is that it depends locally on the vorticity field, which has sharp peaks and rapidly decaying tails for general
  vortices. The $\lambda_{ci}$-criterion, on its turn, is related to the curvature properties of the stream function
	graph, which has much broader peaks and tails, and may lead to poor vortex identification resolution.
    
    The $\lambda_\omega$-criterion can be classified as a higher order derivative vortex identification scheme, since it
  depends on the evaluation of third order derivatives of the velocity field {\black{(in contrast to the $\lambda_{ci}$-criterion, which is defined in terms of first order derivatives)}}. Two decades ago this fact would be probably a
  main objection to its practical use. However, taking into account the present status of optical measurement techniques such
  as particle image velocimetry and the fast increasing computational power of direct numerical simulations, there is an open
  avenue for the investigation of high-order derivative vortex identification methods. A point of great relevance here is that 
	the $\lambda_\omega$-criterion works {\black{efficiently even without the imposition of}} subjective threshold 
	parameters. This brings considerable simplification in the implementation of automated analyses of many-vortex configurations.
  
\begin{figure}[ht]
      \begin{minipage}{0.45\linewidth}
        \includegraphics[width=\linewidth]{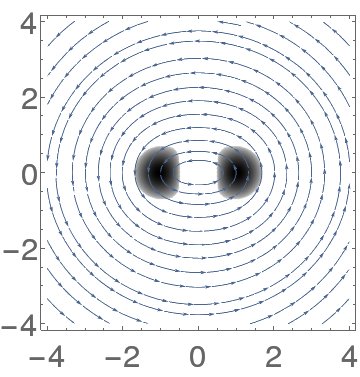}
          \put (-100,85){\makebox[0.05\linewidth][r]{(a)}}
      \end{minipage}
      \quad
      \begin{minipage}{0.45\linewidth}
        \includegraphics[width=\linewidth]{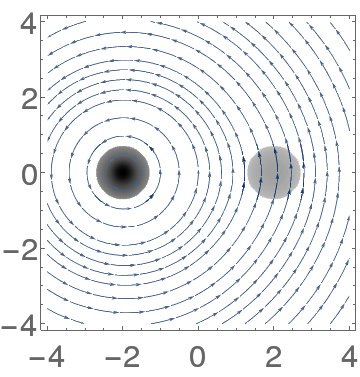}
          \put (-100,85){\makebox[0.05\linewidth][r]{(b)}}
      \end{minipage}
      
      \begin{minipage}{0.45\linewidth}
          \includegraphics[width=\linewidth]{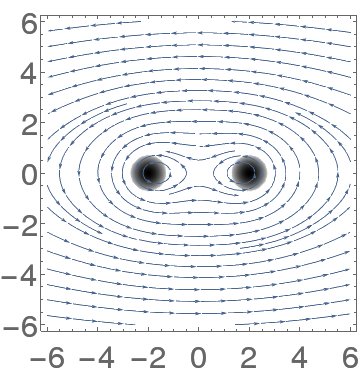}
        \put (-100,75){\makebox[0.05\linewidth][r]{(c)}}
      \end{minipage}
      \quad
      \begin{minipage}{0.45\linewidth}
          \includegraphics[width=\linewidth]{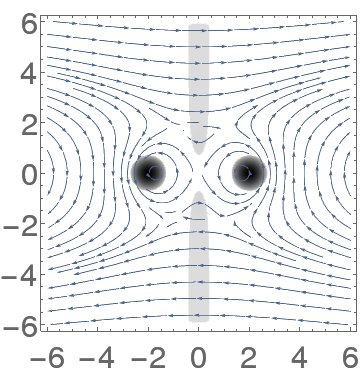}
          \put (-100,75){\makebox[0.05\linewidth][r]{(d)}}
      \end{minipage}

 %   \end{figure}
    
   % \begin{figure}[ht]
   %   \begin{minipage}{0.45\linewidth}
   %    
   %     \includegraphics[width=\linewidth]{fig5a.jpg}
   %          \put (-100,85){\makebox[0.05\linewidth][r]{(a)}}
   %        
   %              \end{minipage}
   %  \quad
     % \begin{minipage}{0.45\linewidth}
     %   \begin{flushright}
      %  \includegraphics[width=\linewidth]{fig5b.jpg}
       %   \put (-100,85){\makebox[0.05\linewidth][r]{(b)}}
      %  \end{flushright}
     % \end{minipage}    
    
    %  \begin{minipage}{0.45\linewidth}
     %   
      %  \includegraphics[width=\linewidth]{fig5c.jpg}
       %  \put (-100,85){\makebox[0.05\linewidth][r]{(c)}}
        %
      % \end{minipage}
     % \quad
     % \begin{minipage}{0.45\linewidth}
      %  \begin{flushright}
      %  \includegraphics[width=\linewidth]{fig5d.jpg}
      %  \put (-100,85){\makebox[0.05\linewidth][r]{(d)}}
      %  \end{flushright}
     % \end{minipage}
      
     % \begin{minipage}{0.45\linewidth}
     %   
      %  \includegraphics[width=\linewidth]{fig5e.jpg}
      %    \put (-100,90){\makebox[0.05\linewidth][r]{(e)}}
      %  
      %\end{minipage}
     % \quad
     % \begin{minipage}{0.45\linewidth}
     %   \begin{flushright}
      %  \includegraphics[width=\linewidth]{fig5f.jpg}
       %   \put (-100,90){\makebox[0.05\linewidth][r]{(f)}}
    %    \end{flushright}
     % \end{minipage}
      
     % \begin{minipage}{0.45\linewidth}
      %  
       % \includegraphics[width=\linewidth]{fig5g.jpg}
        %  \put (-100,90){\makebox[0.05\linewidth][r]{(g)}}
        %
      %\end{minipage}
      %\quad
      %\begin{minipage}{0.45\linewidth}
       % \begin{flushright}
    %    \includegraphics[width=\linewidth]{fig5h.jpg}
     %     \put (-100,90){\makebox[0.05\linewidth][r]{(h)}}
      %  \end{flushright}
      %\end{minipage}
      
      \caption{\black{In (a), (b), (c), and (d), the respective vortex configurations previously studied by means of the 
               $\lambda_{ci}$-criterion in Figs. 2b, 2d, 4a and 4b are now
               reanalysed, taking the $\lambda_\omega$-criterion as the vortex detection tool.}}
      \label{fig:vcurv1}
    \end{figure}

    We re-examine, now under the light of the $\lambda_\omega$-criterion, the relevant vortex identification issues presented
  in the previous section. The results are schematically depicted in in Fig. \ref{fig:vcurv1}.
  
    Without background shear, the $\lambda_\omega$-criterion has, clearly, higher resolution than the standard 
  $\lambda_{ci}$-criterion, since it is able to split coalesced vortices (Fig. \ref{fig:vcurv1}a) that would otherwise 
	be counted as one, and to recover ghost vortices (Fig. \ref{fig:vcurv1}b). With constant background shear, we also find 
	improvements: the vortex shape distortion is considerably reduced and the large, unbounded flaps are completely eliminated
	(Figs. \ref{fig:vcurv1}c and \ref{fig:vcurv1}d) . However, as it can be seen in Fig. \ref{fig:vcurv1}d, there is a couple of relatively small $\lambda_\omega$ spurious regions in the form of vertical stripes, produced for the case where the two vortices have vorticity opposite to the one of the background. This undesirable effect is due to the specific form of the filtering function $\Theta(\omega \tilde \omega)$. 
If a background with constant vorticity $\bar \omega$ is added to the vorticity field $\omega$, the filtering function can 
	be written as $\Theta((\omega + \bar \omega) \tilde \omega)$. Therefore, if $\bar \omega$ and $\omega$ have opposite signs 
	and $|\bar \omega| > |\omega|$, the filtering function may, as a side effect, introduce errors or even hamper the 
	identification of a true vortex. We will have more to say about this issue in Sec. \ref{sec:montecarlo}. 

    \begin{figure}[ht]
		      \begin{minipage}{0.49\linewidth}
        \begin{flushright}
          \includegraphics[width=\linewidth]{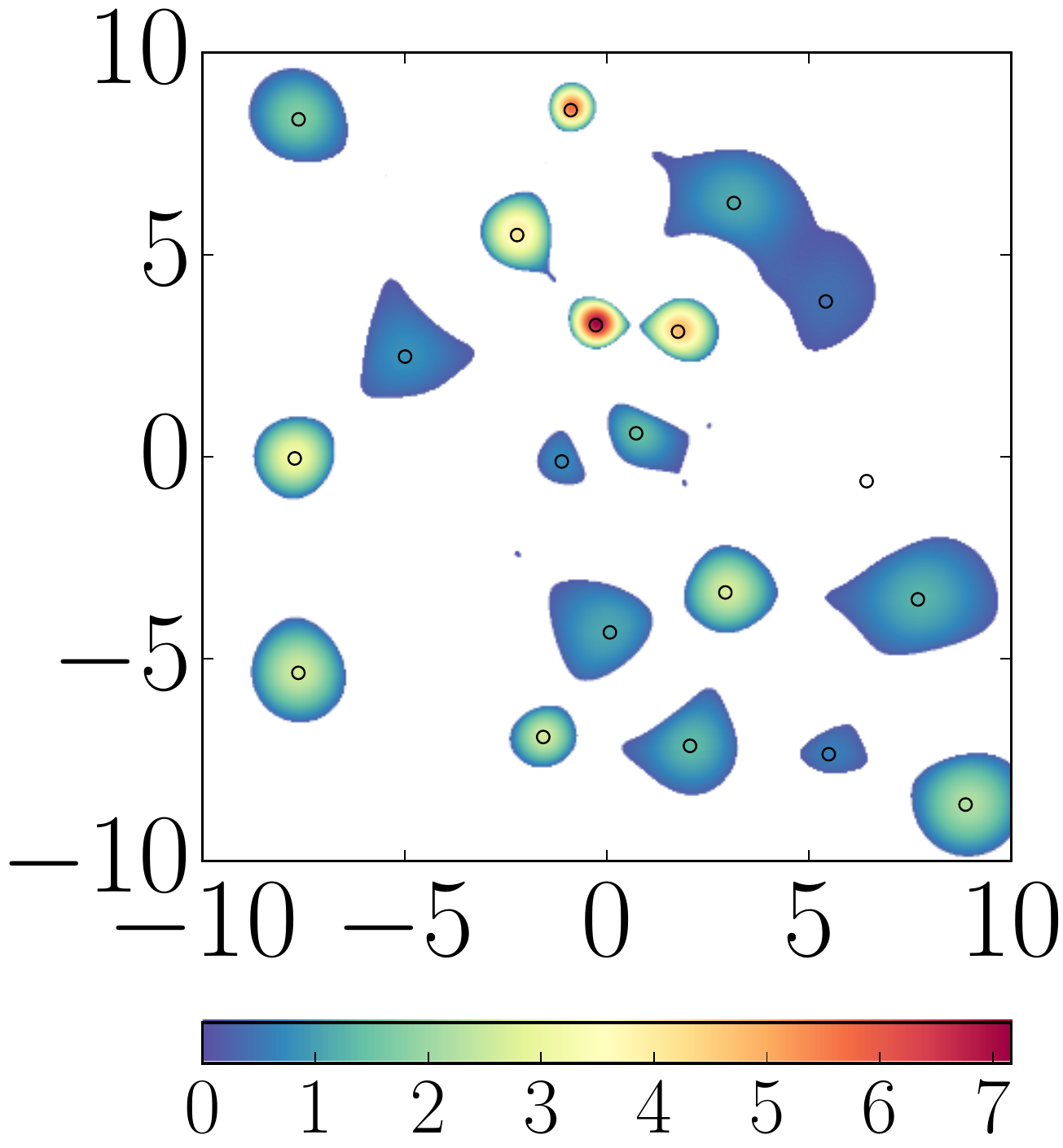}
          \put (-105,100){\makebox[0.05\linewidth][r]{(a)}}
        \end{flushright}
      \end{minipage} 
		%	\end{figure}	
		%	\begin{figure}[ht]
		     \begin{minipage}{0.49\linewidth}        
          \includegraphics[width=\linewidth]{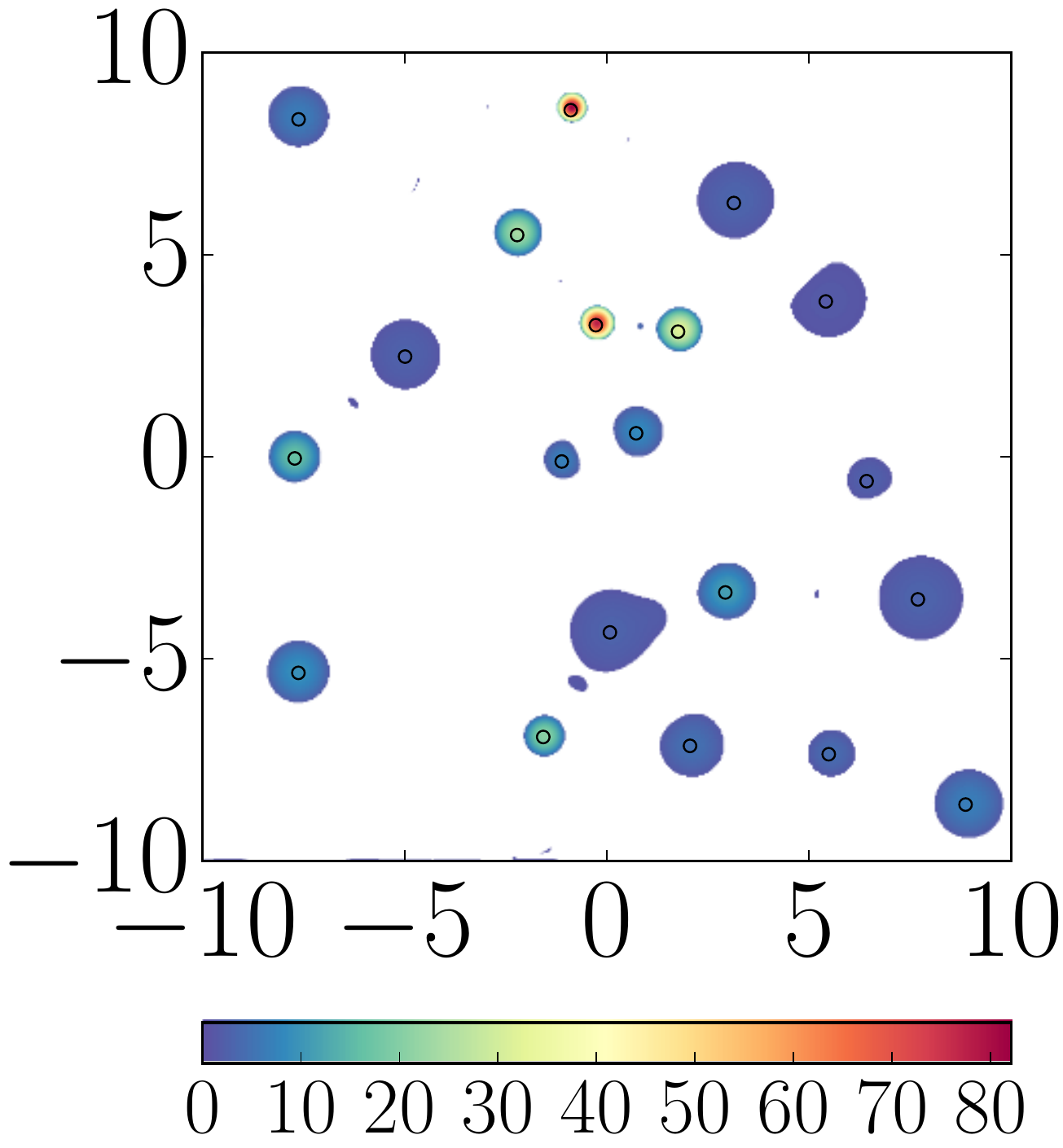}
          \put (-105,100){\makebox[0.05\linewidth][r]{(b)}}
      \end{minipage}
				  \begin{minipage}{0.49\linewidth}        
          \includegraphics[width=\linewidth]{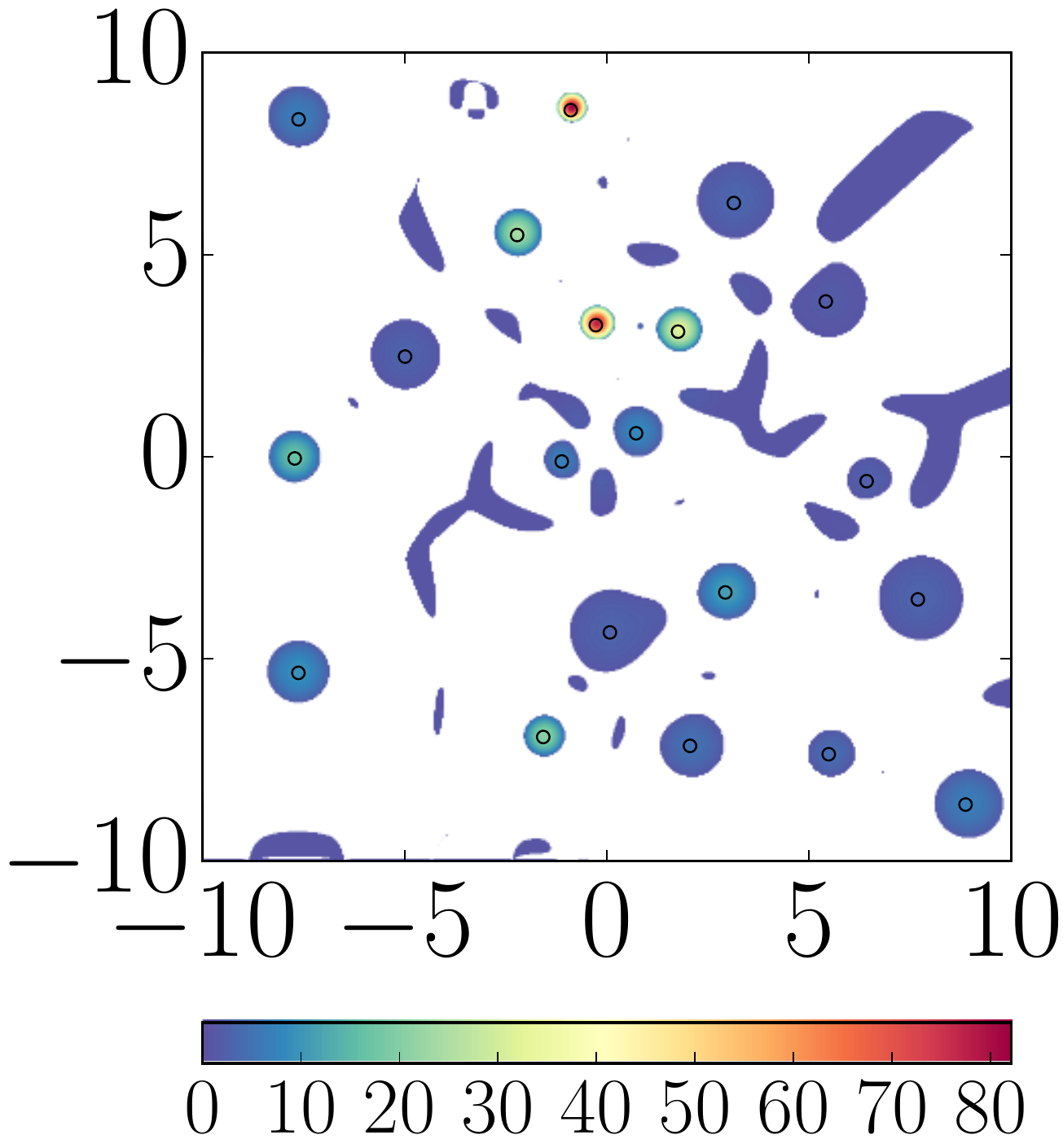}
          \put (-105,100){\makebox[0.05\linewidth][r]{(c)}}
      \end{minipage}
      \caption{Small open circles indicate the positions of 20 randomly distributed vortices. 
			(a) vortex detection via the $\lambda_{ci}$-criterion. The phenomena of vortex coalescence and vortex 
			erasing take place, respectively, in the first and fourth quadrants of the domain; 
			(b) vortex detection via the filtered $\lambda_\omega$-criterion, where all of the original vortices have been identified; 
			(c) inaccurate vortex detection via the unfiltered $\lambda_\omega$-criterion. The color bars represent the $\lambda_{ci}$ 
			and $\lambda_\omega$ fields in arbitrary units.}
     \label{fig:vcurv2}
    \end{figure}
    
    In order to illustrate the crucial importance of the filtering function, and the general improvement gained with the
  $\lambda_\omega$-criterion over the $\lambda_{ci}$-criterion, we show in Fig. \ref{fig:vcurv2} the analysis of
	a sample of 20 Lamb-Oseen vortices with varying radii and circulations, which are randomly distributed in a square domain. 
	While the use of the $\lambda_{ci}$-criterion is unable to avoid the merging of two of the vortices and the disappearance of 
	another one, all of the vortices are recovered with the use of $\lambda_\omega$-criterion, which approximately preserves their 
	original circular shapes. 
    
    If the filtering function were not used, many spurious regions would remain, as evidently pointed out in
  Fig. \ref{fig:vcurv2}c. One notices that a few spurious vortices have survived the screening of the $\lambda_\omega$
  criterion. We have to keep in mind, for proper applications of the $\lambda_\omega$-criterion, that although leading to  
  improvements, it is not free of errors, in the sense that probably any meaningful vortex identification method 
	will eventually break in the analysis of extreme (hopefully unrealistic) flow conditions.
	
  {\black{At this point, it is interesting to briefly discuss the relevance of the Lamb-Oseen vortex as a standard of analysis. The Burgers vortex \cite{batchelor} could be an alternative, having in mind that it is perhaps a more relevant structure for general turbulence modeling, as it has been suggested from turbulent wind tunnel experiments \cite{mouri}, and from the fact that it can play an important role in the theoretical understanding of intermittency in homogeneous and isotropic turbulence \cite{hata}. However, it turns out that if we are actually interested to focus on the performance of vortex identification methods, more than on modeling issues, the Lamb-Oseen vortex is by far the simpler and more convenient choice, leading to equivalent conclusions. More specifically, while the Burgers vortex is defined from four independent parameters (two strain rate eigenvalues, the asymptotic circulation, and its core radius), the Lamb-Oseen vortex is completely determined by its asymptotic circulation and core radius parameters. It is not difficult to show that while variations of the two extra-parameters for the Burgers vortex are rigorously harmless in the context of the $\lambda_\omega$-criterion, they may affect the performance of the $\lambda_{ci}$-criterion in unwanted ways, due to the presence of additional shearing.}}
  
      So far, all of our arguments have been based on the inspection of a few representative analytical vortex configurations.
  Of course, more is needed to validate the $\lambda_\omega$-criterion as a reliable tool. This is our next step, to be
  carried out with the help of extensive Monte Carlo simulations, where we consider, instead, discretized velocity derivatives 
	for the analysis of large ensembles of synthetic many-vortex systems.

\section{Monte-Carlo Study}\label{sec:montecarlo}
  
    To address a comparative study of accuracy for the $\lambda_\omega$ and the $\lambda_{ci}$ criteria, we run Monte Carlo
  tests for large ensembles, where in each sample vortices are randomly distributed over the area of a square domain. The
  velocity field over a discretized grid is recorded and the two vortex identification criteria are applied to investigate 
	how they perform in detecting and also in recovering the properties (circulation, radius, and position) of the original 
	vortices.
  
    In all of the synthetic samples, evaluations of the velocity gradient, pseudo-velocity and pseudo-velocity gradient have
  been done with five-point weighted finite diferences, which in the worst situations (the ones involving three derivatives of
  the velocity field) have precision of $O(\delta^2)$ in the grid spacing $\delta$. Integrations rely on bilinear
  interpolations, which are also precise to $O(\delta^2)$. The connected regions where vortices are detected are individualized
  in the grid with the use of a connected component labeling algorithm \cite{shapiro}. For each connected region 
  $R_k$ ($k=1,2,...$) we compute
    \begin{linenomath*} \bea
      &&A_k \equiv \pi \bar r^2 = \int_{R_k} d^2 \vec r \ , \ \label{Ak} \\
      &&\bar \Gamma_k = \int_{R_k} \omega(\vec r ) d^2 \vec r  \ , \ \label{gammak} \\
      &&(x_k,y_k) \equiv \frac{\int_{R_k} (x,y) \ \omega^2(\vec r) d^2 \vec r  }{ \int_{R_k}  \omega^2(\vec r) d^2 \vec r} \ . \ \label{rk}
    \eea \end{linenomath*} 
    Eqs. (\ref{Ak}) and (\ref{gammak}) allow us to infer, respectively, with the help of Eq. (\ref{alpha-beta}), the real
  radius $r_k$ and circulation $\Gamma_k$ vortex parameters. While for the $\lambda_{ci}$-criterion $\alpha$ and $\beta$ are
  already known from Eqs. (\ref{alpha}) and (\ref{beta}), a similar and straightforward analysis for the $\lambda_
  \omega$-criterion yields the analougous pair of parameters $(\alpha,\beta) = (\sqrt{2},1/(1-1/\sqrt{e})) \simeq (1.41 , 2.54)$. 
	Additionally, Eq. (\ref{rk}) gives the ``center of enstrophy" coordinates for the position of the identified vortex. {\black{The $\alpha$ parameter for vortex core radius conversion is, in the $\lambda_\omega$-criterion, about 1.6 times greater than the one for the $\lambda_{ci}$-criterion. This is a casual, but nevertheless very helpful fact, since it improves the resolution of the detected structures, as it could have already been noticed from the former's section results.}}
    
    We have worked, for a set of flow configurations of interest, with $N = 10^5$ Monte Carlo samples, each one containing $N_v=20$
  randomly distributed vortices, on a $[-9,9]^2$ square (arbitrary length scale). The velocity field is exactly defined at 
  the sites of a $N_x \times N_y = 200^2 $ grid, which models the square box $[-10,10]^2$. When sampled, vortex centers are
  always separated by distances greater than $1.2$ times the sum of their radii \cite{charita}. Circulations and vortex radii 
	are sampled with uniform random distribution in the domains given, respectively, by $1 \leq |\Gamma| \leq 20$ 
	(or $-20 \leq \Gamma \leq -1$) and $0.5 \leq r_c \leq 1.5$. 
	    \begin{table}[ht]
      \centering
      \begin{tabular}{|l|l|}                                   \hline
       Number of Samples & $N = 10^5$                       \\ \hline
       Number of Vortices/Sample  & $N_v = 20$              \\ \hline
       System's Dimensions  &  $(L_x,L_y) = (20,20)$        \\ \hline 
       Vortex Positions  & $-9 \leq x,y \leq 9$             \\ \hline
       Grid Size & $200 \times 200$                         \\ \hline
       Vortex Circulations  & $\Gamma \in \pm [1,20]$       \\ \hline
       Vortex Core Radii & $r_c \in [0.5,1.5]$              \\ \hline 
       Acceptance Cutoff & $\Gamma_0=0.5$           \\ \hline
       Vortex Pair Separation & $d_{ij}> 1.2 \times (r_{ci}+r_{cj})$ \\ \hline
      \end{tabular}
      \caption{General definitions for the Monte Carlo simulations of the synthetic many-vortex two-dimensional systems.}
      \label{tab:mc1}
    \end{table}
    
	As a way to get rid of spurious vortices, we furthermore prescribe that $R_k$ is accepted as vortex only if 
	$|\Gamma_k| \geq \Gamma_0$, for some small circulation scale $\Gamma_0$. Note that this cutoff prescription is conceptually distinct 
	from the imposition of a threshold, where the main worry is not exactly on the existence of spurious vortices as individual objects, 
	but on specific - noise contaminated - regions of the flow. The circulation cutoff for vortex acceptance is defined as 
	$\Gamma_0 = 0.5$. The Monte Carlo simulation definitions are summarized in Table \ref{tab:mc1}.
    
    Motivated by the distribution of spanwise vortices observed in streamwise/wall normal planes of turbulent boundary layers 
		\cite{adrian_mein_tom,camussi,wu,jeff,herpin_a,herpin} we have considered, in our Monte Carlo simulations, five 
		distinct flow patterns, denoted by latin capital letters 
	  from A to E, described in Table \ref{flow_regimes}.
  
    To define the weak and strong shear regimes referred to in Table \ref{flow_regimes}, observe, as it can be derived from
  (\ref{det}), that a vortex with peak vorticity $\omega_p$ disappears from swirling strength detection if the vorticity of
  the background shear is $|\bar \omega| > |\omega_p|/2$, with $-\bar \omega \omega_p < 0$. Recalling that for a Lamb-Oseen
  vortex, $\omega_p = \Gamma/ \pi r_c^2$, and that in our Monte Carlo samples, $|\Gamma| \leq 20$ and $|r_c-1| \leq 0.5$, we
  take, as representative parameters, $\Gamma = 10$ and $r_c=1$, which lead to $\omega_p/2 \simeq 1.6$. Weak and strong
  regimes are then defined as the ones which have {\black{background velocity field components}} given, respectively, 
  by $(v_x,v_y) = (0.35y,0)$ and $(v_x,v_y) = (1.6y,0)$. Note that for flow patterns with either weak or strong background shear,
	the background vorticity is negative. 
  
    In the following, we organize the large lists of input and output vortex parameters (circulation, radius, and position) in
  the form of histograms that indicate how the $\lambda_{ci}$ and $\lambda_\omega$ vortex identification criteria perform in
  the automated analysis of Monte Carlo ensembles.
  
    \begin{table}[ht]
      \centering
      \begin{tabular}{|l|l|l|} \hline
       Flow Pattern & Vortex Circulations & Background Shear \\ \hline
       A & $1 \leq |\Gamma| \leq 20$ & No Background        \\ \hline
       B & $1 \leq |\Gamma| \leq 20$ & Weak                 \\ \hline
       C & $1 \leq |\Gamma| \leq 20$ & Strong               \\ \hline
       D & $-20 \leq \Gamma \leq -1$ & No Background        \\ \hline
       E & $-20 \leq \Gamma \leq -1$ & Strong               \\ \hline
      \end{tabular}
      \caption{The five flow patterns considered in our Monte Carlo simulations.}
      \label{flow_regimes}
    \end{table}
    
\begin{figure}[ht]
      \begin{minipage}{0.49\linewidth}
        \includegraphics[width= \linewidth]{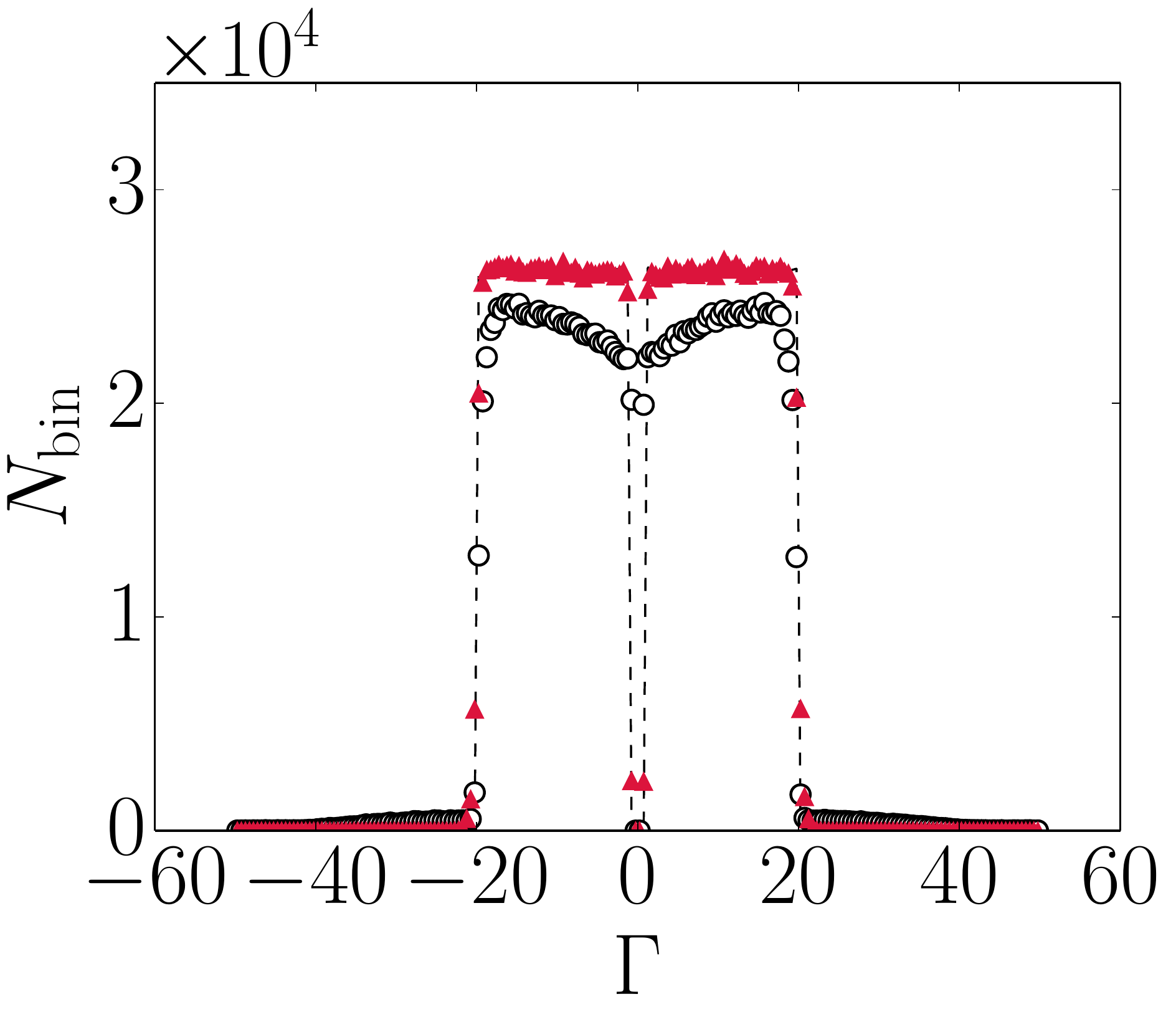}
         \put (-77,77){\makebox[0.05\linewidth][r]{(a)}}
      \end{minipage}
     \begin{minipage}{0.49\linewidth}
        \includegraphics[width=\linewidth]{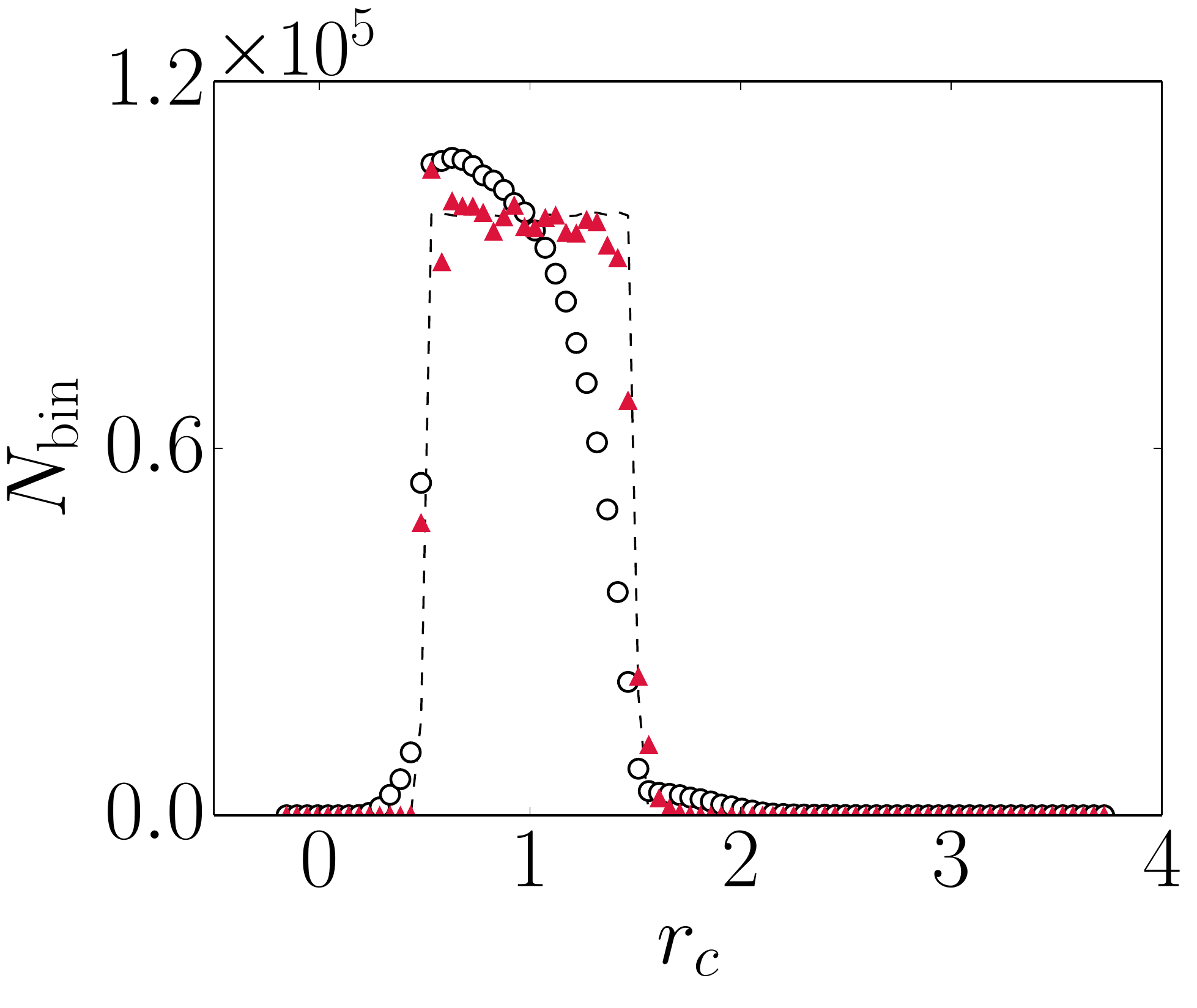}
      \put (-77,75){\makebox[0.05\linewidth][r]{(b)}}
     \end{minipage}
      \begin{minipage}{0.49\linewidth}
        \includegraphics[width=\linewidth]{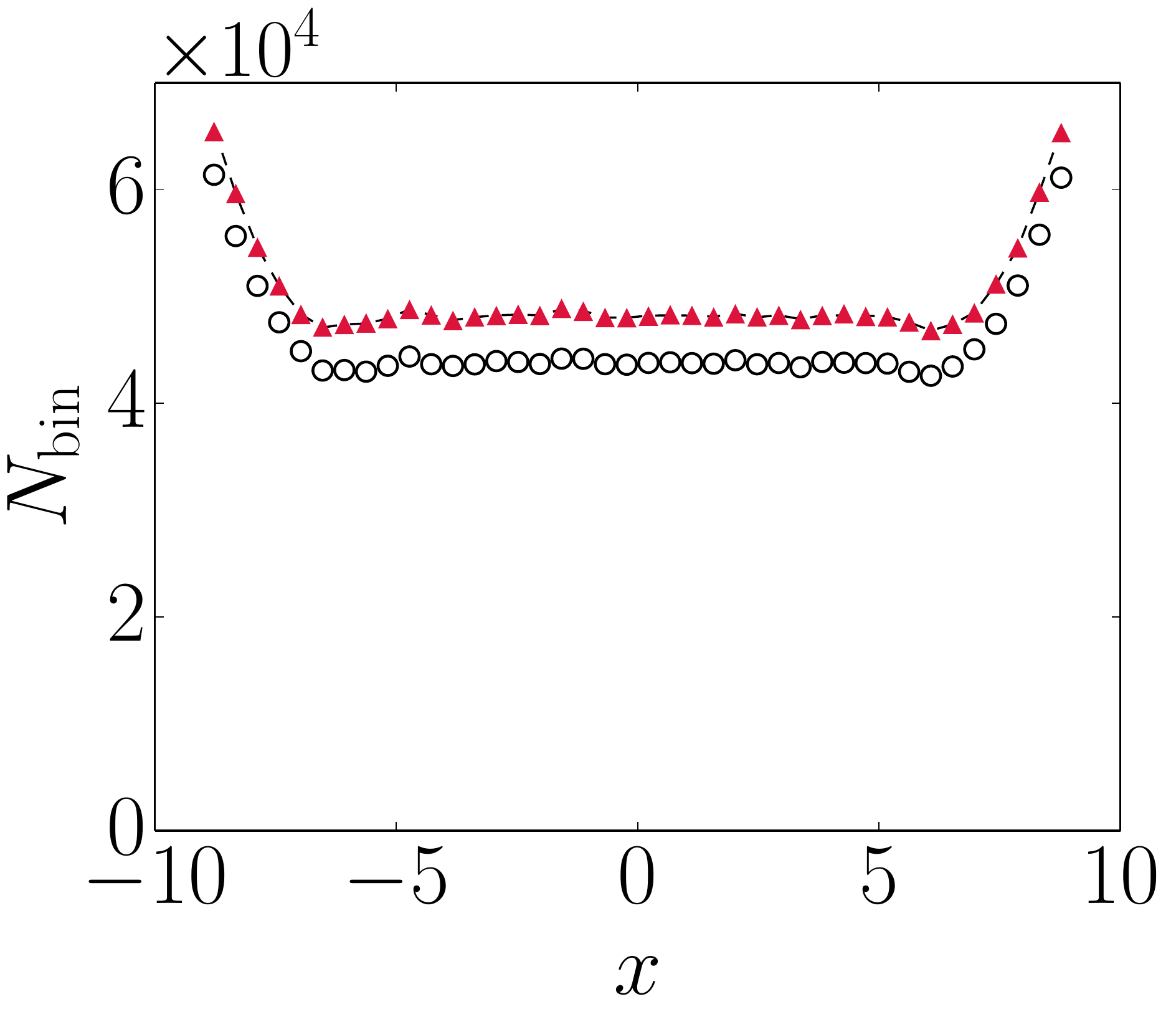}
        \put (-77,75){\makebox[0.05\linewidth][r]{(c)}}
      \end{minipage}
      \begin{minipage}{0.49\linewidth}
        \includegraphics[width=\linewidth]{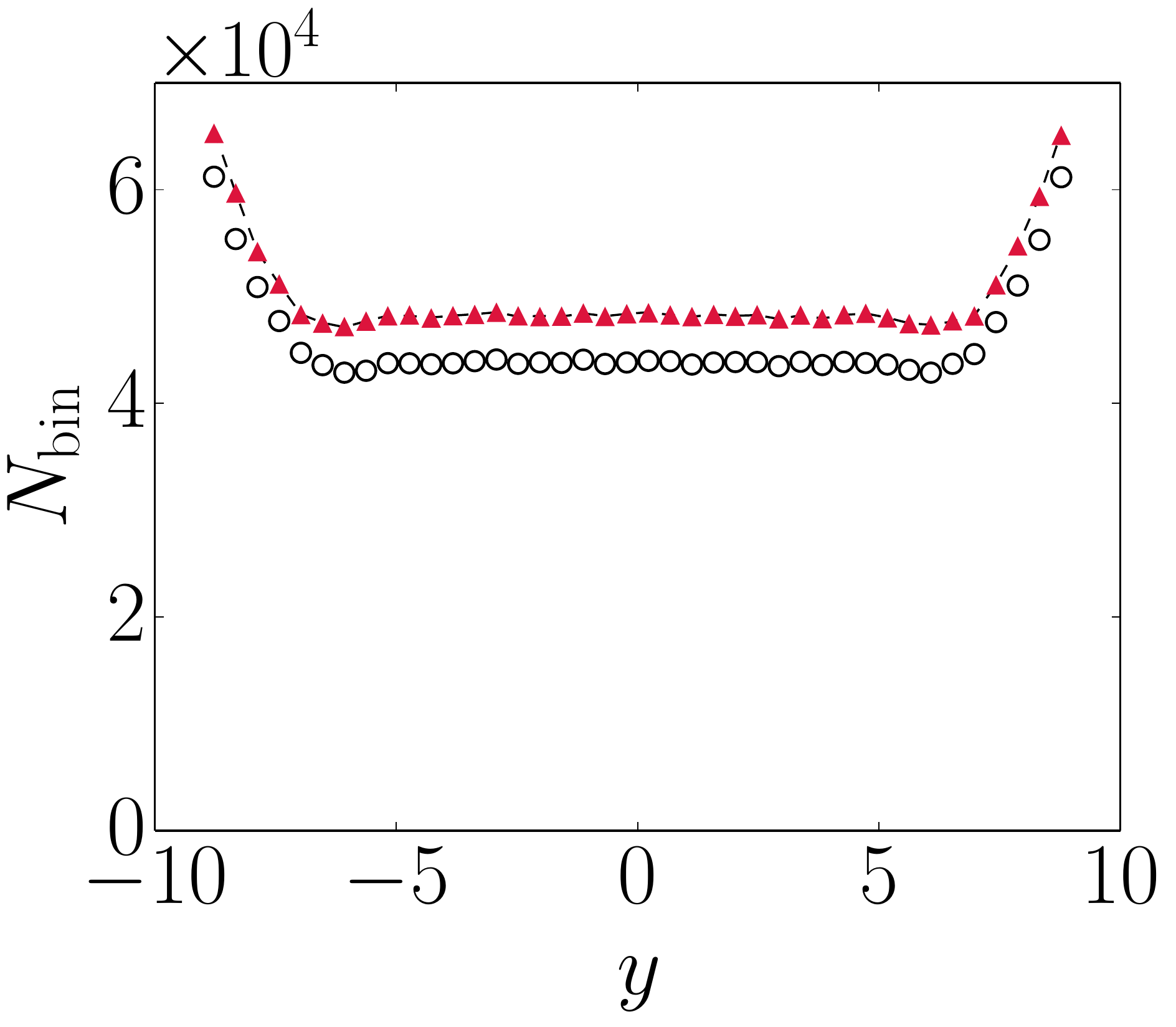}
        \put (-77,75){\makebox[0.05\linewidth][r]{(d)}}
      \end{minipage}
      \caption{Flow pattern A. Histograms for performance comparison between the $\lambda_\omega$-criterion (triangles) and the
               $\lambda_{ci}$-criterion (circles), in the evaluation of vortex parameters. (a) circulations; (b) radii; 
               (c) $x$ coordinates; (d) $y$ coordinates. The dashed lines are the histograms for the input data.}
               \label{fig:mc2}      
    \end{figure}

    Results for the flow pattern A are given in Fig. \ref{fig:mc2}. The $\lambda_\omega$-criterion has an excellent
  performance, while the $\lambda_{ci}$-criterion is mainly affected by vortex coalescence, which explains why the counting
  is reduced for the larger vortices, and why so many non-existent structures with circulation $|\Gamma| > 20$ have been
  artificially produced. One can note, from Figs. \ref{fig:mc2}c and \ref{fig:mc2}d that there are boundary effects in the 
	distribution of vortices. This is actually due to the fact that by definition they ``avoid each other" in the bounded 
	domain. The same feature is observed in all of the other flow patterns.
    
\begin{figure}[ht]
      \begin{minipage}{0.49\linewidth}
        \includegraphics[width= \linewidth]{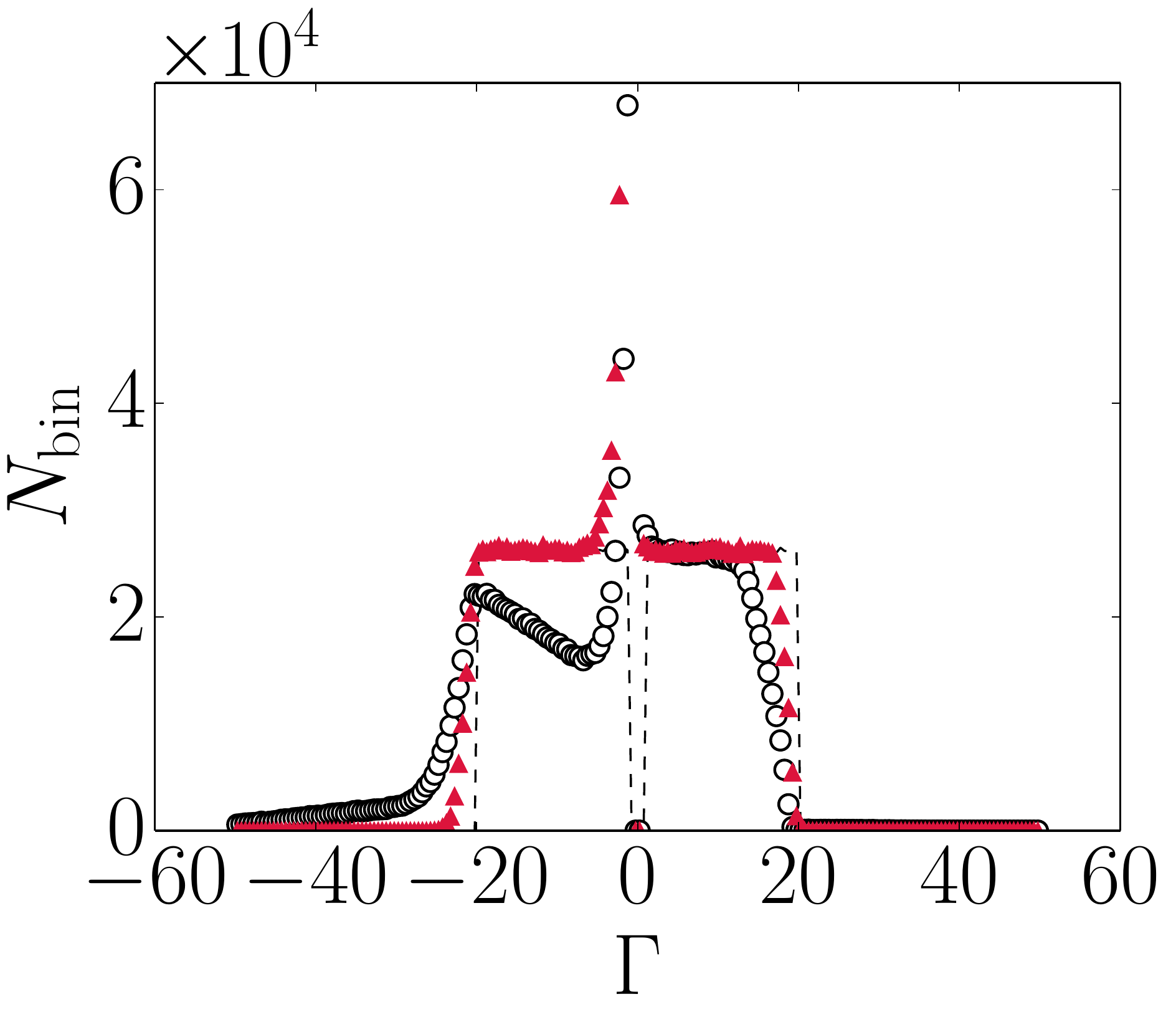}
        \put (-77,77){\makebox[0.05\linewidth][r]{(a)}}
      \end{minipage}
      \begin{minipage}{0.49\linewidth}
        \includegraphics[width=\linewidth]{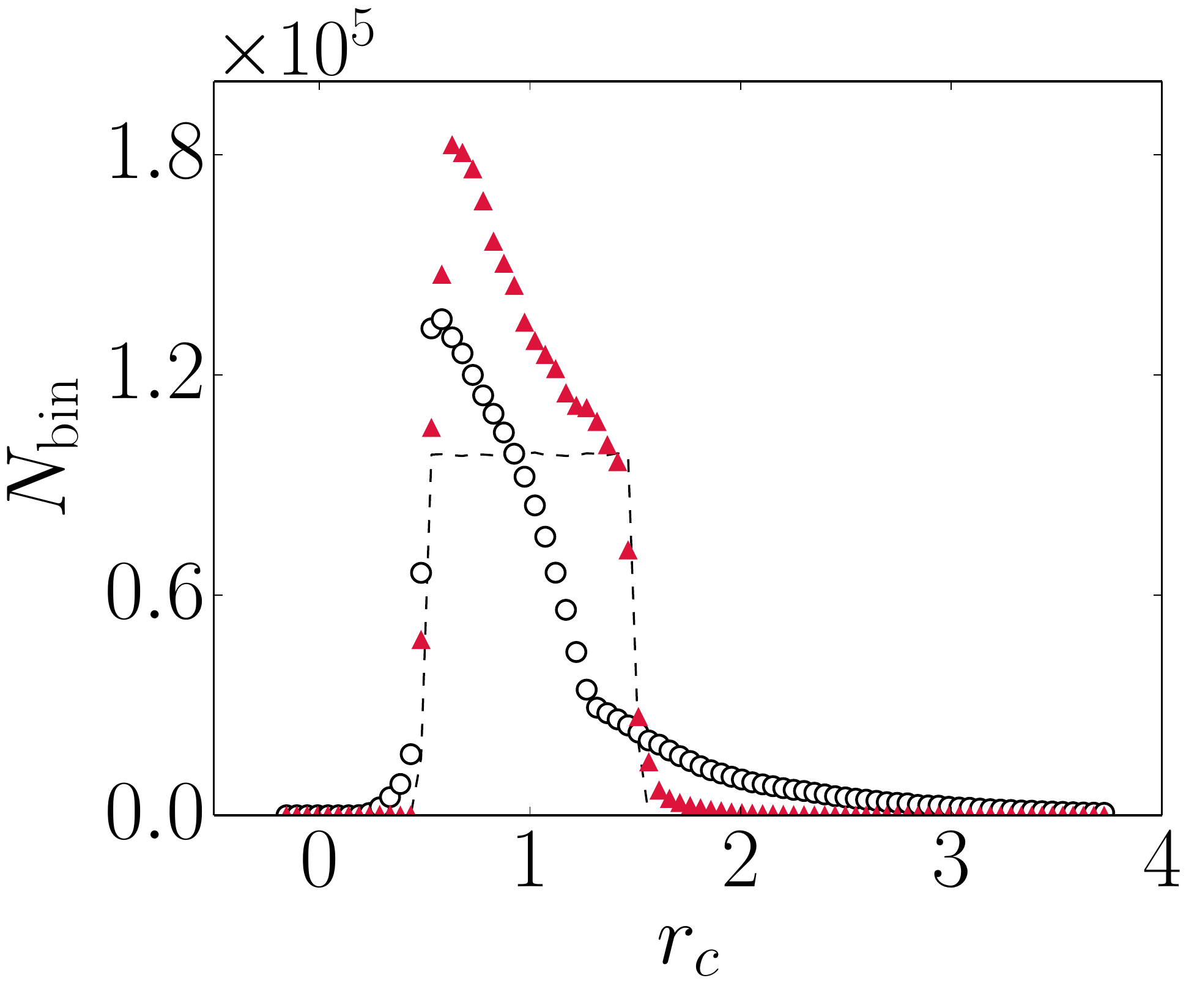}
        \put (-77,75){\makebox[0.05\linewidth][r]{(b)}}
      \end{minipage}
      
      \begin{minipage}{0.49\linewidth}
        \includegraphics[width=\linewidth]{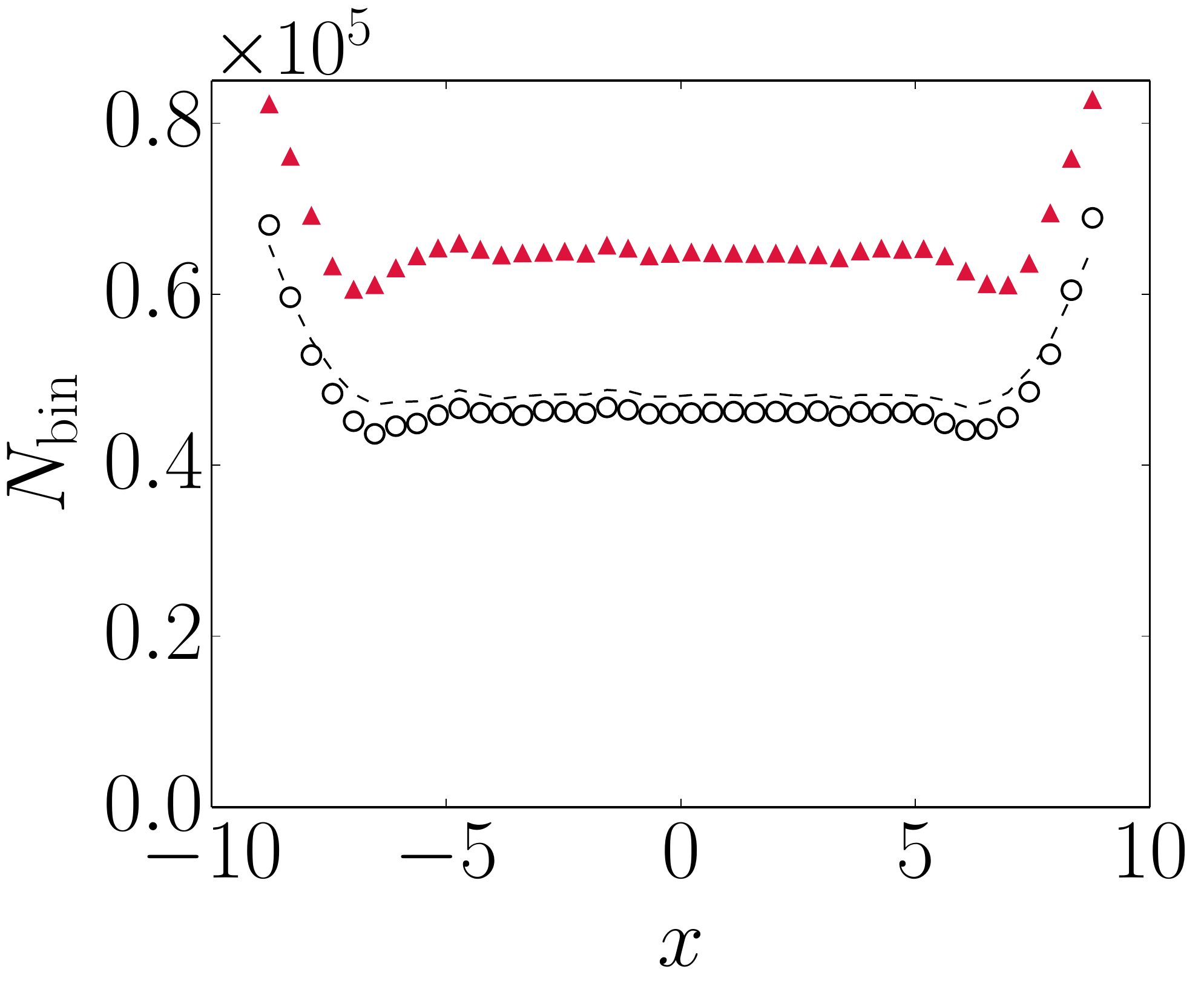}
         \put (-77,75){\makebox[0.05\linewidth][r]{(c)}}
      \end{minipage}
      \begin{minipage}{0.49\linewidth}
        \includegraphics[width=\linewidth]{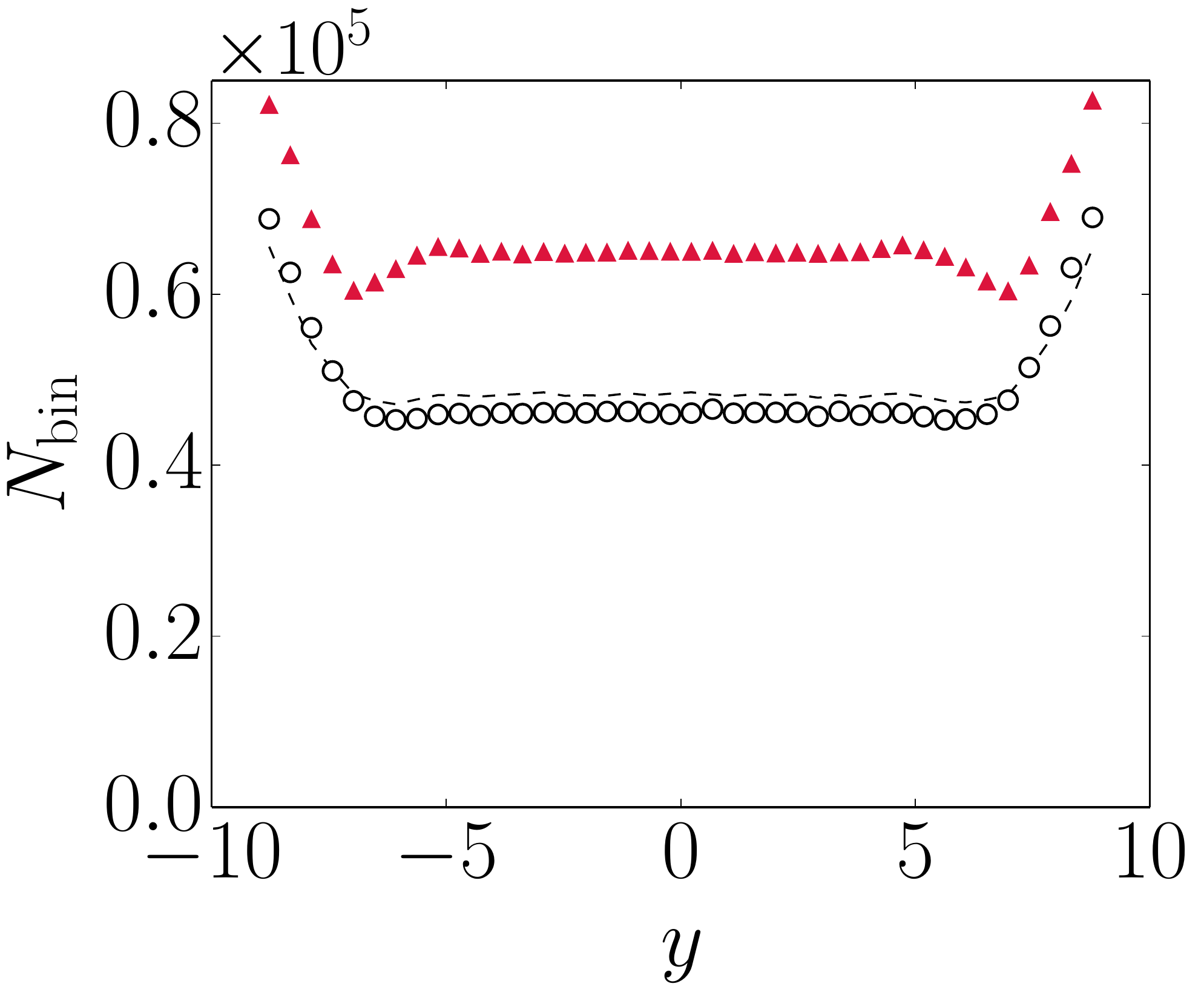}
         \put (-77,75){\makebox[0.05\linewidth][r]{(d)}}
      \end{minipage}
      \caption{Flow pattern B. All the rest as in the caption of Fig. \ref{fig:mc2}.}
               \label{fig:mc3}      
    \end{figure}

\begin{figure}[ht]
      \begin{minipage}{0.49\linewidth}
        \includegraphics[width= \linewidth]{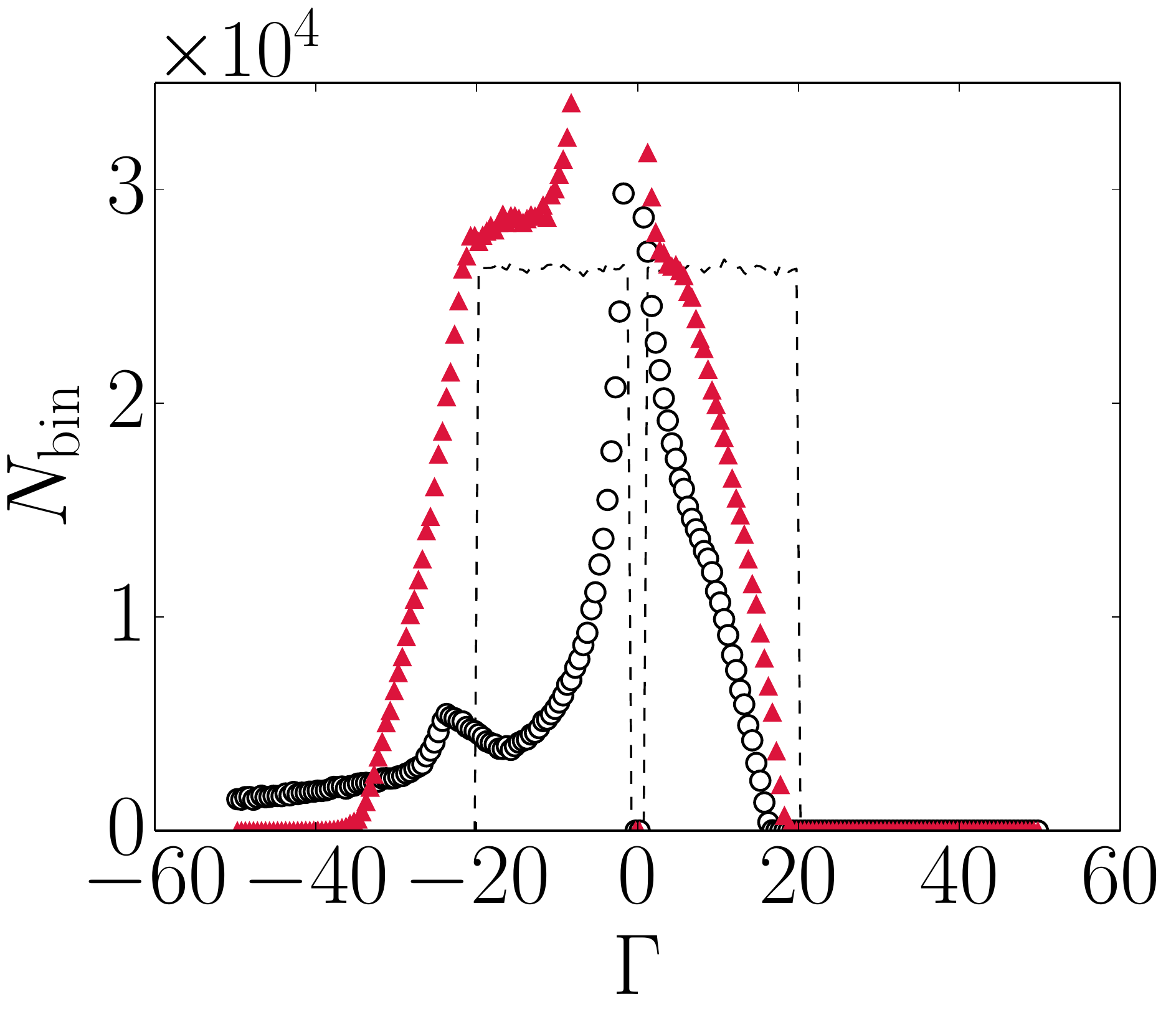}
        \put (-77,77){\makebox[0.05\linewidth][r]{(a)}}
      \end{minipage}
      \begin{minipage}{0.49\linewidth}
        \includegraphics[width=\linewidth]{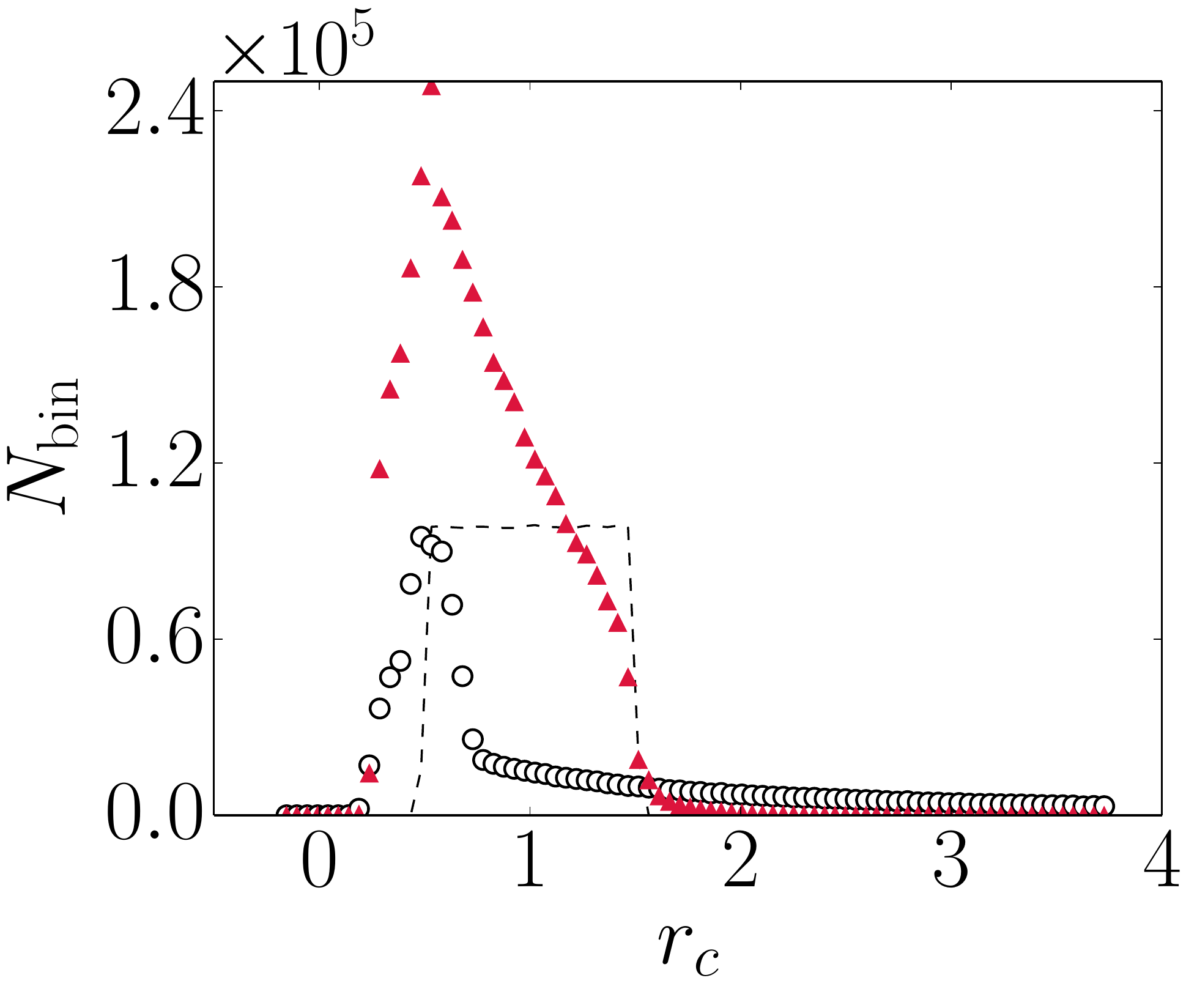}
        \put (-77,75){\makebox[0.05\linewidth][r]{(b)}}
      \end{minipage}
      
      \begin{minipage}{0.49\linewidth}
        \includegraphics[width=\linewidth]{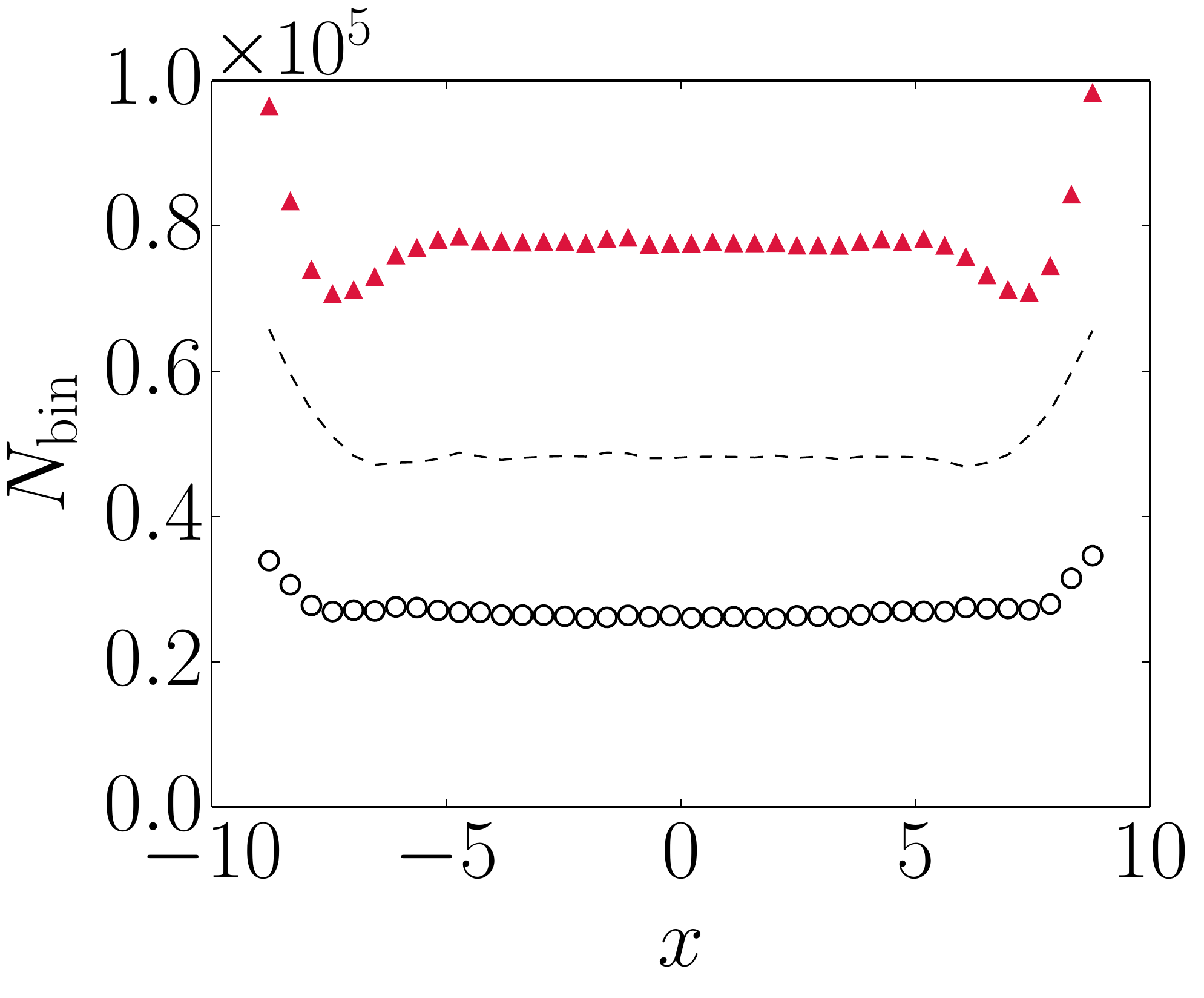}
        \put (-77,75){\makebox[0.05\linewidth][r]{(c)}}
      \end{minipage}
      \begin{minipage}{0.49\linewidth}
        \includegraphics[width=\linewidth]{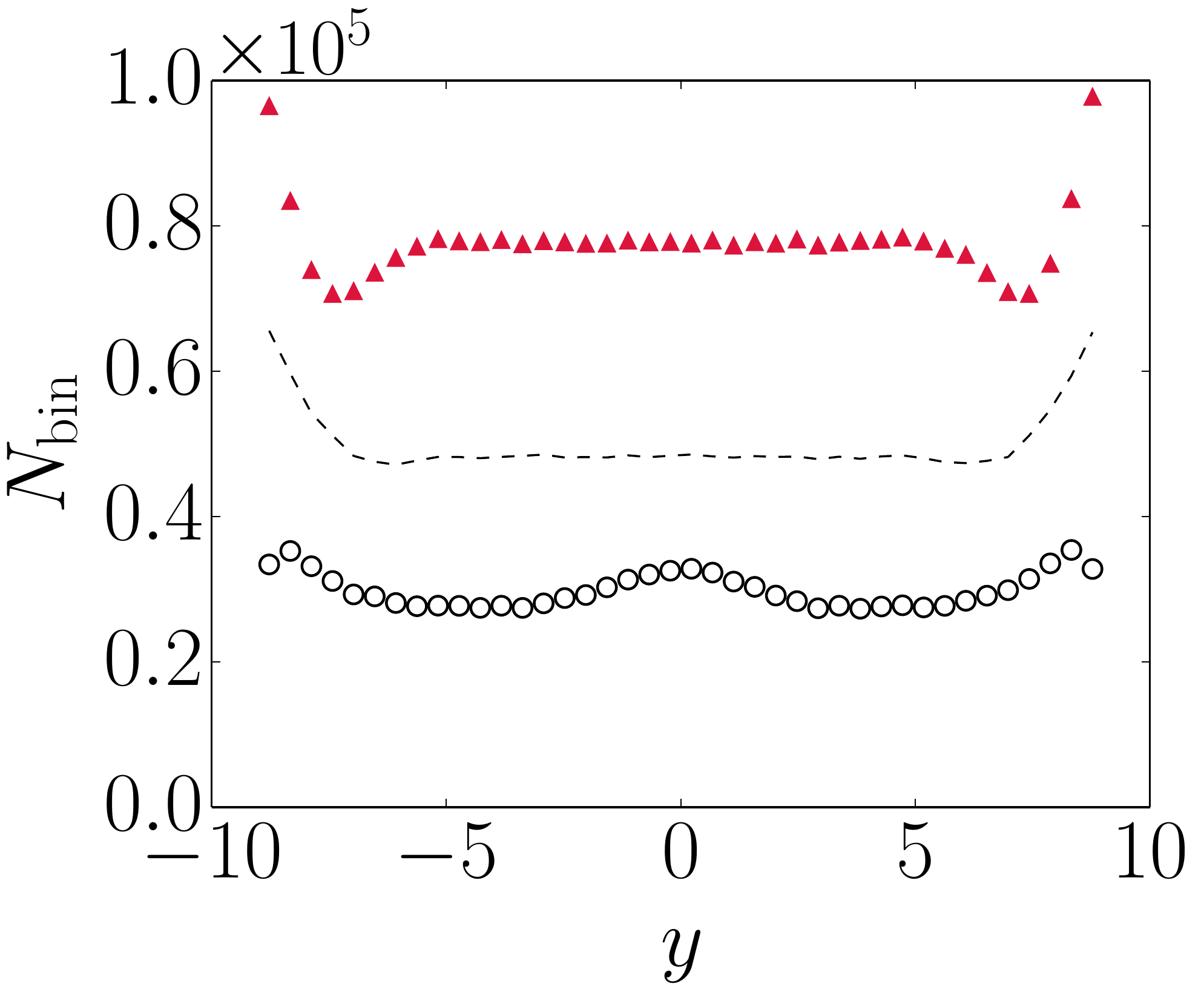}
        \put (-77,75){\makebox[0.05\linewidth][r]{(d)}}
      \end{minipage}
      \caption{Flow pattern C. All the rest as in the caption of Fig. \ref{fig:mc2}.}
               \label{fig:mc4}      
    \end{figure}

    For the flow patterns B and C,  which have weak and strong background shear, respectively, the related histograms are given
  in Figs. \ref{fig:mc3} and \ref{fig:mc4}. In the flow pattern B, as shown in Fig. \ref{fig:mc3}, the $\lambda_{ci}$-criterion
  yields a small and uniform supression of vortices in the samples, but the circulation and radius countings are actually
  close to the ones found for the flow pattern A. The $\lambda_\omega$-criterion is still the better choice, despite the fact that
	vortex counting is strongly affected by the addition of spurious vortices of small circulation and artificial structures like the 
	stripes previously observed in Fig. \ref{fig:vcurv1}d. Actually, as we will see in a moment, the $\lambda_\omega$-criterion is 
	able to capture the input vortices in this case, which are more precisely counted when background shearing effects are 
	removed.
  
		    Driving our attention now to the flow pattern C, Fig. \ref{fig:mc4} tells us that both the $\lambda_{ci}$ and the
  $\lambda_\omega$ criteria perform badly. It turns out that strong external shearing introduces, in general, relevant effects
  in vortex identification that demand improvement. 
	
\begin{figure}[ht]
      \begin{minipage}{0.49\linewidth}
        \includegraphics[width=\linewidth]{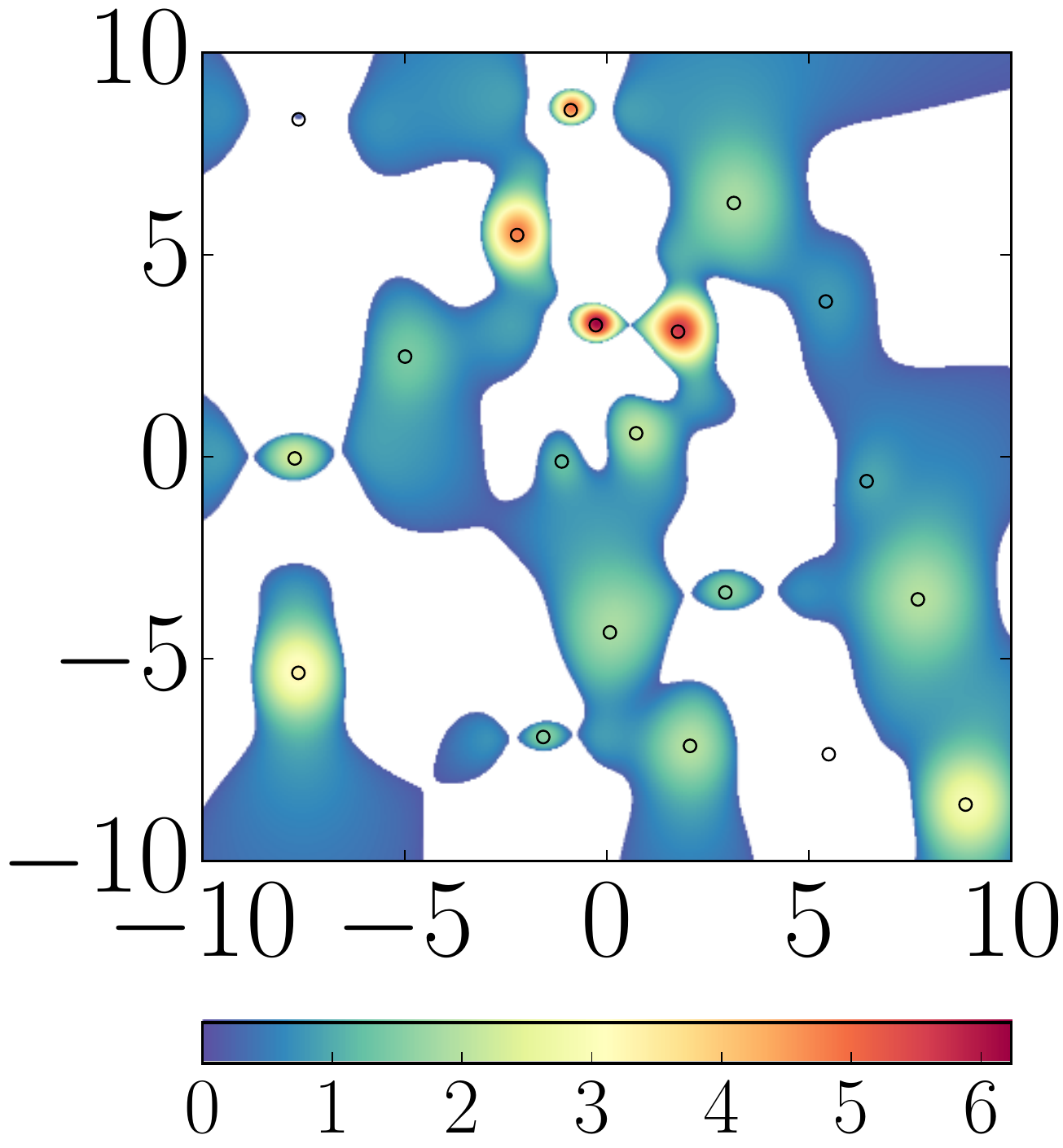}
          \put (-105,100){\makebox[0.05\linewidth][r]{(a)}}
      \end{minipage}
      \begin{minipage}{0.49\linewidth}
        \includegraphics[width=\linewidth]{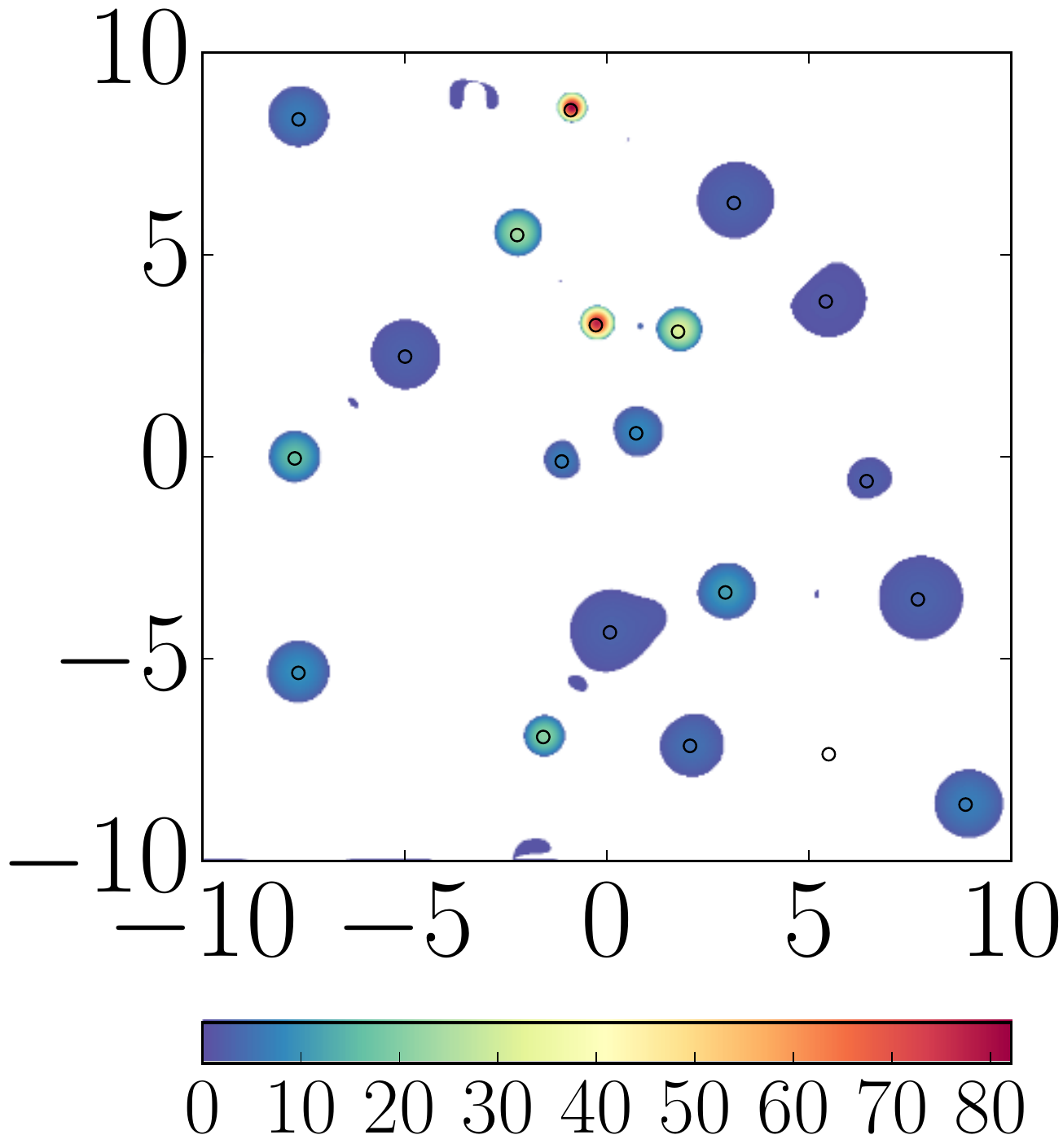}
          \put (-105,100){\makebox[0.05\linewidth][r]{(b)}}
      \end{minipage}
      \caption{A sample of 20 vortices - the same as in Fig. \ref{fig:vcurv2}, now in the presence of strong background shear
               (flow pattern C), investigated through the (a) swirling strength and (b) the vorticity curvature fields. The color bars
			represent the $\lambda_{ci}$ and $\lambda_\omega$ fields in arbitrary units.}
      \label{fig:mc5}
    \end{figure}

	The visualization of a typical Monte Carlo sample of the flow pattern C is
  given in Fig. \ref{fig:mc5}, where we see, as a dominant effect, coalescence percolation of flaps and vortices in the
  application of the $\lambda_{ci}$-criterion. On the other hand, the image associated to the $\lambda_\omega$-criterion looks
  qualitatively different, and although most of the input vortices have been retrieved from the sample, they are surrounded by
  several spurious structures that can spoil histograms, like the ones we consider here. 
		
    In order to deal with the shortcomings associated with shearing/vorticity backgrounds, we put forward an improved computational
    strategy, based on the subtraction of the background velocity field, sample by sample, from individual velocity field
  realizations. This is, of course, nothing more than the method of Reynolds decomposition, {\black{which, actually, has been already employed in previous studies of coherent structure identification, as in Ref. \cite{tanashi}.}} The idea, thus, is to revisit our
  previous analyses, by just replacing the original velocity field {\black{components}} $v_i (\vec r) $ by its fluctuations 
  over the background, that is, 
    \begin{linenomath*} \be
      \delta v_i(\vec r) = v_i(\vec r) - \langle v_i(\vec r) \rangle \ , \ \label{back_remov}
    \ee \end{linenomath*}
    where $\langle v_i(\vec r) \rangle$ stands for the expectation value of the velocity field taken over the ensemble of
  configurations. Furthermore, as an important prescription, in order to avoid additional spurious effects, we assign a given point 
  in the flow to a vortex if it is detected in the vortex identification screening carried out with and without the 
  background subtraction procedure.

    We compare, in the next six sets of histograms, the performance of the $\lambda_{ci}$ and the $\lambda_\omega$ criteria, both 
		with background subtraction procedure for the flow patterns B and C, while analogous comparisons are done for the 
		flow patterns D and E, with and without background subtraction. We do not report here the additional background
		subtraction analysis of the flow pattern A, since (as expected) we find that both criteria work again as in 
		Fig. \ref{fig:mc2}, due to the fact that the balanced mixing of vortices with positive and negative circulations produces 
		a very small background.

    \begin{figure}[ht]
      \begin{minipage}{0.49\linewidth}
        \includegraphics[width=\linewidth]{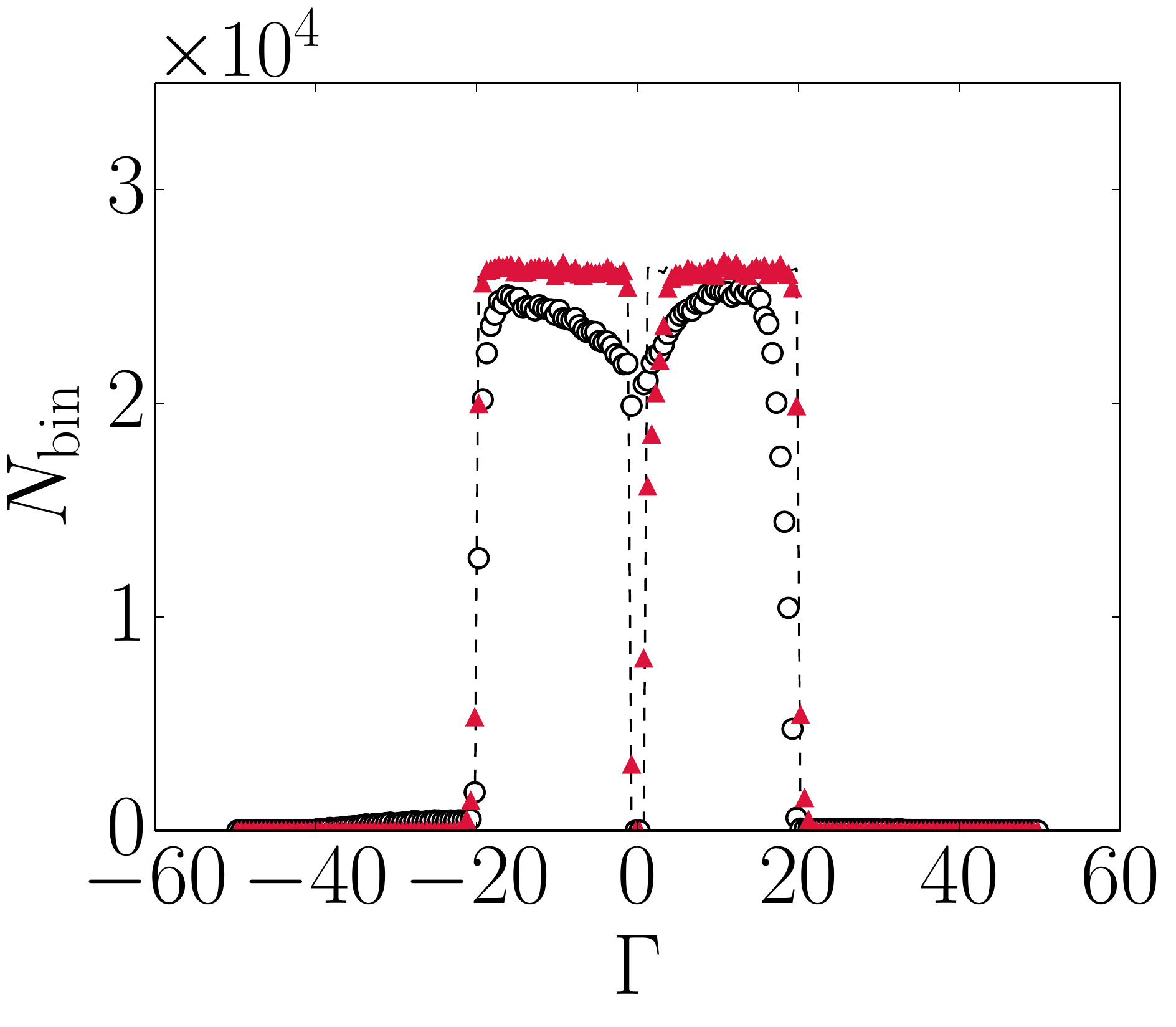}
        \put (-77,77){\makebox[0.05\linewidth][r]{(a)}}
      \end{minipage}
      \begin{minipage}{0.49\linewidth}
        \includegraphics[width=\linewidth]{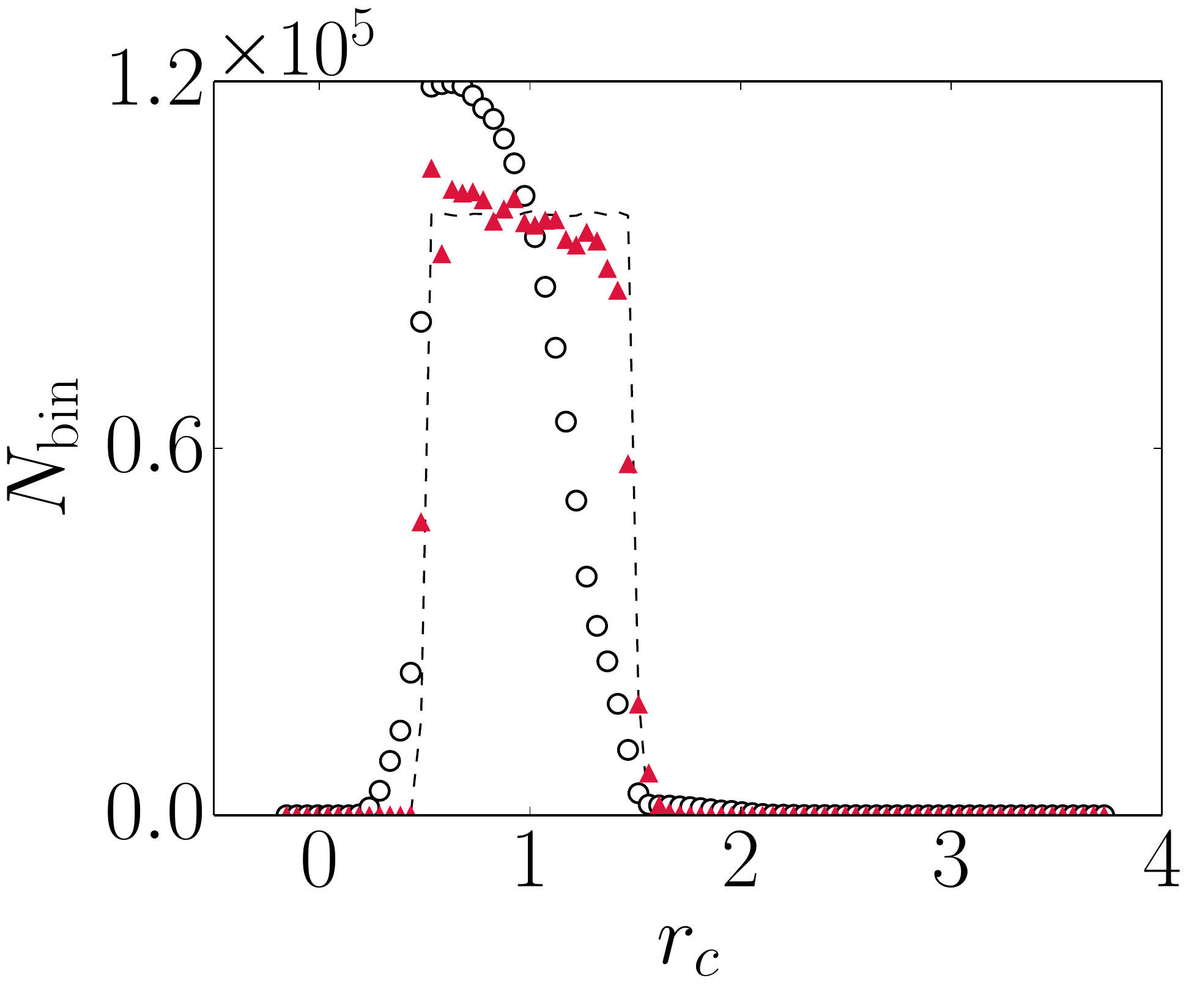}
        \put (-77,75){\makebox[0.05\linewidth][r]{(b)}}
      \end{minipage}
      
      \begin{minipage}{0.49\linewidth}
        \includegraphics[width=\linewidth]{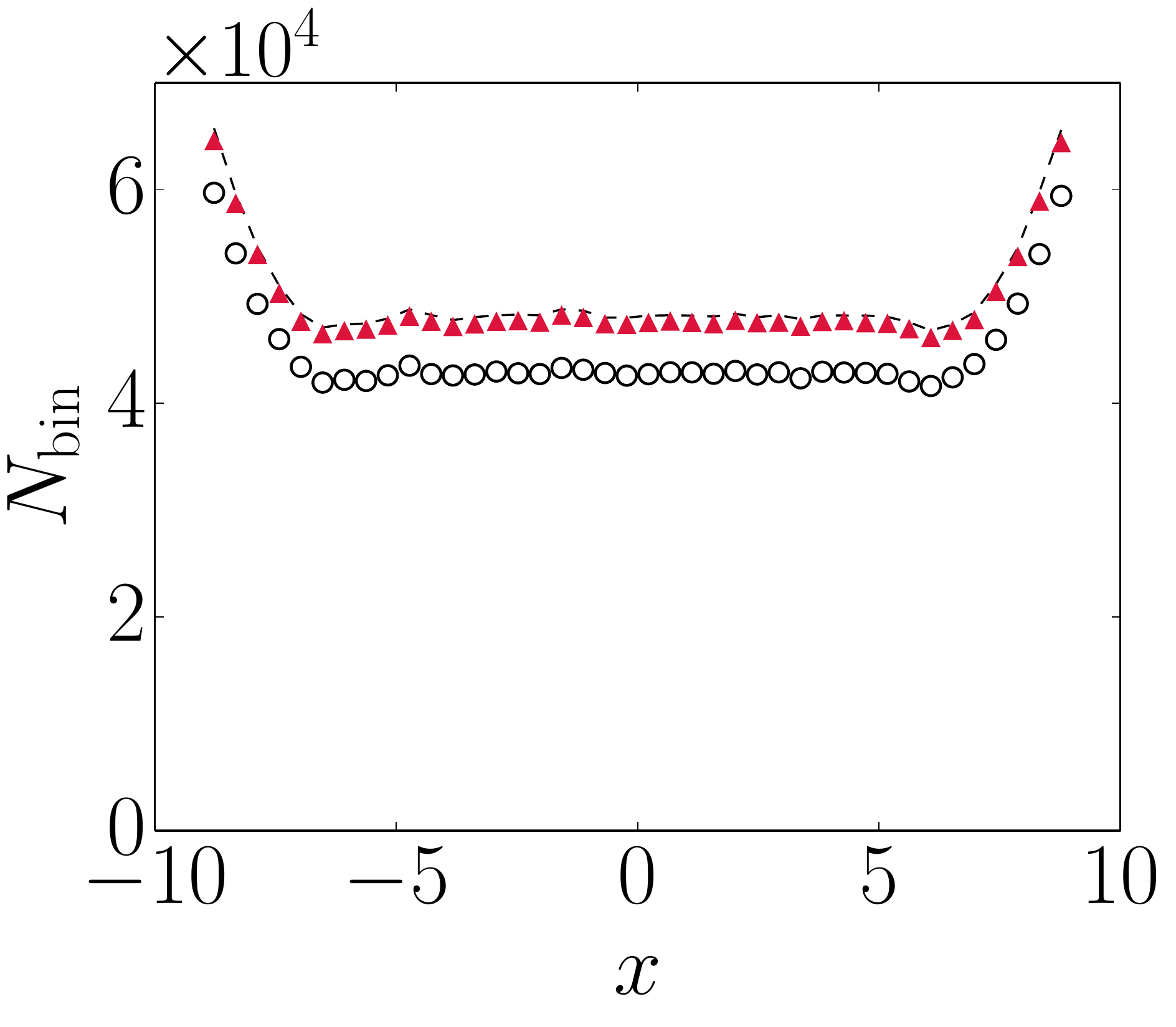}
        \put (-77,75){\makebox[0.05\linewidth][r]{(c)}}
      \end{minipage}
      \begin{minipage}{0.49\linewidth}
        \includegraphics[width=\linewidth]{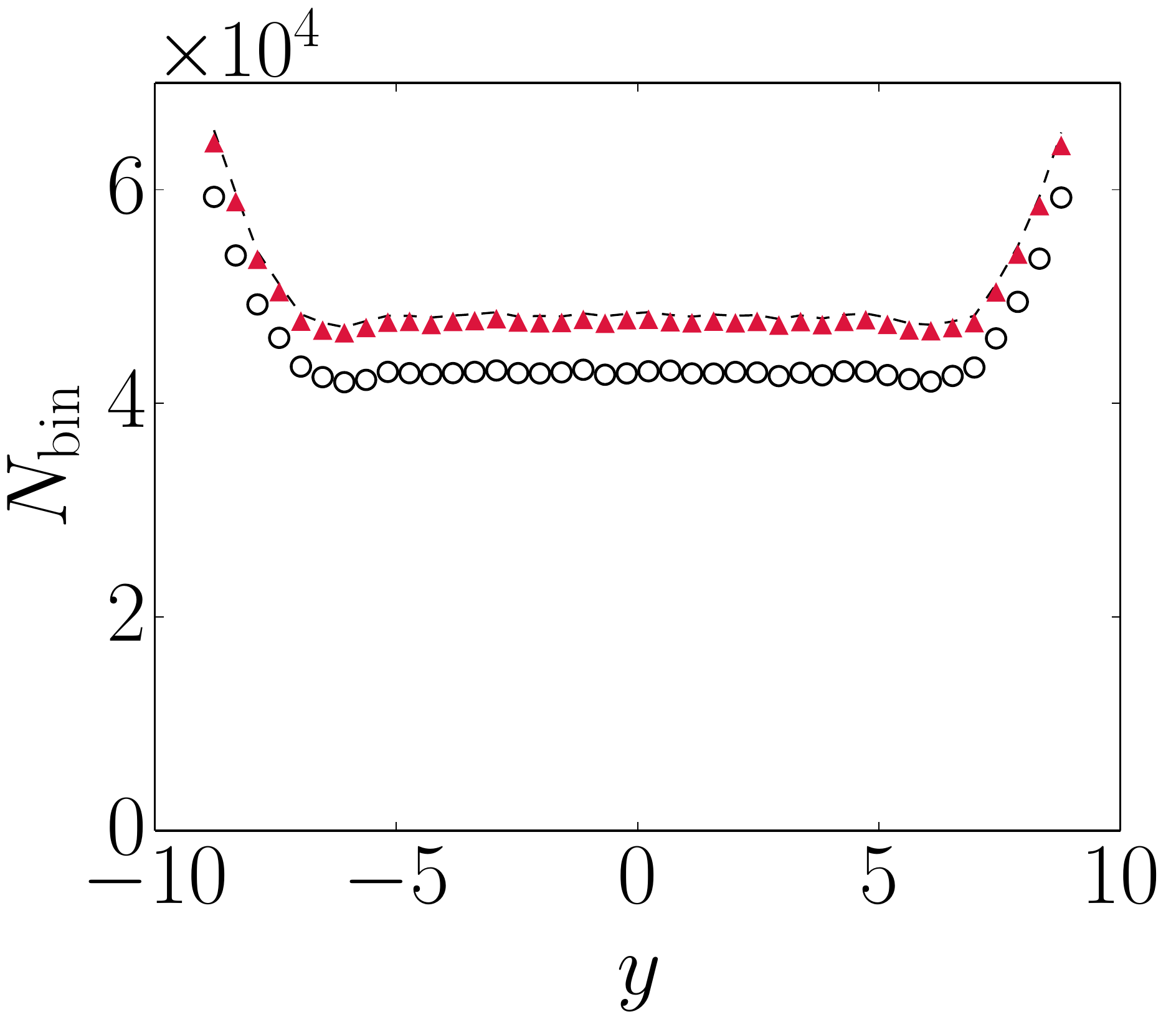}
        \put (-77,75){\makebox[0.05\linewidth][r]{(d)}}
      \end{minipage}
      \caption{Analysis of flow pattern B, with background subtraction. All the rest as in the caption of Fig. \ref{fig:mc2}.}
               \label{fig:b5}      
    \end{figure}
		    The weak shear case, flow pattern B, is given in Fig. \ref{fig:b5}, where both the $\lambda_{ci}$ and $\lambda_\omega$ criteria
		are noted to improve in their performances, with a clear advantage for the latter.
		    \begin{figure}[ht]
      \begin{minipage}{0.49\linewidth}
        \includegraphics[width=\linewidth]{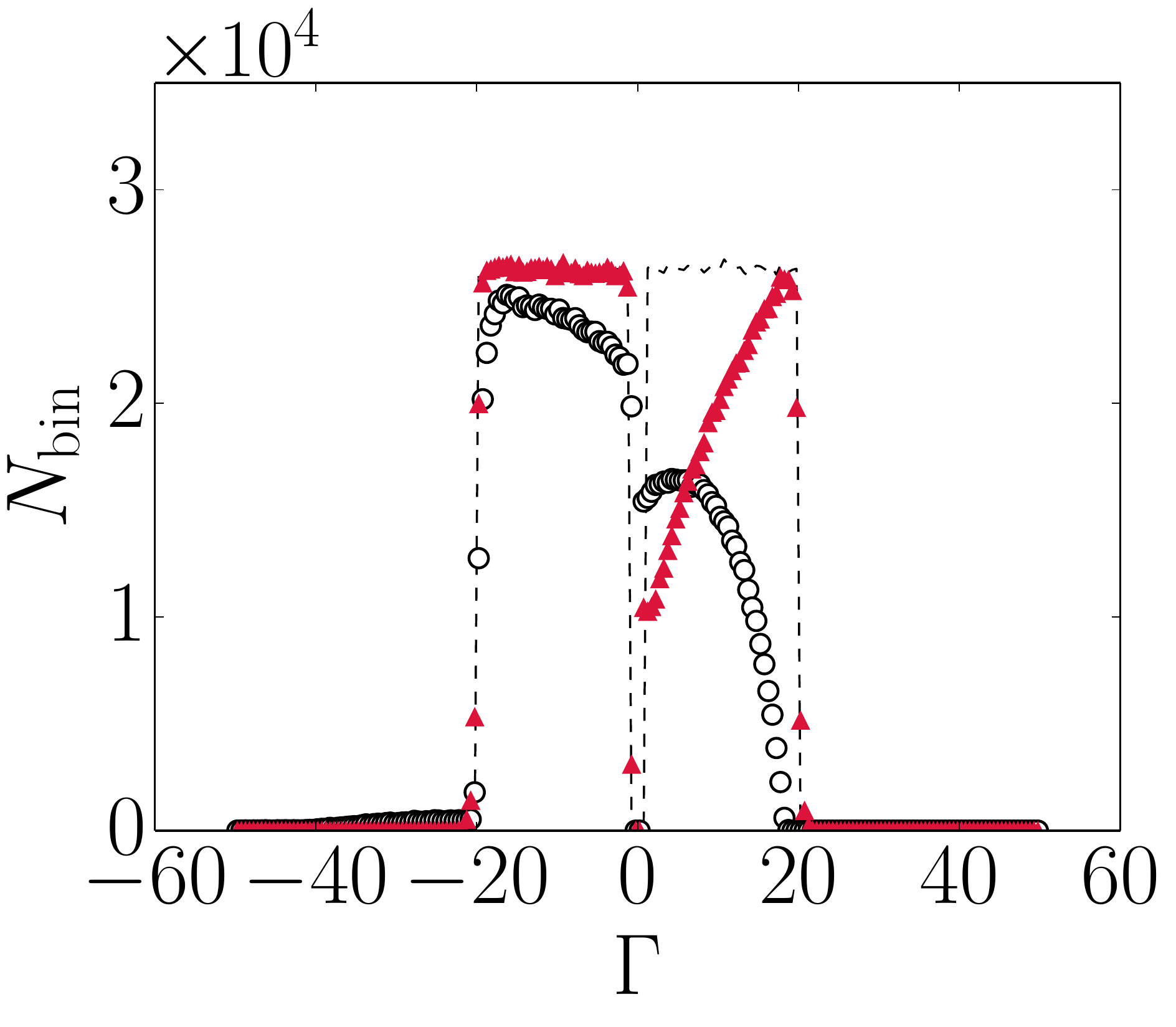}
        \put (-77,77){\makebox[0.05\linewidth][r]{(a)}}
      \end{minipage}
      \begin{minipage}{0.49\linewidth}
        \includegraphics[width=\linewidth]{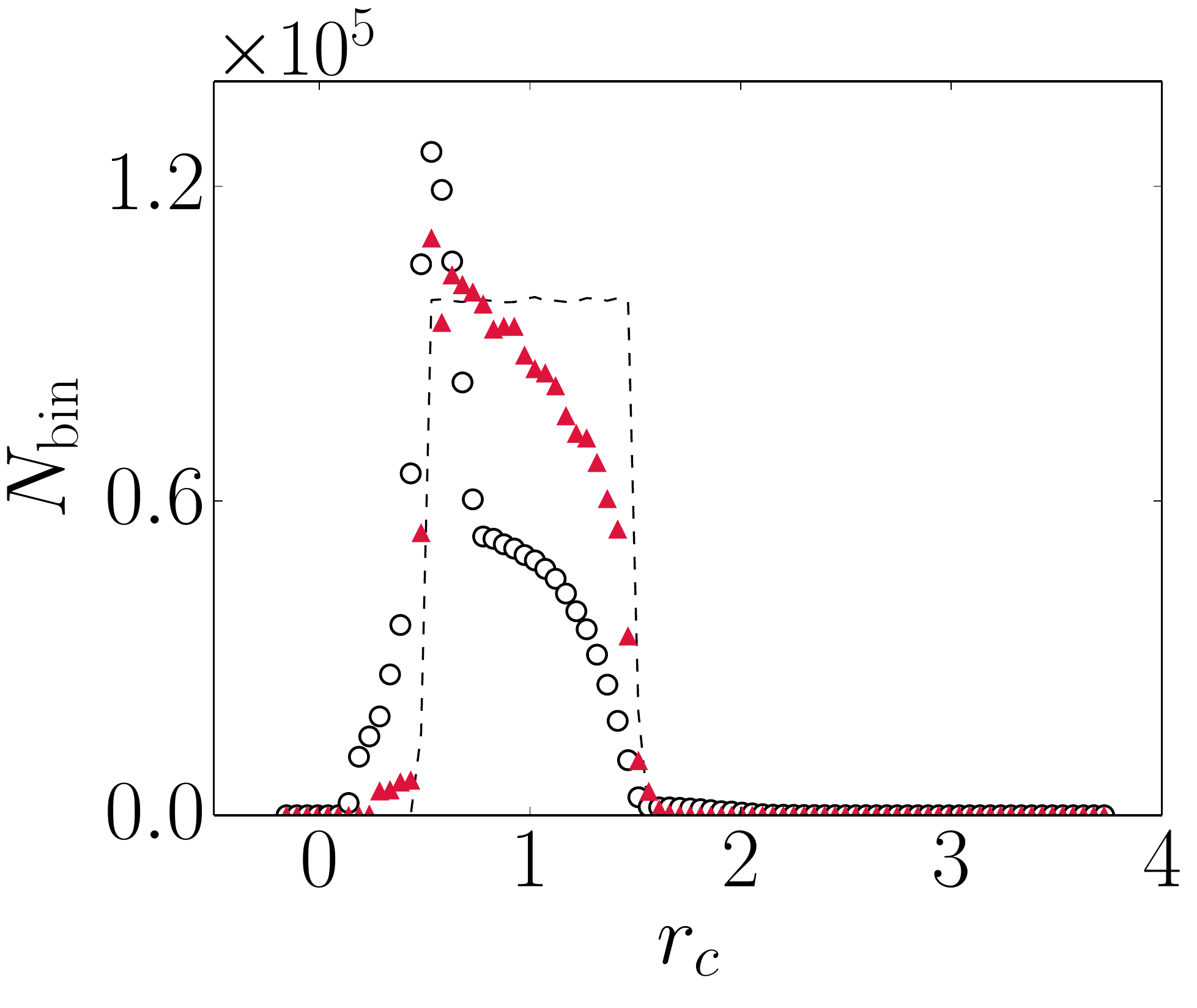}
        \put (-77,75){\makebox[0.05\linewidth][r]{(b)}}
      \end{minipage}
      
      \begin{minipage}{0.49\linewidth}
        \includegraphics[width=\linewidth]{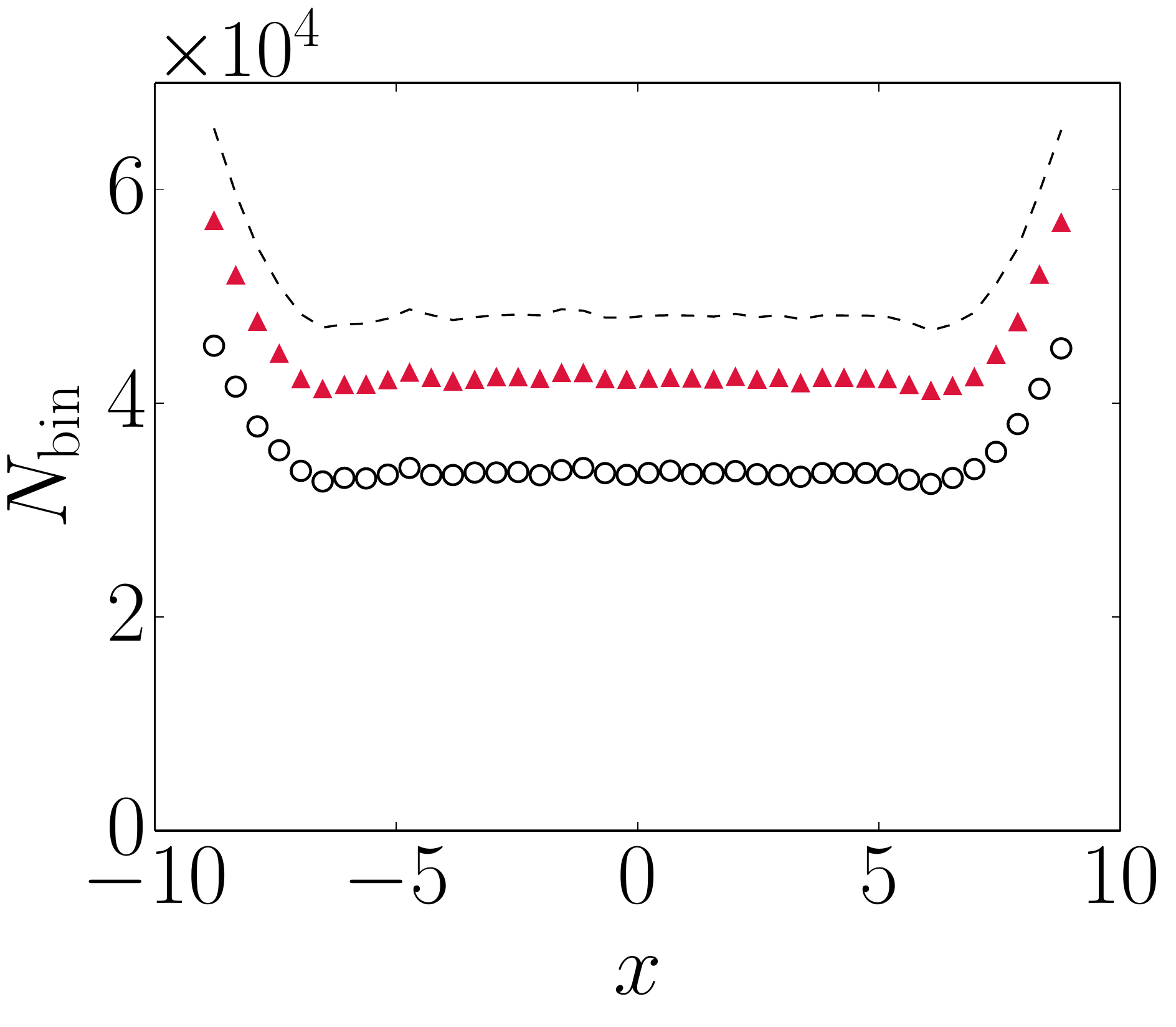}
        \put (-77,75){\makebox[0.05\linewidth][r]{(c)}}
      \end{minipage}
      \begin{minipage}{0.49\linewidth}
        \includegraphics[width=\linewidth]{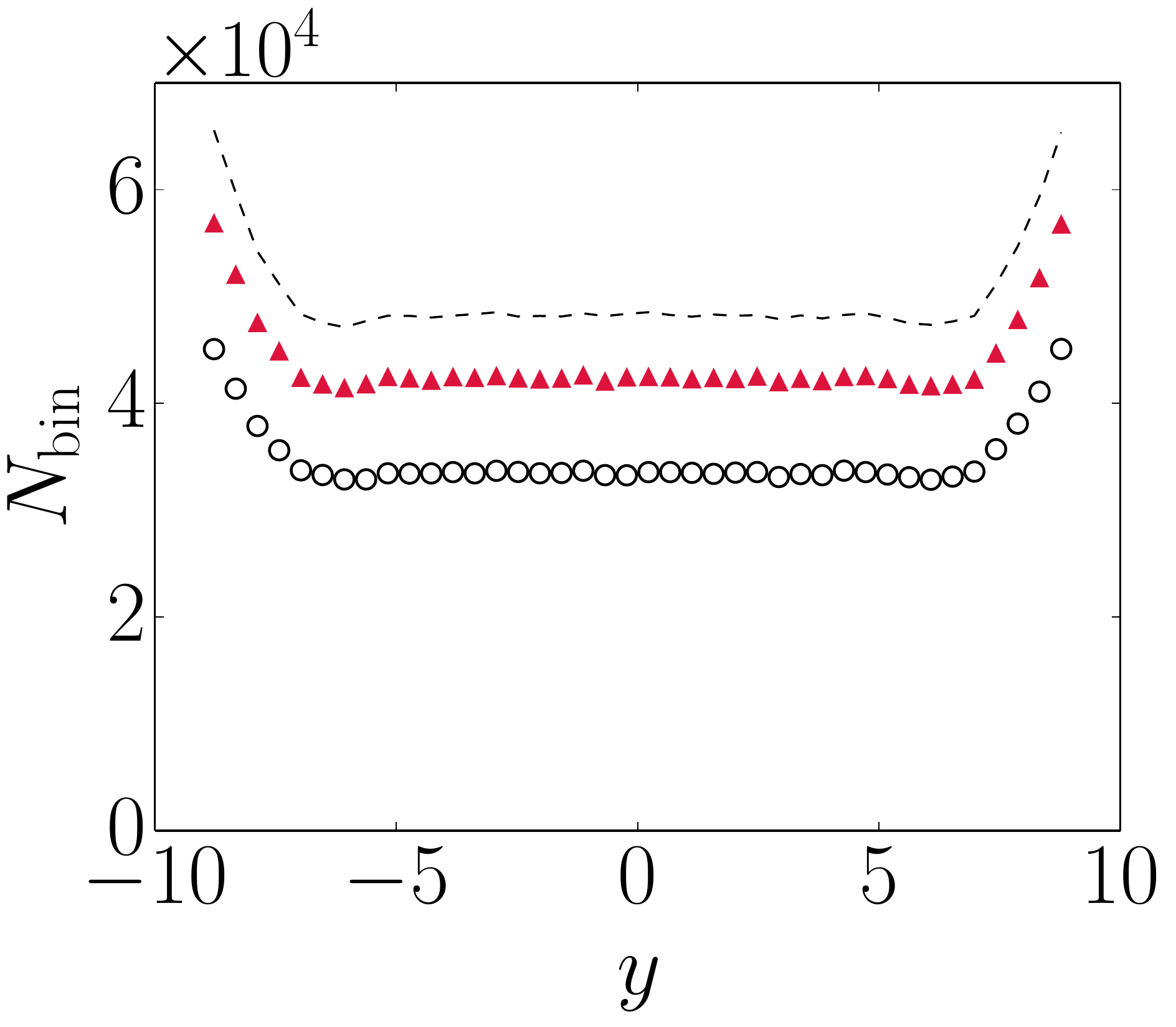}
        \put (-77,75){\makebox[0.05\linewidth][r]{(d)}}
      \end{minipage}
      \caption{Analysis of flow pattern C, with background subtraction. All the rest as in the caption of Fig. \ref{fig:mc2}.}
               \label{fig:b6}      
    \end{figure}	
		
					    \begin{figure}[]
       \begin{minipage}{0.49\linewidth}
        \includegraphics[width=\linewidth]{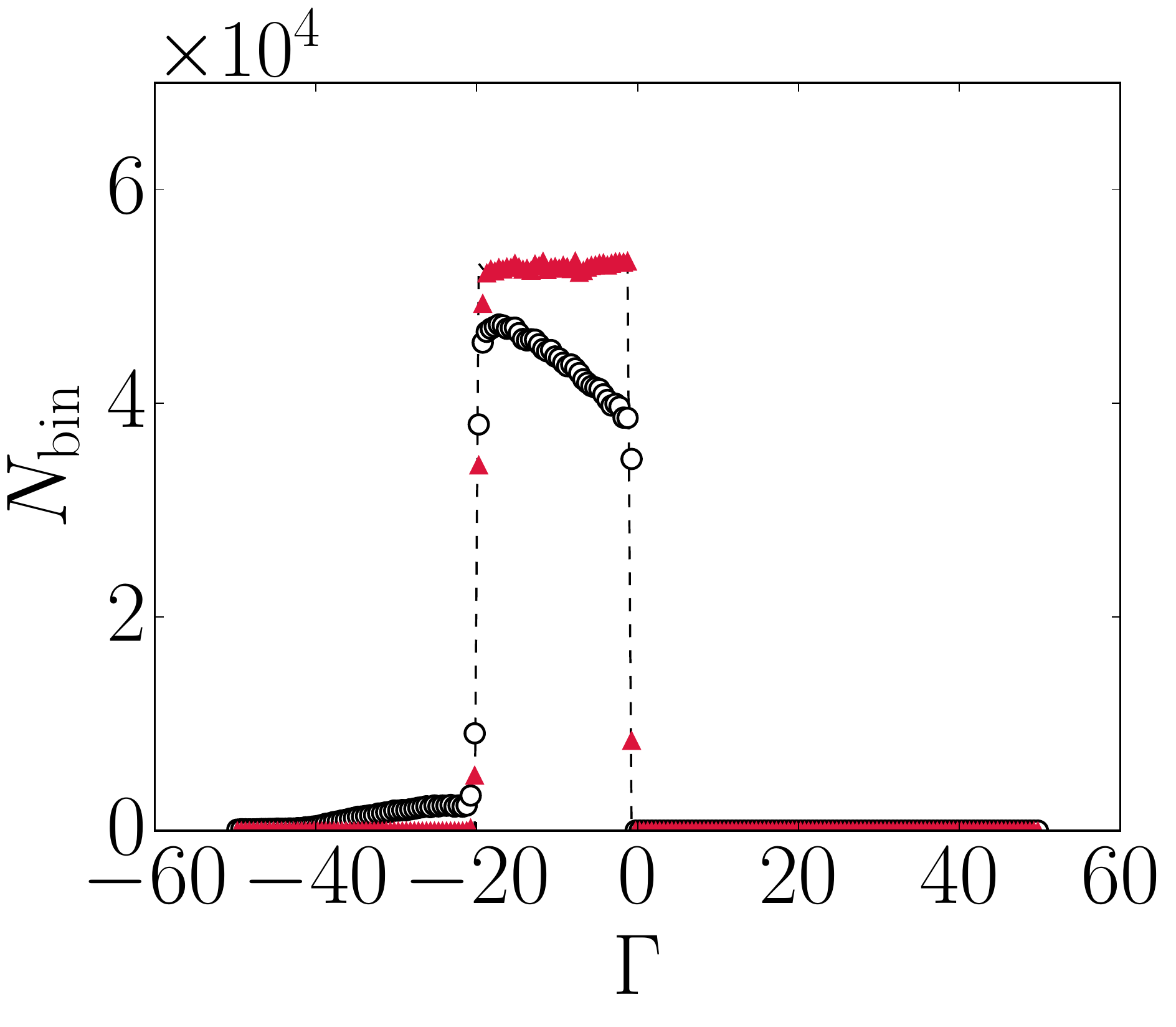}
        \put (-77,77){\makebox[0.05\linewidth][r]{(a)}}
      \end{minipage}
      \begin{minipage}{0.49\linewidth}
        \includegraphics[width=\linewidth]{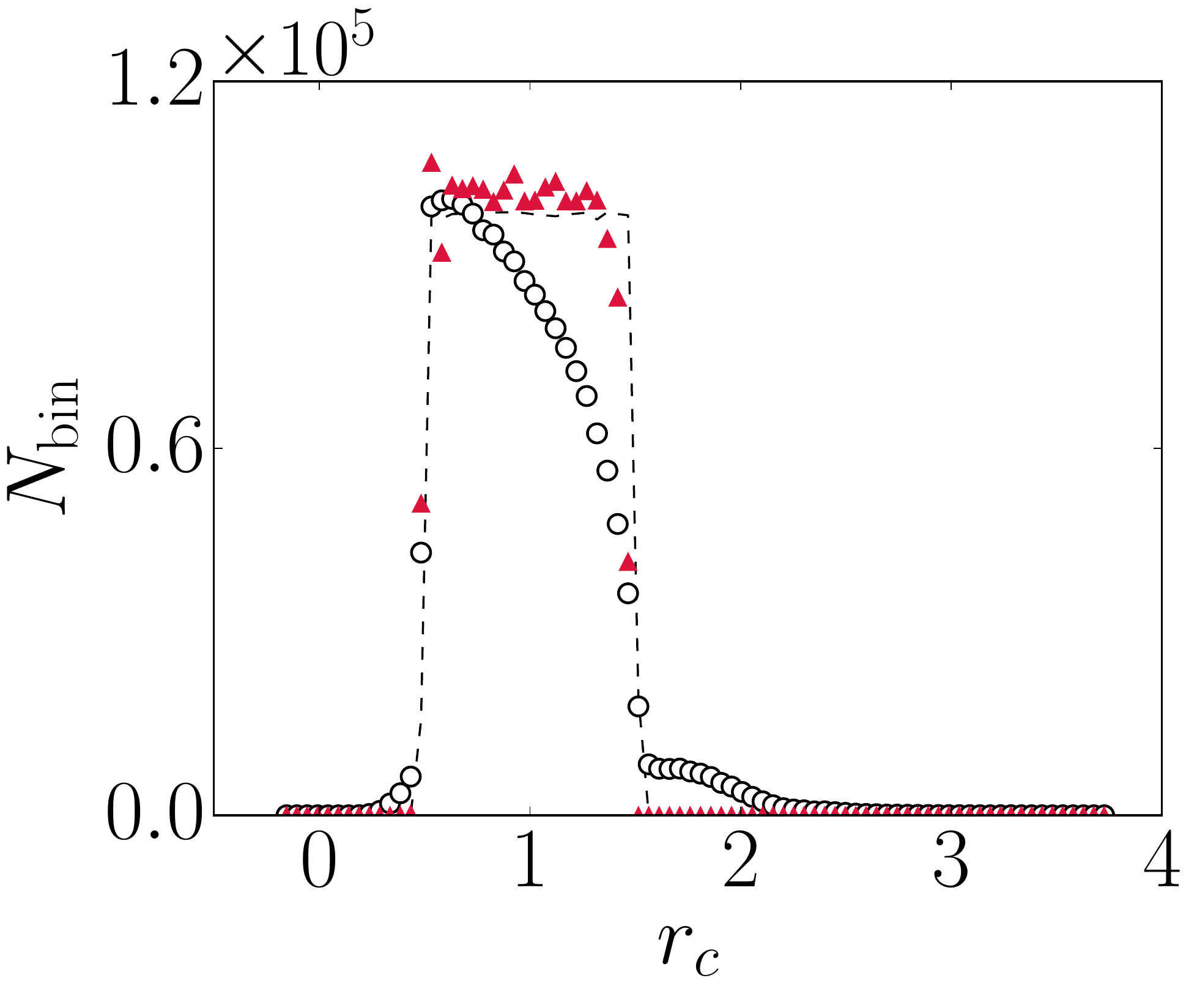}
        \put (-77,75){\makebox[0.05\linewidth][r]{(b)}}
      \end{minipage}
      
      \begin{minipage}{0.49\linewidth}
        \includegraphics[width=\linewidth]{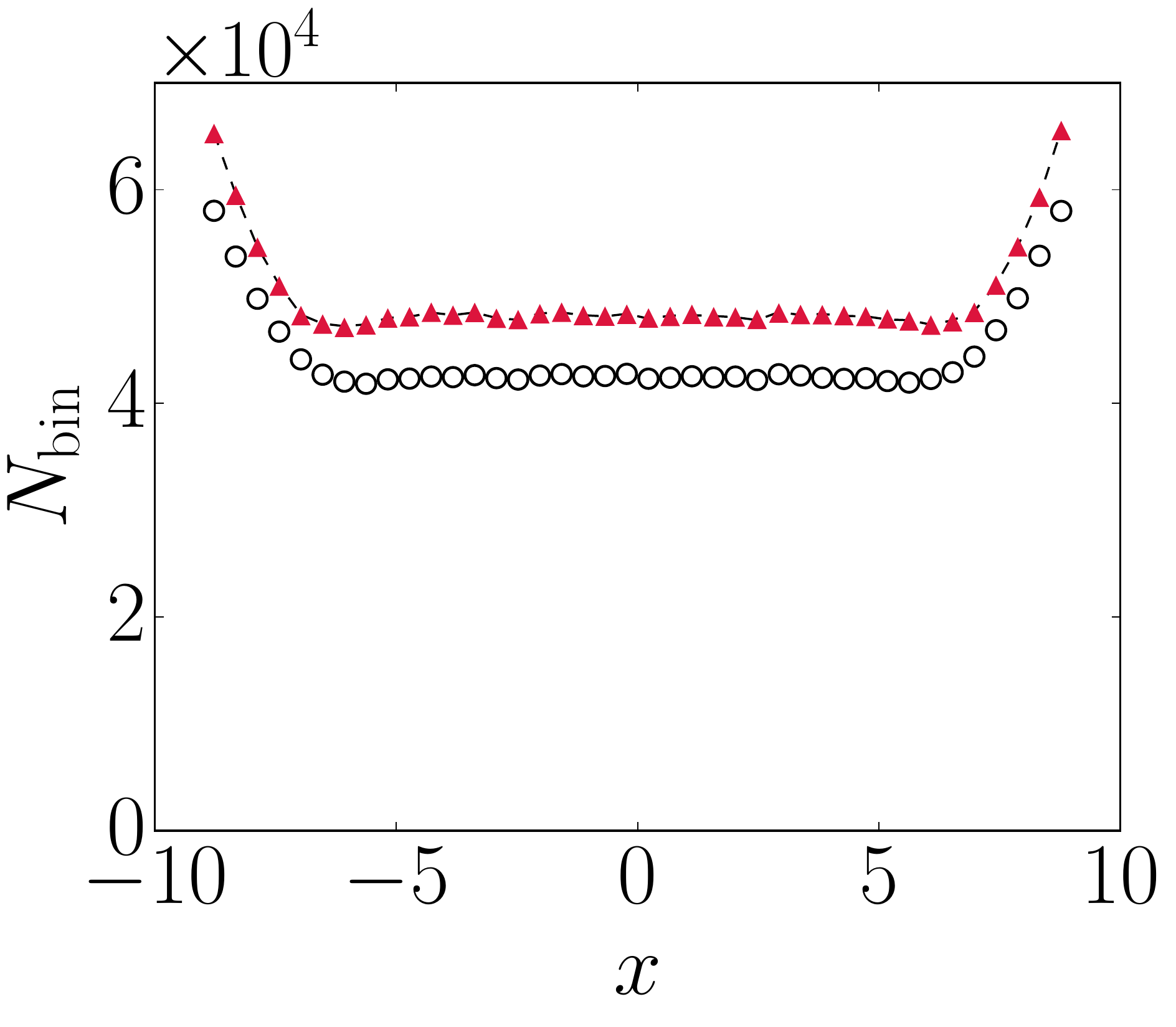}
        \put (-77,75){\makebox[0.05\linewidth][r]{(c)}}
      \end{minipage}
      \begin{minipage}{0.49\linewidth}
        \includegraphics[width=\linewidth]{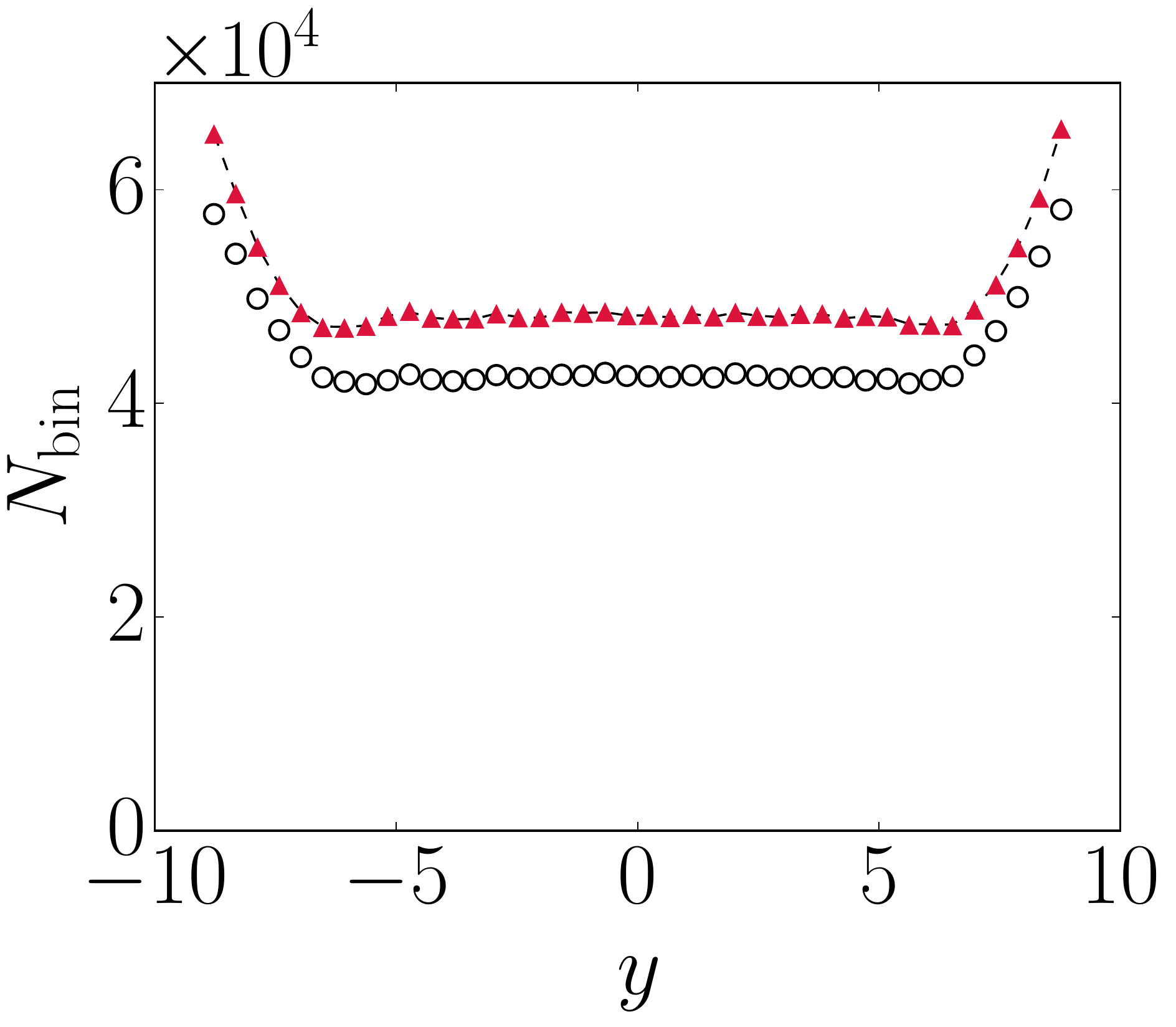}
        \put (-77,75){\makebox[0.05\linewidth][r]{(d)}}
      \end{minipage}
      \caption{Analysis of flow pattern D, without background subtraction. All the rest as in the caption of Fig. \ref{fig:mc2}.}
      \label{patternD}      
    \end{figure} 
		
				           \begin{figure}[ht]
      \begin{minipage}{0.49\linewidth}
        \includegraphics[width=\linewidth]{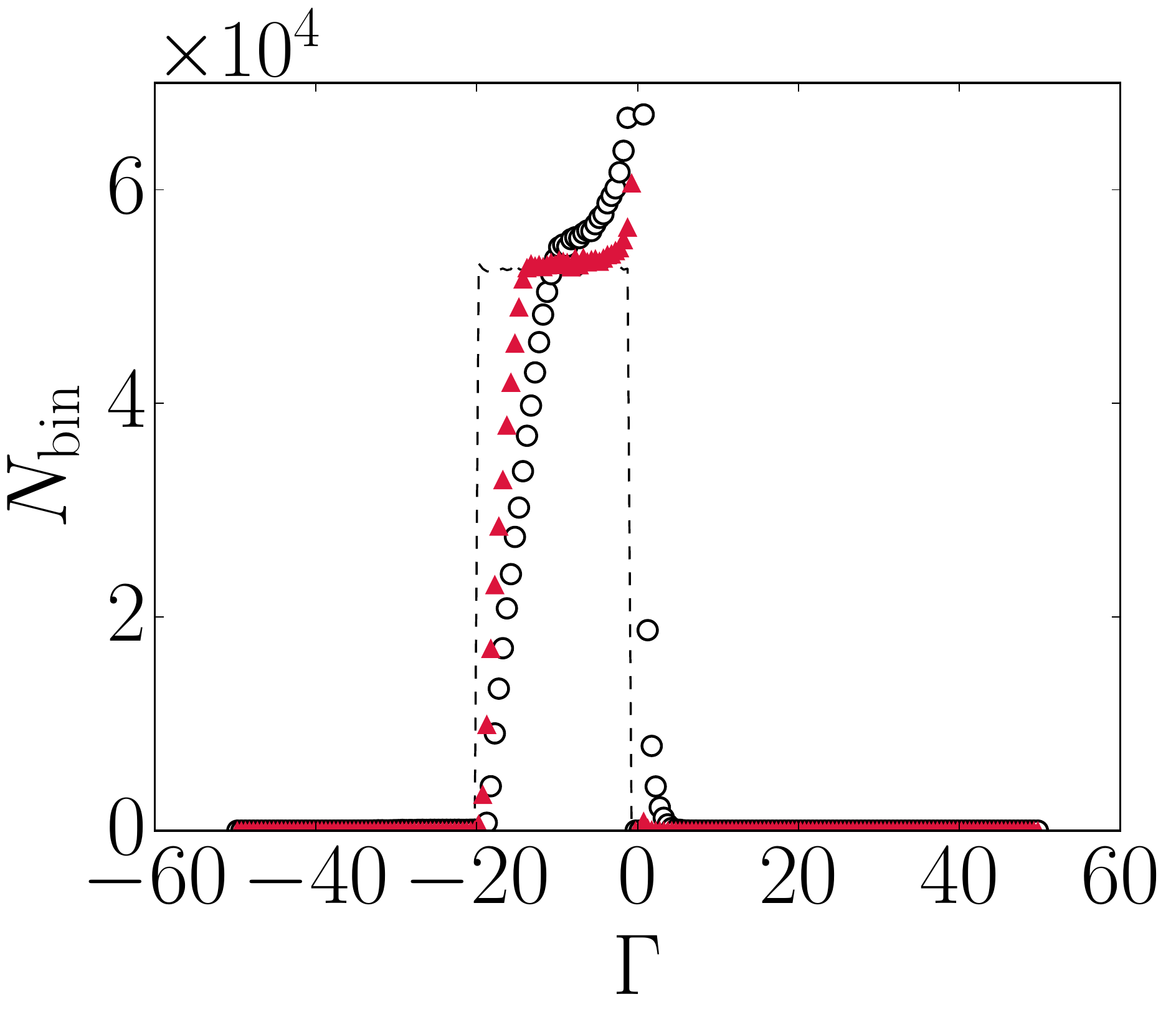}
        \put (-77,77){\makebox[0.05\linewidth][r]{(a)}}
      \end{minipage}
      \begin{minipage}{0.49\linewidth}
        \includegraphics[width=\linewidth]{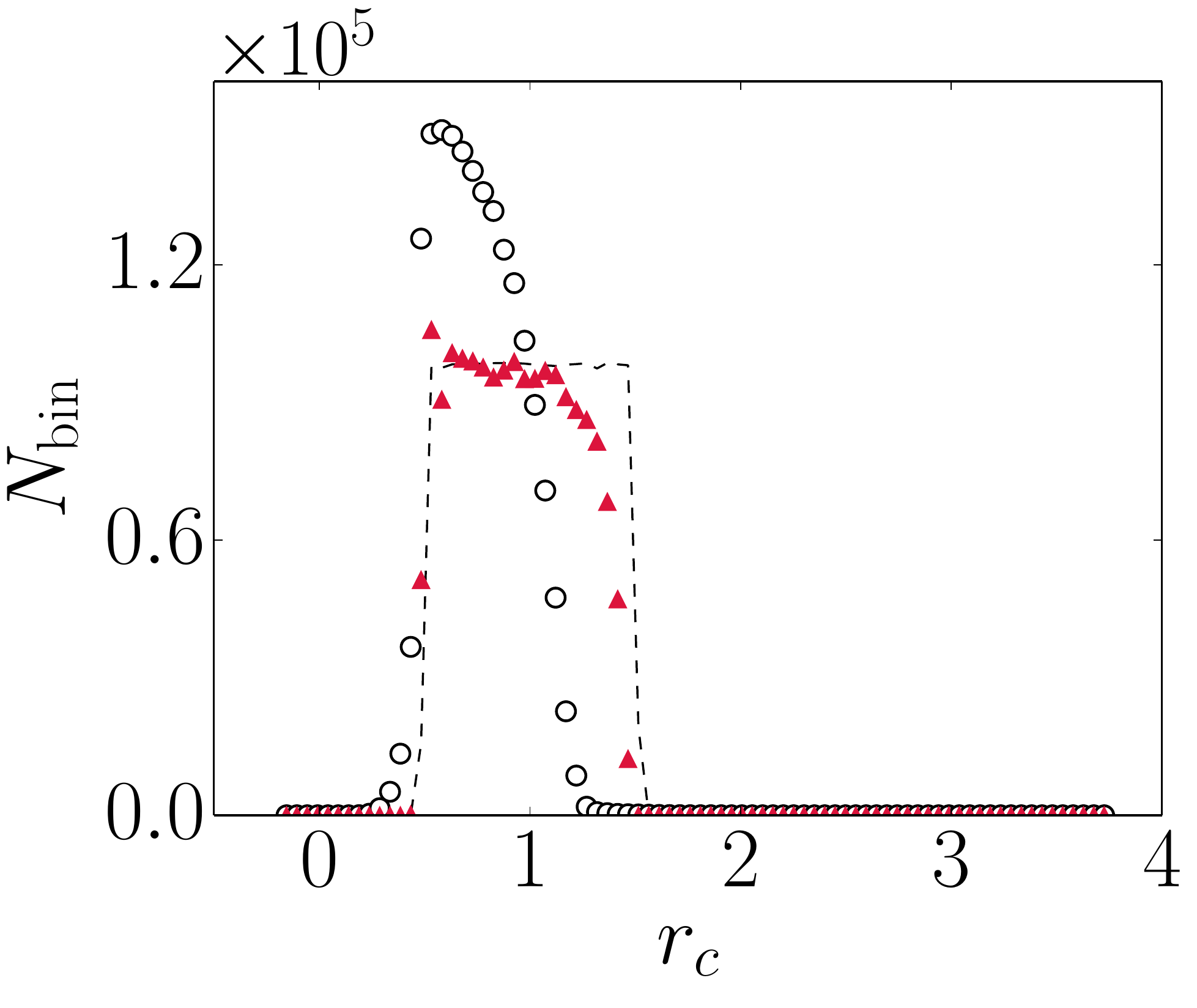}
        \put (-77,75){\makebox[0.05\linewidth][r]{(b)}}
      \end{minipage}
      
      \begin{minipage}{0.49\linewidth}
        \includegraphics[width=\linewidth]{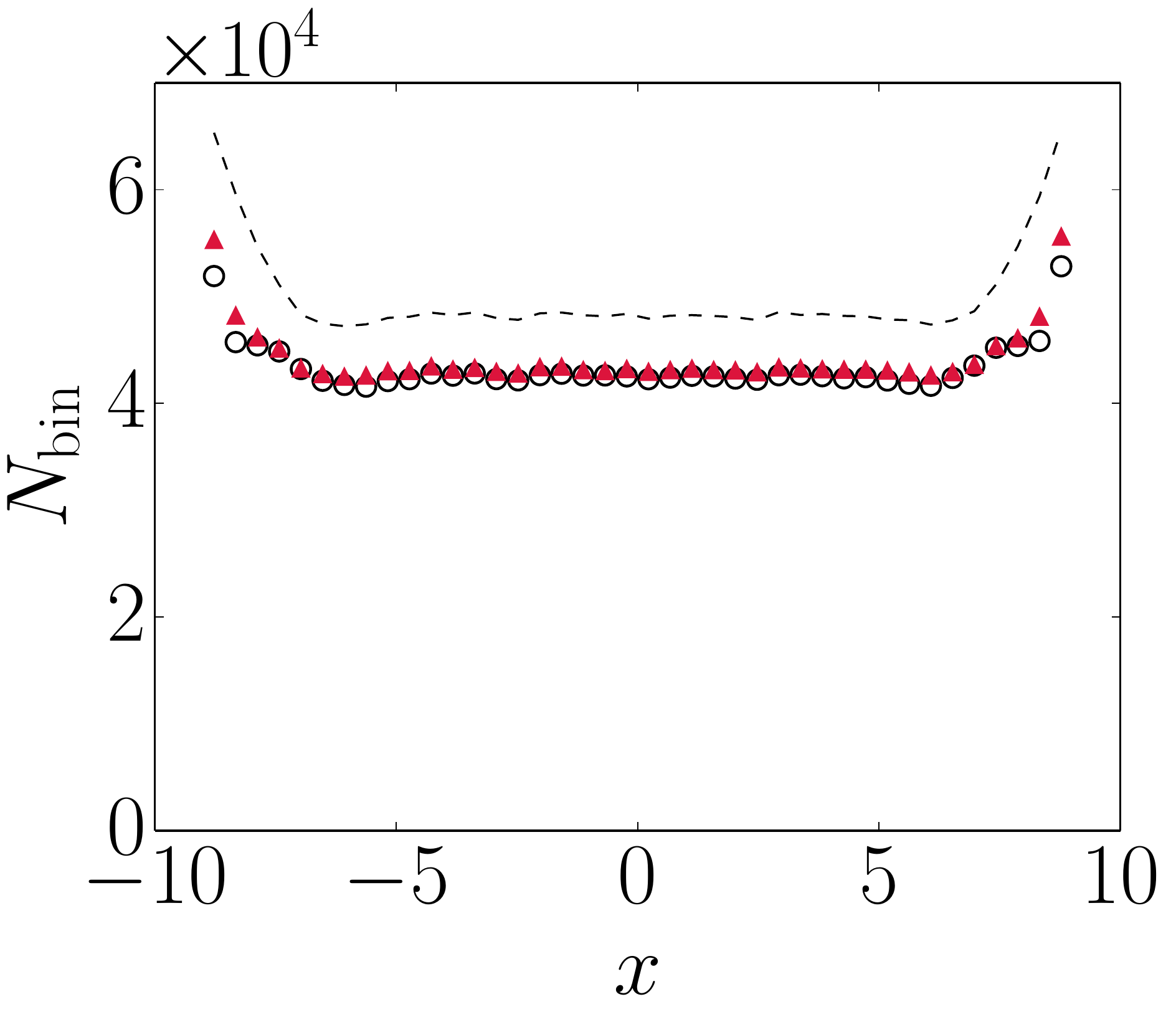}
        \put (-77,75){\makebox[0.05\linewidth][r]{(c)}}
      \end{minipage}
      \begin{minipage}{0.49\linewidth}
        \includegraphics[width=\linewidth]{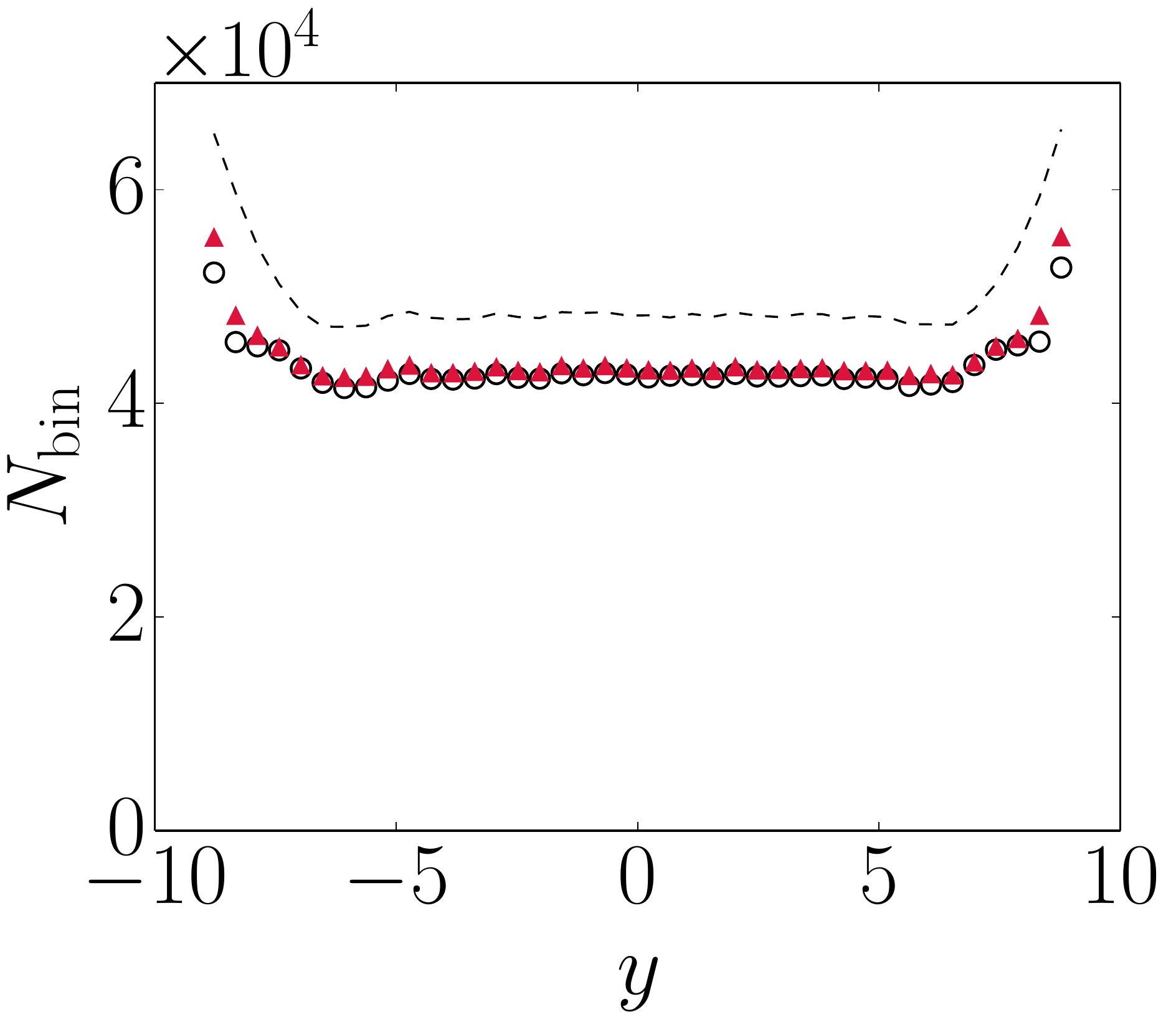}
        \put (-77,75){\makebox[0.05\linewidth][r]{(d)}}
      \end{minipage}
      \caption{Analysis of flow pattern D, with background subtraction. All the rest as in the caption of Fig. \ref{fig:mc2}.}
      \label{fig:b3}      
    \end{figure} 
		
										\begin{figure}[ht]
       \begin{minipage}{0.49\linewidth}
        \includegraphics[width=\linewidth]{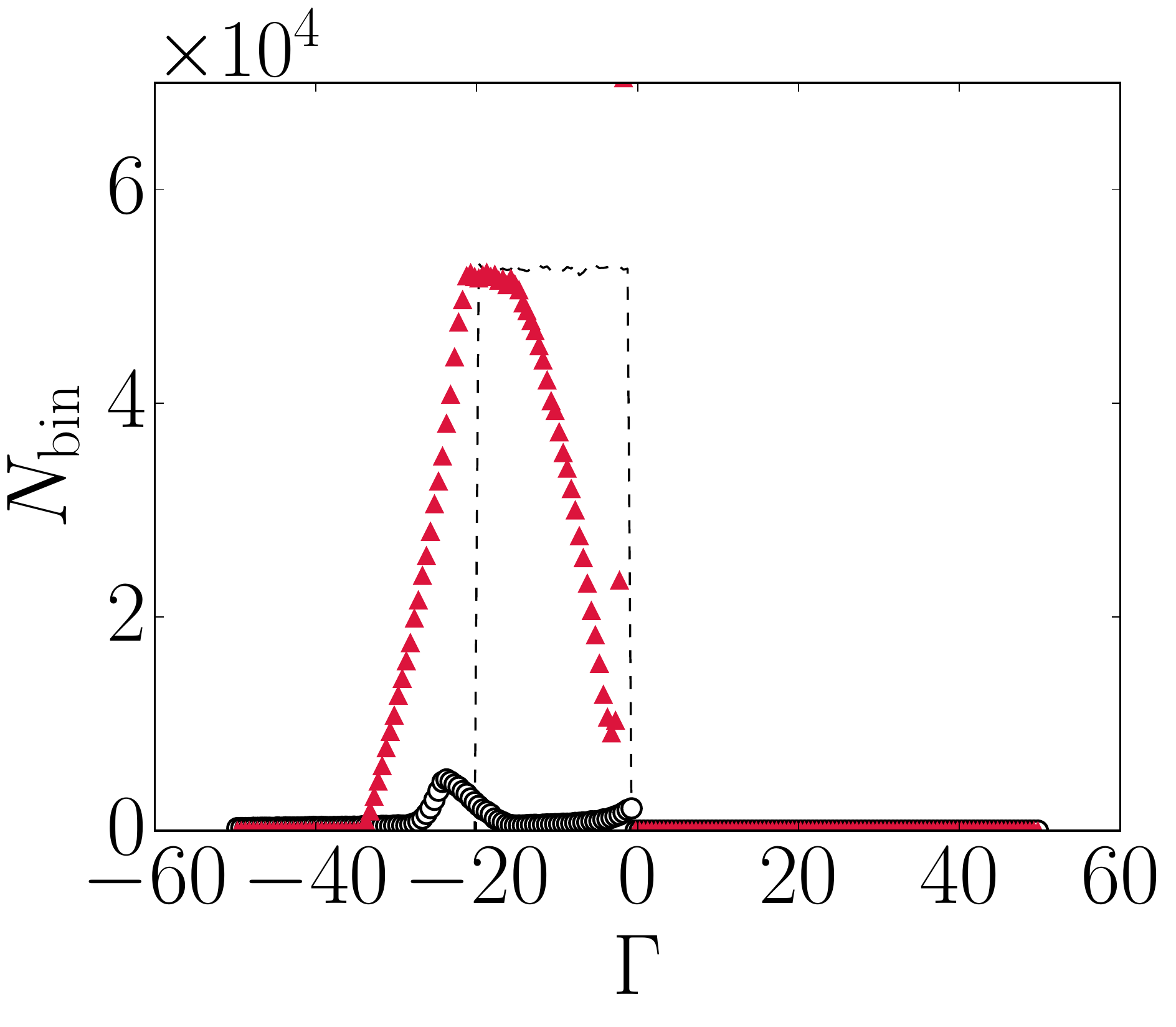}
        \put (-77,77){\makebox[0.05\linewidth][r]{(a)}}
      \end{minipage}
      \begin{minipage}{0.49\linewidth}
        \includegraphics[width=\linewidth]{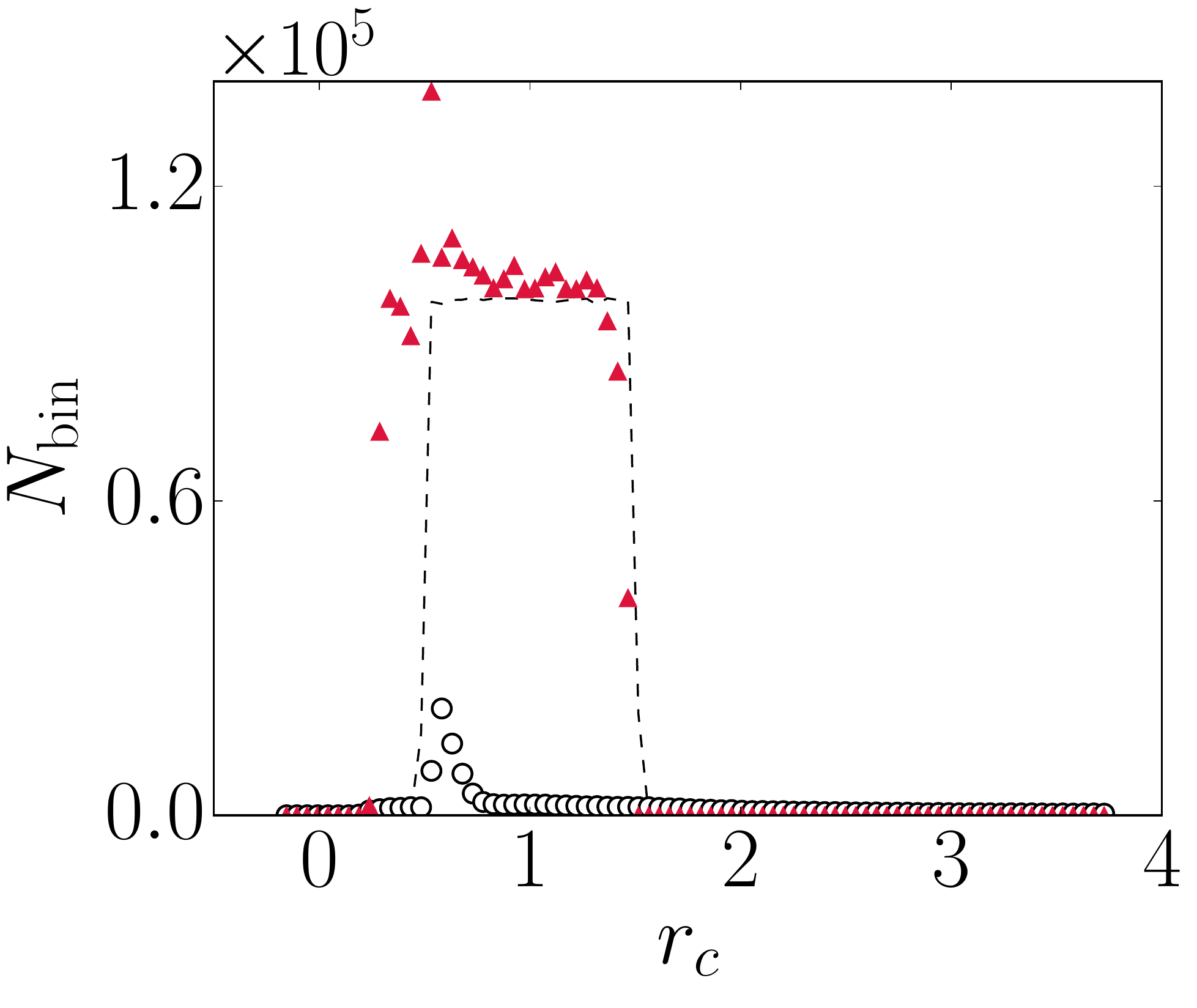}
        \put (-77,75){\makebox[0.05\linewidth][r]{(b)}}
      \end{minipage}
      
      \begin{minipage}{0.49\linewidth}
        \includegraphics[width=\linewidth]{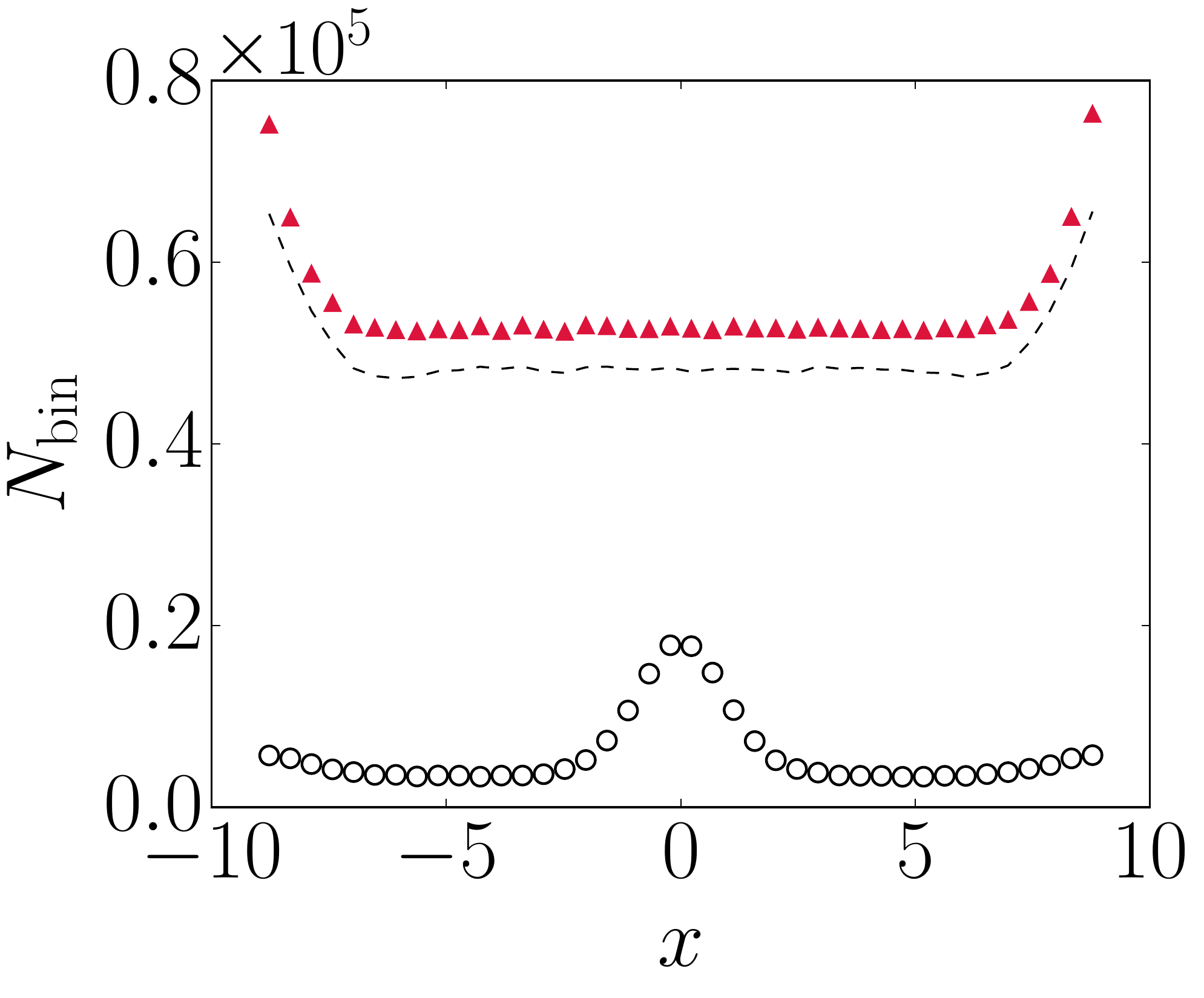}
        \put (-77,75){\makebox[0.05\linewidth][r]{(c)}}
      \end{minipage}
      \begin{minipage}{0.49\linewidth}
        \includegraphics[width=\linewidth]{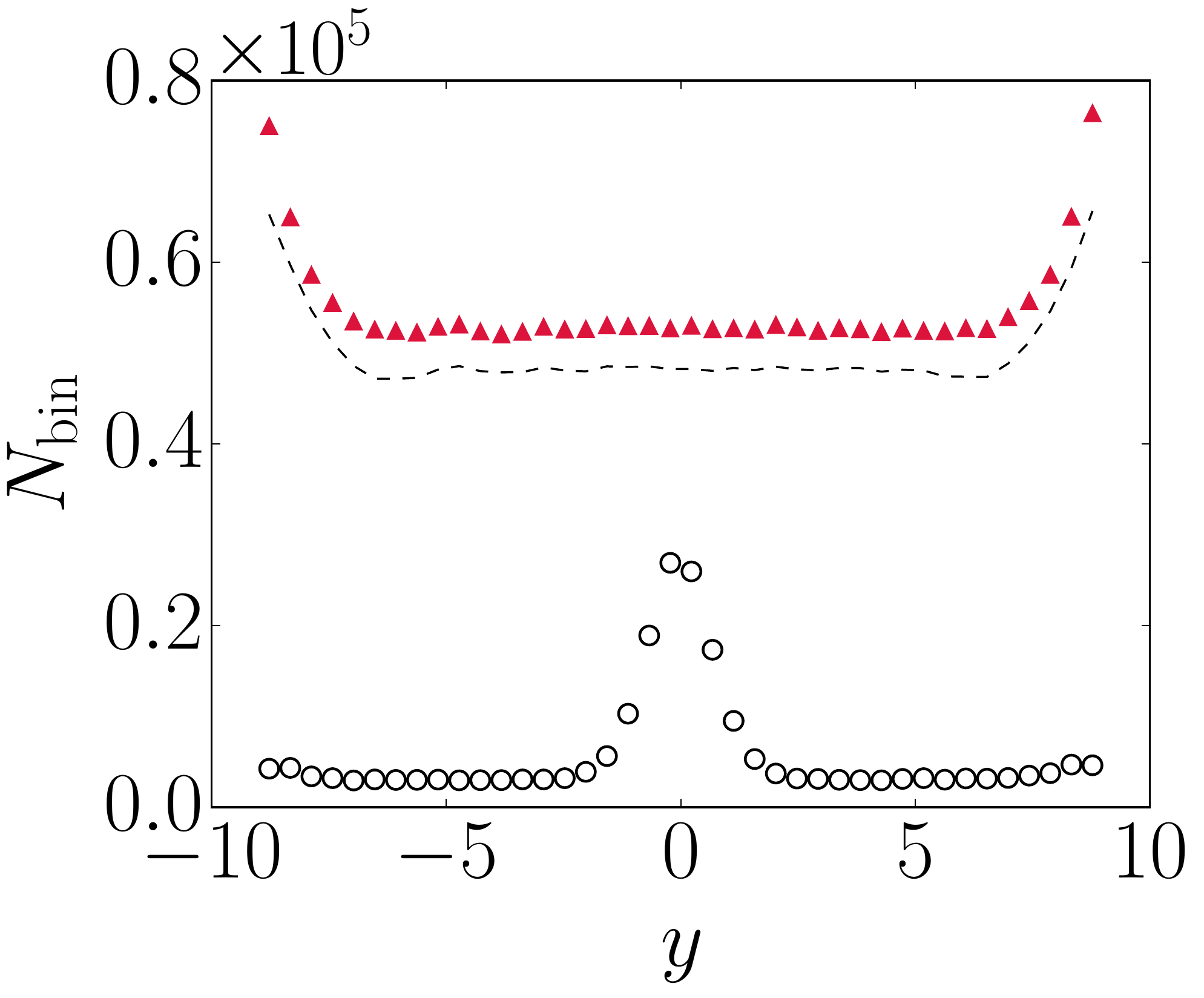}
        \put (-77,75){\makebox[0.05\linewidth][r]{(d)}}
      \end{minipage}
      \caption{Analysis of flow pattern E, without background subtraction. All the rest as in the caption of Fig. \ref{fig:mc2}.}
      \label{patternE}      
    \end{figure}
		
    For the flow pattern C, we conclude, from Figs. \ref{fig:mc4} and \ref{fig:b6}, that the background subtraction procedure considerably 
  improves the performance of the $\lambda_\omega$-criterion, which now becomes valid as a method of vortex 
  identification. Its only residual deficiency is the suppression of vortices which have relatively large radii and 
  small positive circulations. This is, very clearly, a side effect of the Heaviside filtering function, which erases 
  positive-circulation vortices that are completely ``submerged" in the negative vorticity background.
    
				    As a way to losely mimic some of the turbulence boundary layer characteristics found in streamwise/wall normal planes, where the
  background vorticity has the same sign as most of the viscous layer vortices \cite{wu,herpin_a,herpin}, we have devised the 
	flow regimes D and E. Note that in the flow pattern D, there is no external background, $\bar \omega = 0$, but there is an 
  essentially uniform negative vorticity background produced by the many-vortex system, because $\langle v_i(\vec r) \rangle \neq 0$. 
	Curiously, as it can be seen from Figs. \ref{patternD} and \ref{fig:b3}, the 
	$\lambda_\omega$-criterion is acceptable in both cases, but it works a bit better, for the flow pattern D, if the background 
	were not subtracted. This has to do, this time, with the existence of vortices that are placed in regions of the flow where the 
	local vorticity background is momentarily greater, due to the effect of fluctuations, than the mean self-induced vorticity background.
		
\begin{figure}[ht]
       \begin{minipage}{0.49\linewidth}
        \includegraphics[width=\linewidth]{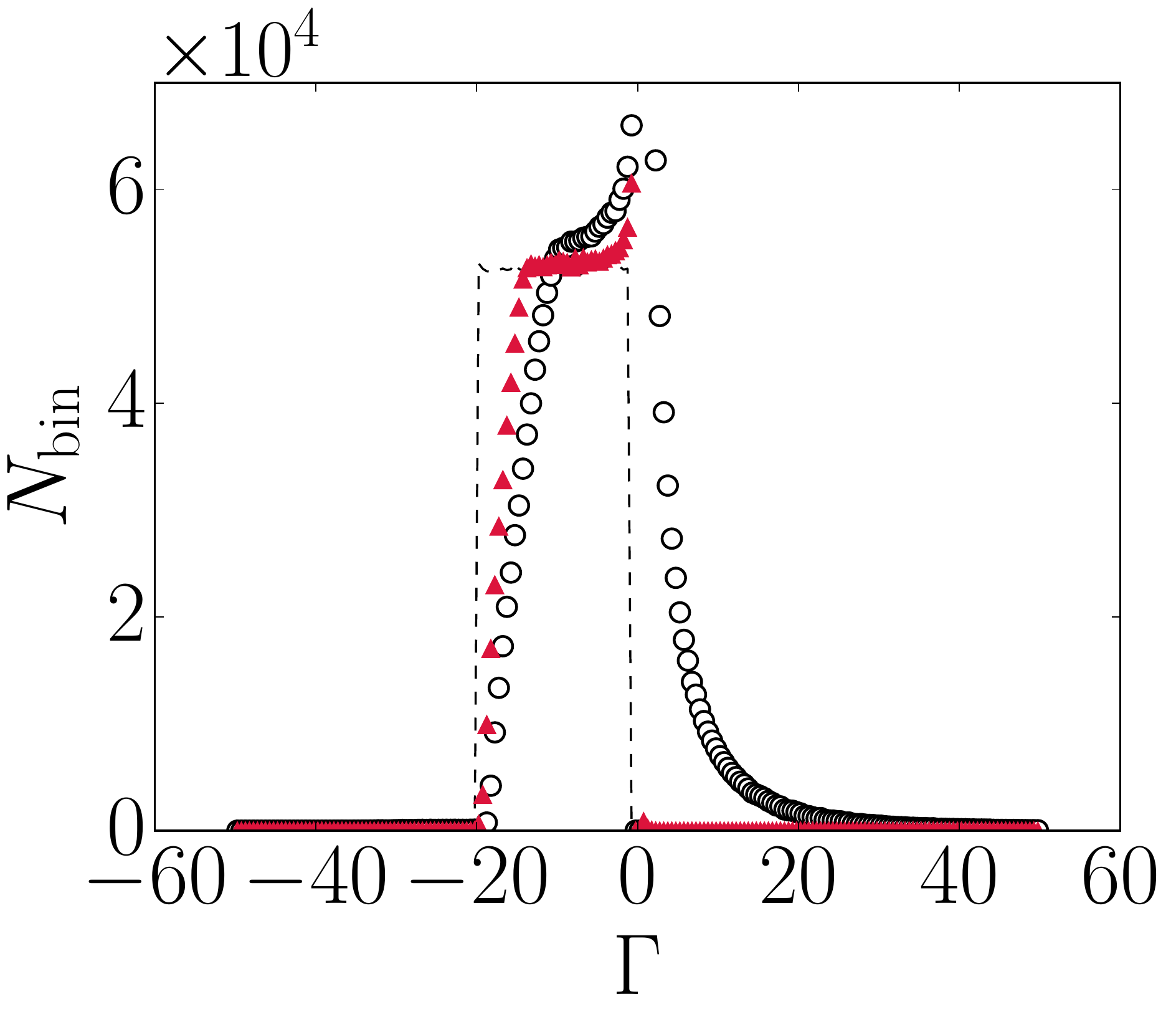}
        \put (-77,77){\makebox[0.05\linewidth][r]{(a)}}
      \end{minipage}
      \begin{minipage}{0.49\linewidth}
        \includegraphics[width=\linewidth]{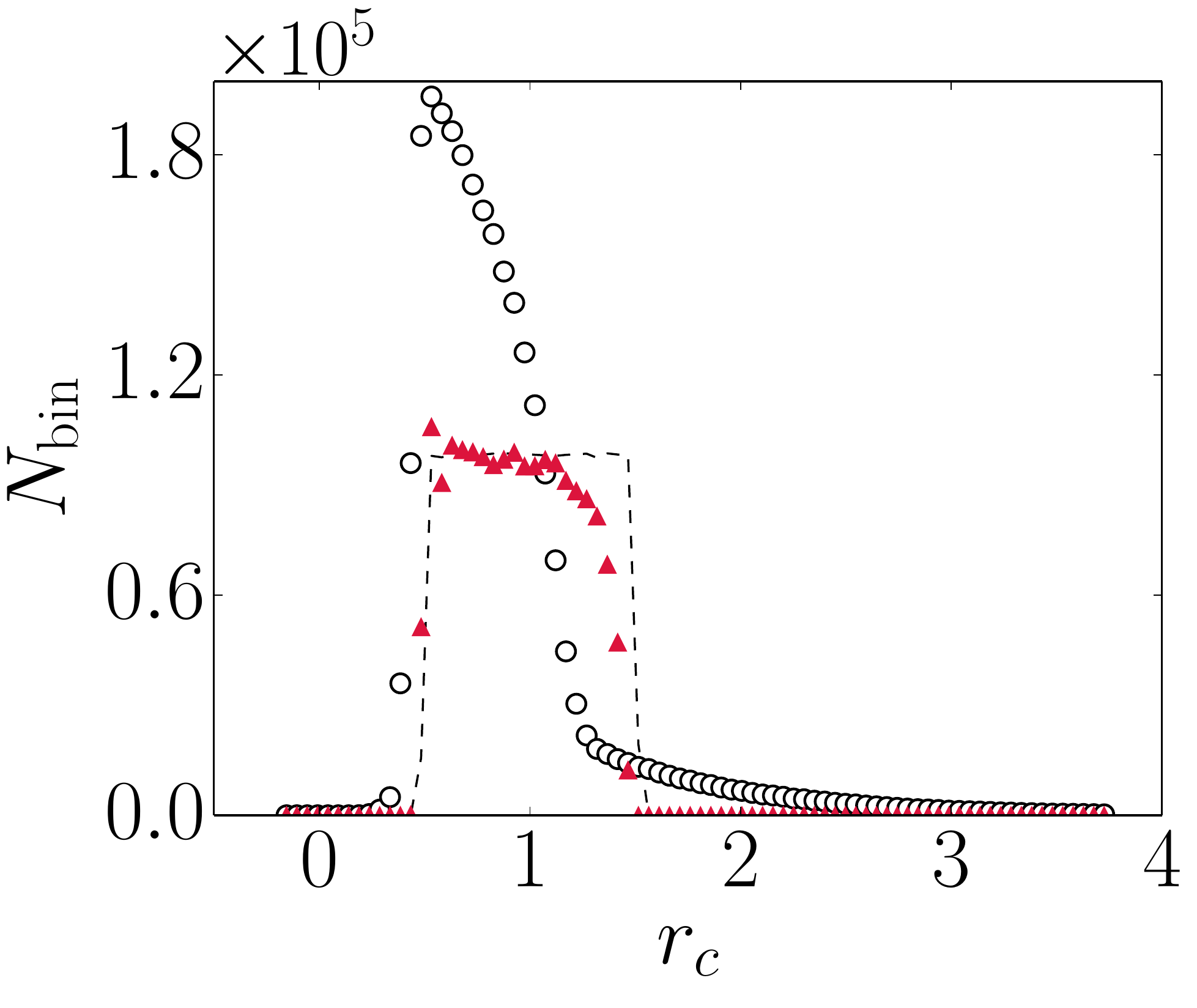}
        \put (-77,75){\makebox[0.05\linewidth][r]{(b)}}
      \end{minipage}
      
      \begin{minipage}{0.49\linewidth}
        \includegraphics[width=\linewidth]{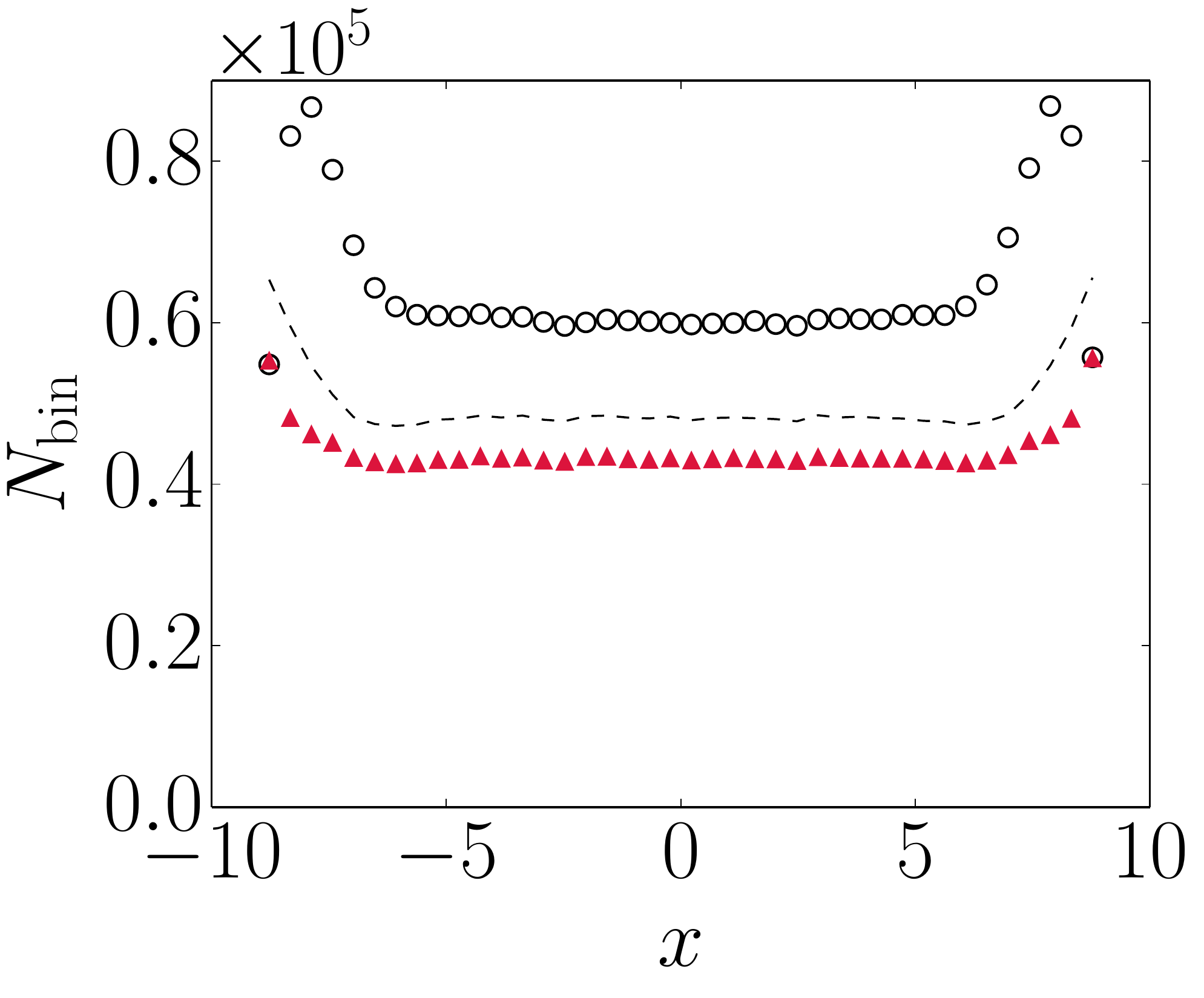}
        \put (-77,75){\makebox[0.05\linewidth][r]{(c)}}
      \end{minipage}
      \begin{minipage}{0.49\linewidth}
        \includegraphics[width=\linewidth]{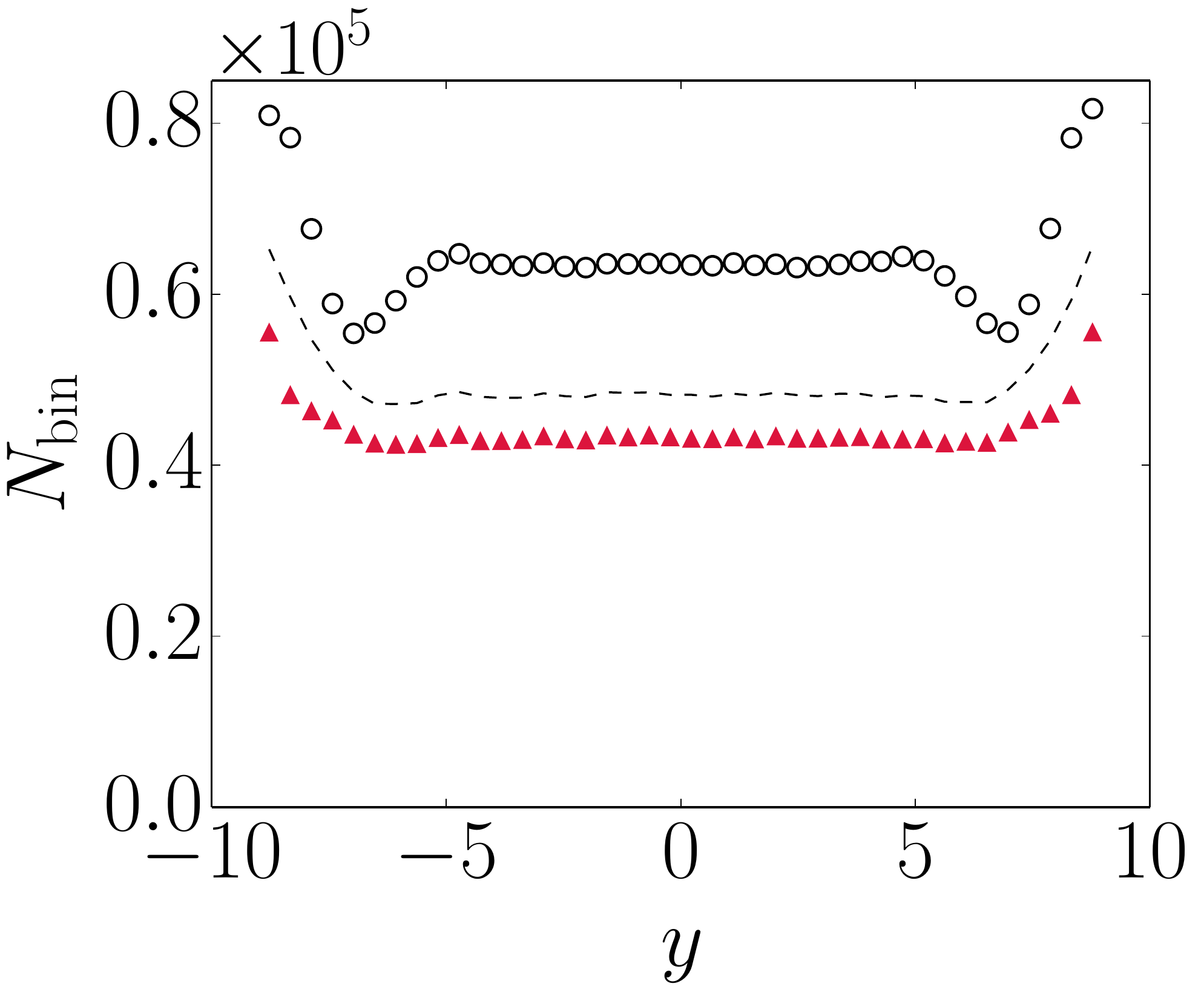}
        \put (-77,75){\makebox[0.05\linewidth][r]{(d)}}
      \end{minipage}
      \caption{Analysis of flow pattern E, with background subtraction. All the rest as in the caption of Fig. \ref{fig:mc2}.}
      \label{fig:b4}      
    \end{figure}

								    For the strong background case, flow pattern E, it turns out, as indicated from Figs. \ref{patternE} and \ref{fig:b4}, 
						that the background subtraction procedure leads to improvement, mainly in recovering circulation statistics, which brings 
						the quality of vortex identification back to the reasonably good standards observed in the analysis of flow pattern D.
						
		    The above benchmarking Monte Carlo study shows that the $\lambda_\omega$-criterion, enhanced by the background subtraction
  procedure, provides an appropriate identification prescription for the investigation of two-dimensional vortex systems.
  With the confidence acquired from the numerical experiments carried out with synthetic samples, we focus now on the
  analysis of a more realistic flow situation.
        
\section{Application to a Turbulent Channel Flow}\label{sec:dns}

    Cross sections of spanwise vortices, interpreted as heads of hairpin vortices, have been usually observed in streamwise/wall normal
  plane sections of wall-bounded flows \cite{adrian_mein_tom,camussi,wu,dennis_nickels,jeff,herpin_a,herpin}. We have investigated 
	the statistical properties of such two-dimensional vortex flow patterns by means of the $\lambda_{ci}$ and the $\lambda_\omega$
  criteria, for a turbulent channel flow DNS. 
        
  \begin{figure*}[ht]
    \includegraphics[width=0.97\linewidth]{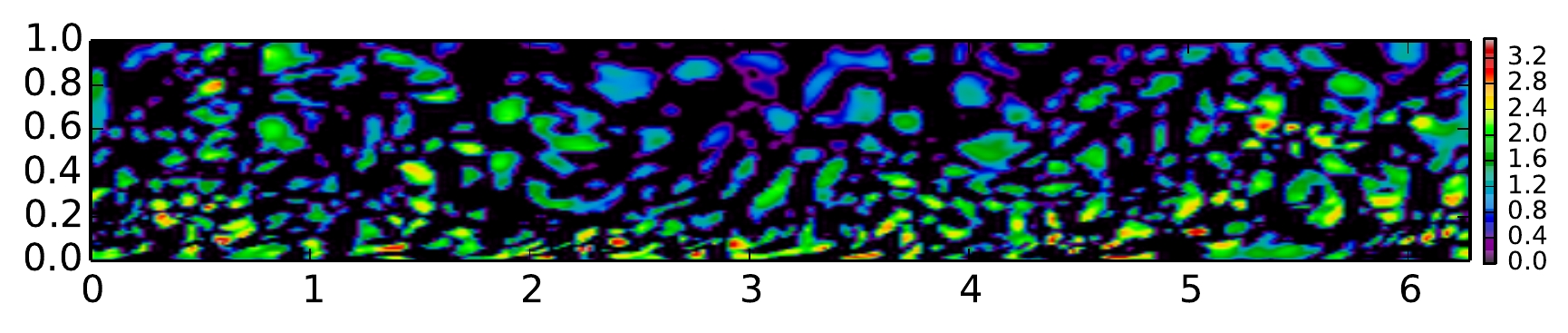}
    \put (-475,80){\makebox[0.05\linewidth][r]{(a)}}
		\put (-480,50){\makebox[0.05\linewidth][r]{{\large{\bf{y}}}}}
		\put (-240,0){\makebox[0.05\linewidth][r]{{\large{\bf{x}}}}}
    
		\includegraphics[width=0.97\linewidth]{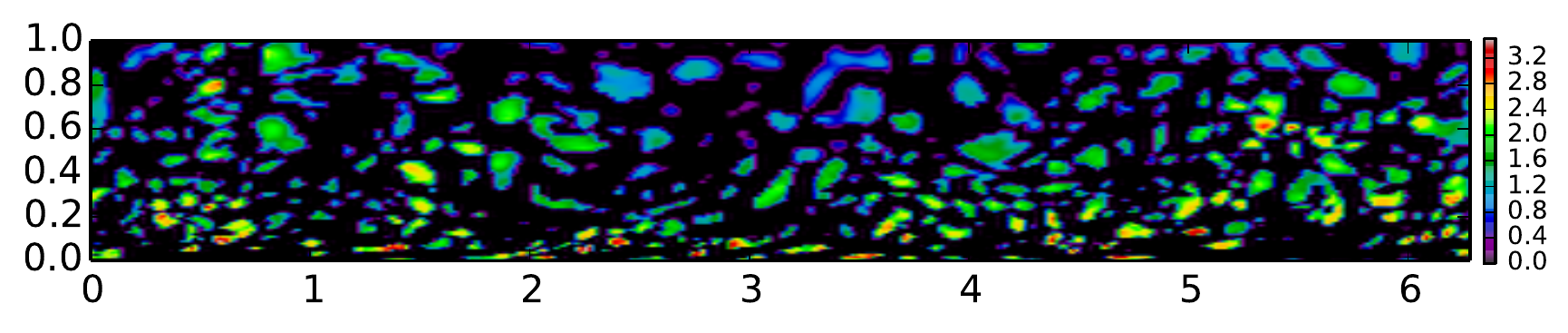}
    \put (-475,80){\makebox[0.05\linewidth][r]{(b)}}
		\put (-480,50){\makebox[0.05\linewidth][r]{{\large{\bf{y}}}}}
		\put (-240,0){\makebox[0.05\linewidth][r]{{\large{\bf{x}}}}}
		
 	  \includegraphics[width=0.97\linewidth]{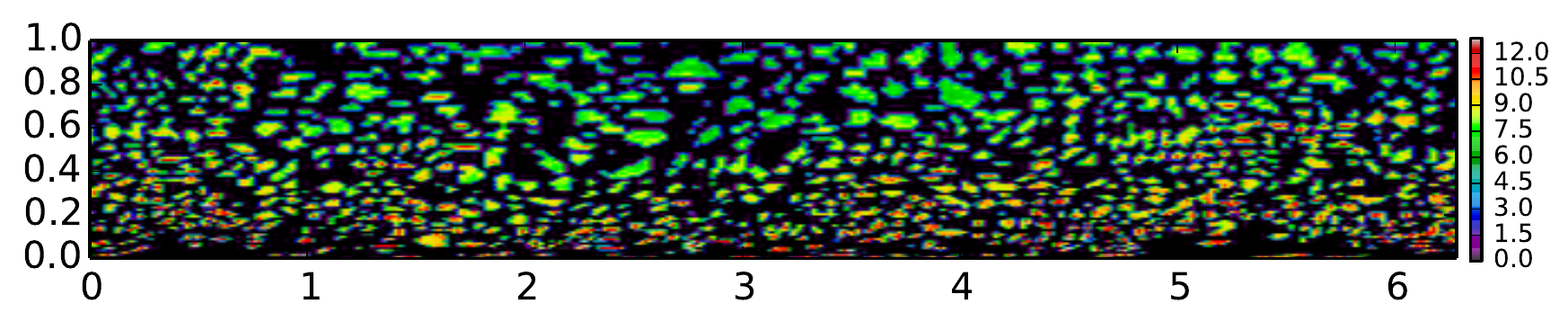}
    \put (-475,80){\makebox[0.05\linewidth][r]{(c)}}
		\put (-480,50){\makebox[0.05\linewidth][r]{{\large{\bf{y}}}}}
		\put (-240,0){\makebox[0.05\linewidth][r]{{\large{\bf{x}}}}}

    \caption{Density plots of the  $\lambda_{ci}$, [figures (a) and (b)] and the $\lambda_\omega$  [figure (c)] fields in a
             streamwise/wall normal plane for the DNS of a turbulent channel flow, for all the channel extension and from the bottom wall
             up to the mid-channel height. No treshold is used in the vortex identification analyses. The background
             subtraction procedure is implemented only in figures (b) and (c). The color bars represent the $\lambda_{ci}$ and the $\lambda_\omega$ fields in linear and logarithmic scales, respectively.}
    \label{fig:dns2}
  \end{figure*}
  
  	  	   \begin{figure}[ht]
      \includegraphics[width=0.92\linewidth]{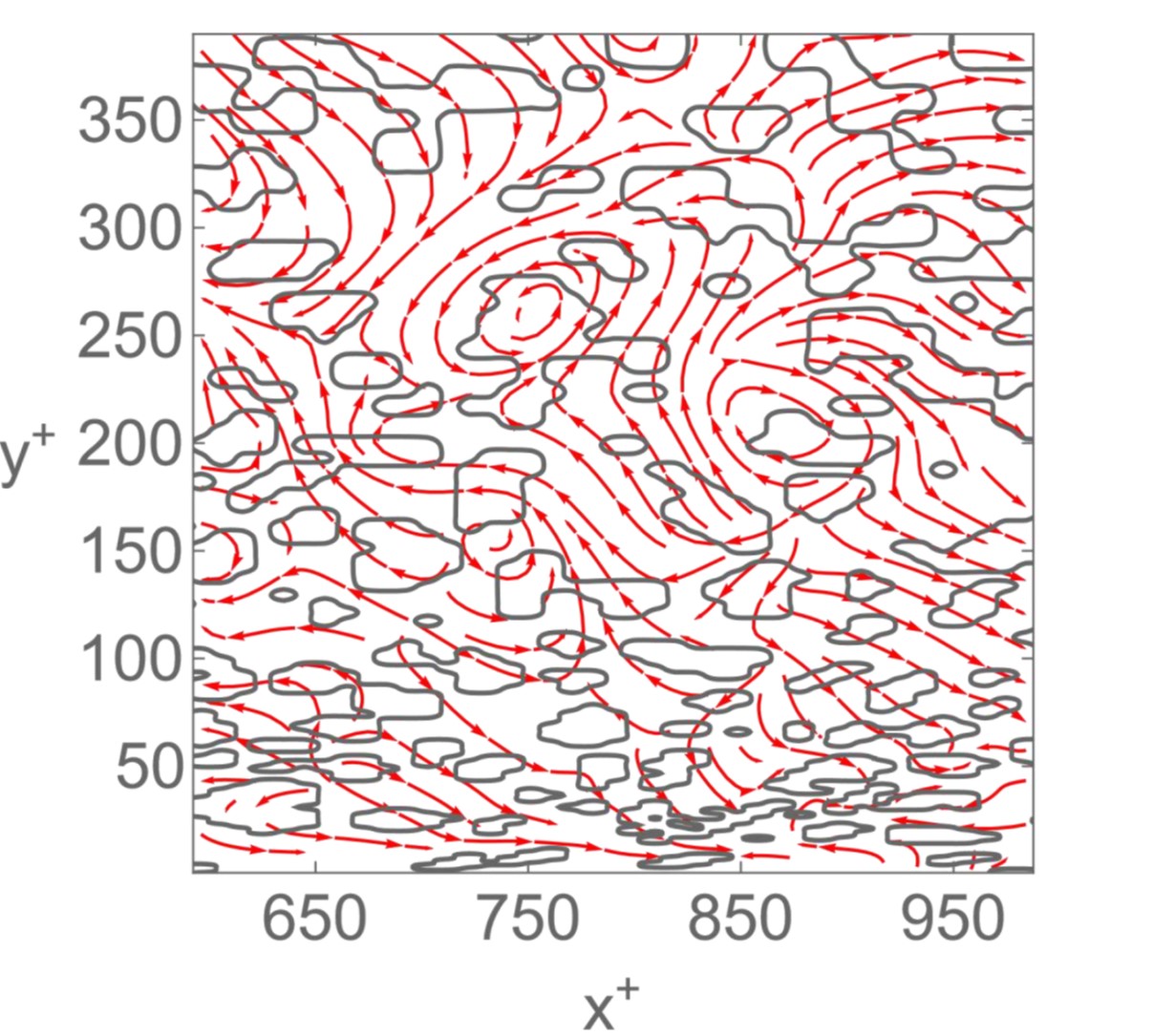}
   \caption{{\black{Streamlines (red lines) for the velocity fluctuations around the mean flow and the closed 
   contours (black lines) of vortices identified through the vorticity curvature criterion}}, in the 
   region of wall units $0 \leq y^+ \leq 395$ and $ 590 \leq  x^+ \leq 990 $ (corresponding to
   $0 \leq y \leq 1$ and $1.5 \leq x \leq 2.5$ in Fig. 17c).}
    \label{fig:zoom}
  \end{figure}
    
    The turbulent channel flow simulation has friction Reynolds number $Re_\tau \simeq 395$ and setup parameters described in
  Table \ref{DNS}. We follow here the simulation guidelines put foward by Kim, Moin and Moser \cite{kim_moin_moser}.
  The streamwise, normal to the wall, and spanwise coordinates are, respectively, $x$, $y$, and $z$; periodic boundary conditions 
	are imposed along the streamwise and spanwise directions; the grid is not uniform, with enhanced resolution near the walls, so 
	that the viscous sublayer can be resolved with approximately one viscous length per lattice spacing. The simulation has been 
	validated by standard tests, like the reproduction of the law of the wall and of statistical moments.
    
    We have recorded, at every ten timesteps in the turbulent stationary regime, the projection of the velocity field of three
  parallel streamwise/wall normal planes $z=0$, $z=\pi/3$ and $z=2\pi/3$. The ensemble defined in
  this way has a total number of $5268$ flow configuration snapshots, which are, then, studied as two-dimensional velocity
  fields.
  
    \begin{table}[ht]
      \centering
      \begin{tabular}{|l|l|} \hline
       System's Dimensions & $ (L_x,L_y,L_z) = (2 \pi, 2, \pi)$ \\ \hline
       Grid Size & $256 \times 192 \times 192 $ \\ \hline
       Kinematic Viscosity & $\nu \simeq  8.6 \times 10^{-4}$ \\ \hline
       Kinematic Pressure Gradient  & $dP/dx = 0.11$ \\ \hline
       Simulation Time Step & $\Delta t = 1.2 \times 10^{-3}$\\ \hline
      \end{tabular}
      \caption{Parameters for the DNS of a turbulent channel flow.}
      \label{DNS}
    \end{table}
    
    We show, in Fig. \ref{fig:dns2}, vortex identification images for one representative snaphost,
		analysed in three different ways. Figs. \ref{fig:dns2}a and \ref{fig:dns2}b give the results obtained from the application of the
  $\lambda_{ci}$-criterion without and with the use of the background subtraction procedure, respectively. Fig. \ref{fig:dns2}c is
	the analogous result associated to the use of the $\lambda_\omega$-criterion with background subtraction; no circulation 
	cutoff has been used in the identification of vortices. 
	
	There are expressive qualitative differences between the two images 
	produced by the $\lambda_{ci}$-criterion, for regions which are closer to the wall, where shear effects become more relevant. The 
	$\lambda_\omega$-criterion leads, on the other hand, to much better vortex resolution, but the background subtraction procedure does not lead, 
	in visual terms, to expressive modifications - that's why we have not shown the picture associated to the application of the 
	$\lambda_\omega$-criterion without background subtraction. This, in fact, suggests that the flow takes place in weak background shear conditions. 
	There are, however, small but meaningful improvements from the use of the background subtraction procedure that become evident only through
	histogram analysis, as we will show below. 
        
    As a practical remark to be emphasized here, we note that as it is a higher order derivative method, the 
  $\lambda_\omega$-criterion is related to identification fields that typically fluctuate over a much wider range of values
  than the ones associated to the $\lambda_{ci}$-criterion. This justifies our use of the logarithmic scale in the
  elaboration of the image given in \ref{fig:dns2}c. Fixing attention on the $\lambda_\omega$-criterion, 
	the natural application of the logarithmic scale implies, furthermore, that an optional use of thresholds is somewhat delicate for 
	the case of turbulent (intermittent) flows: in fact, if the threshold is defined, for instance, as 20$\%$ of the maximum value 
	of the logarithm of the $\lambda_\omega$ field, then its effects are likely to be irrelevant, since only structures with very low
  kinetic energy would be discarded; alternatively, if an analogous definition of the threshold is given in a linear scale, 
  it is not difficult to see that almost all of the vortex structures would be erased in this way.
  
  {\black{A closer look at the the structures identified by the $\lambda_\omega$-criterion is given in Fig. \ref{fig:zoom}, where we plot their contours and the surrounding streamlines, computed for the velocity field fluctuations around their mean values. The streamwise and wall normal coordinates are defined in wall units. From this picture, we can have a hint on some known important features of boundary layer flows, as (i) the larger aspect ratios and typical inclination of structures below the onset of the logarithmic layer ($ y^+ < 30 $), (ii) the scaling of structure sizes with their distances to the wall, (iii) the presence of strong vortices which dominate the local velocity flucutations (there at least, two of these in the picture), and (iv) the fact that zones of quasi-uniform momentum are correlated with vortex regions \cite{mein_adrian}, which in our specific example is particularly clear from the organization of streamlines in the upper region of the sample ($y^+ > 300$).}}
            
\begin{figure}[ht]
         \begin{minipage}{0.48\linewidth}
           \includegraphics[width=\linewidth]{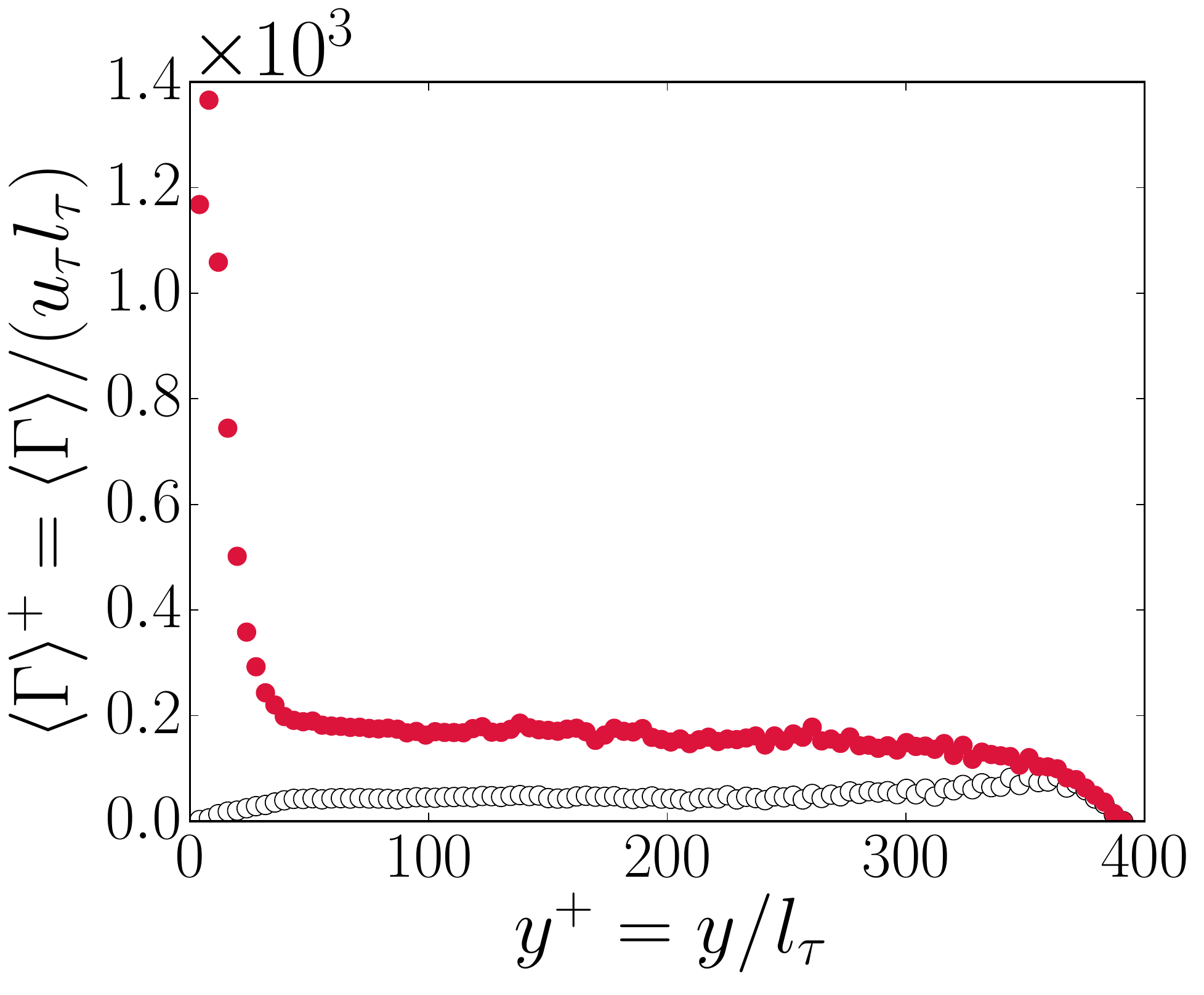}
            \put (-80,72){\makebox[0.05\linewidth][r]{(a)}}
						\end{minipage}
						         \begin{minipage}{0.48\linewidth}
           \includegraphics[width=\linewidth]{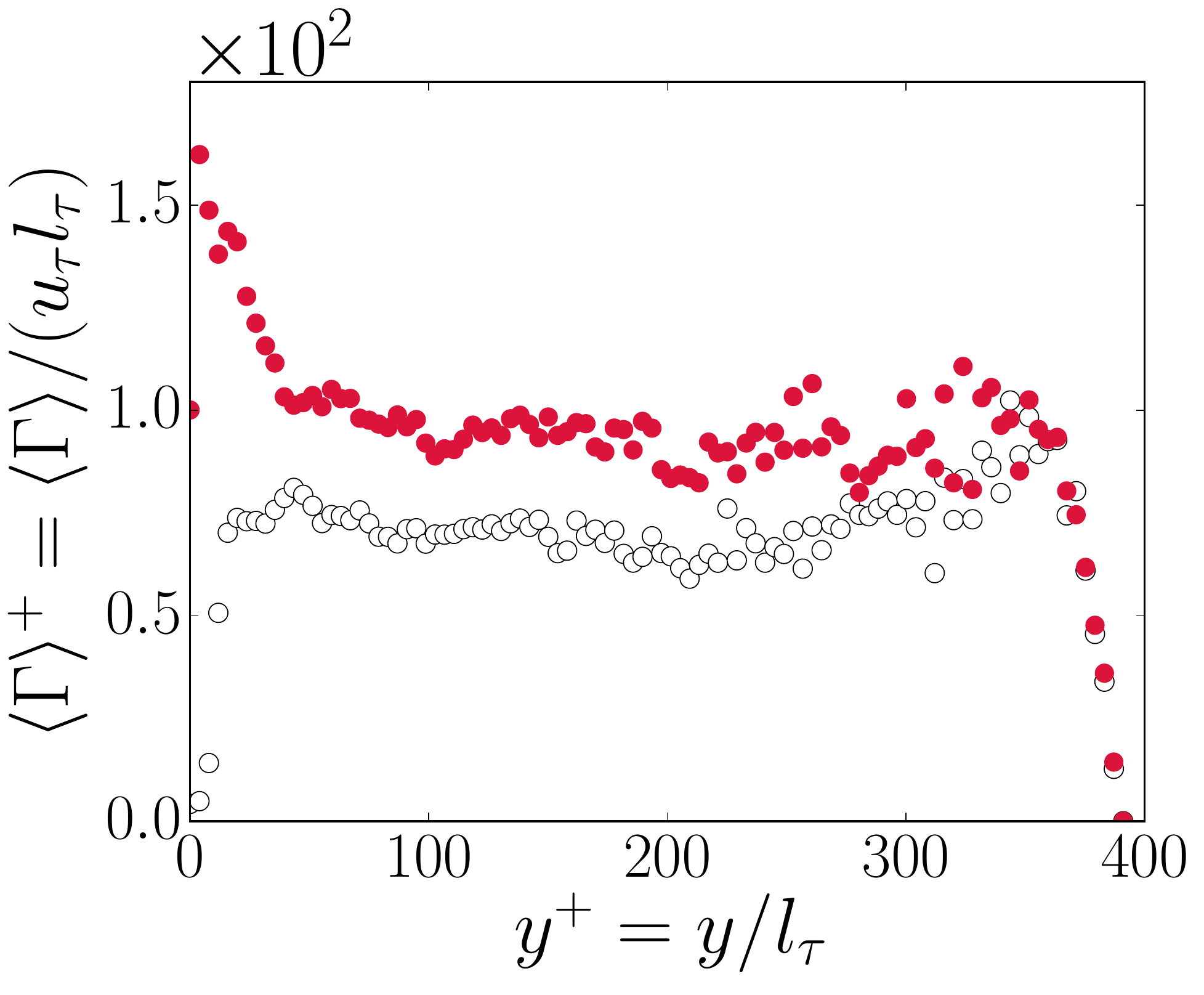}
            \put (-80,72){\makebox[0.05\linewidth][r]{(b)}}
						\end{minipage}
						\begin{minipage}{0.48\linewidth}
           \includegraphics[width=\linewidth]{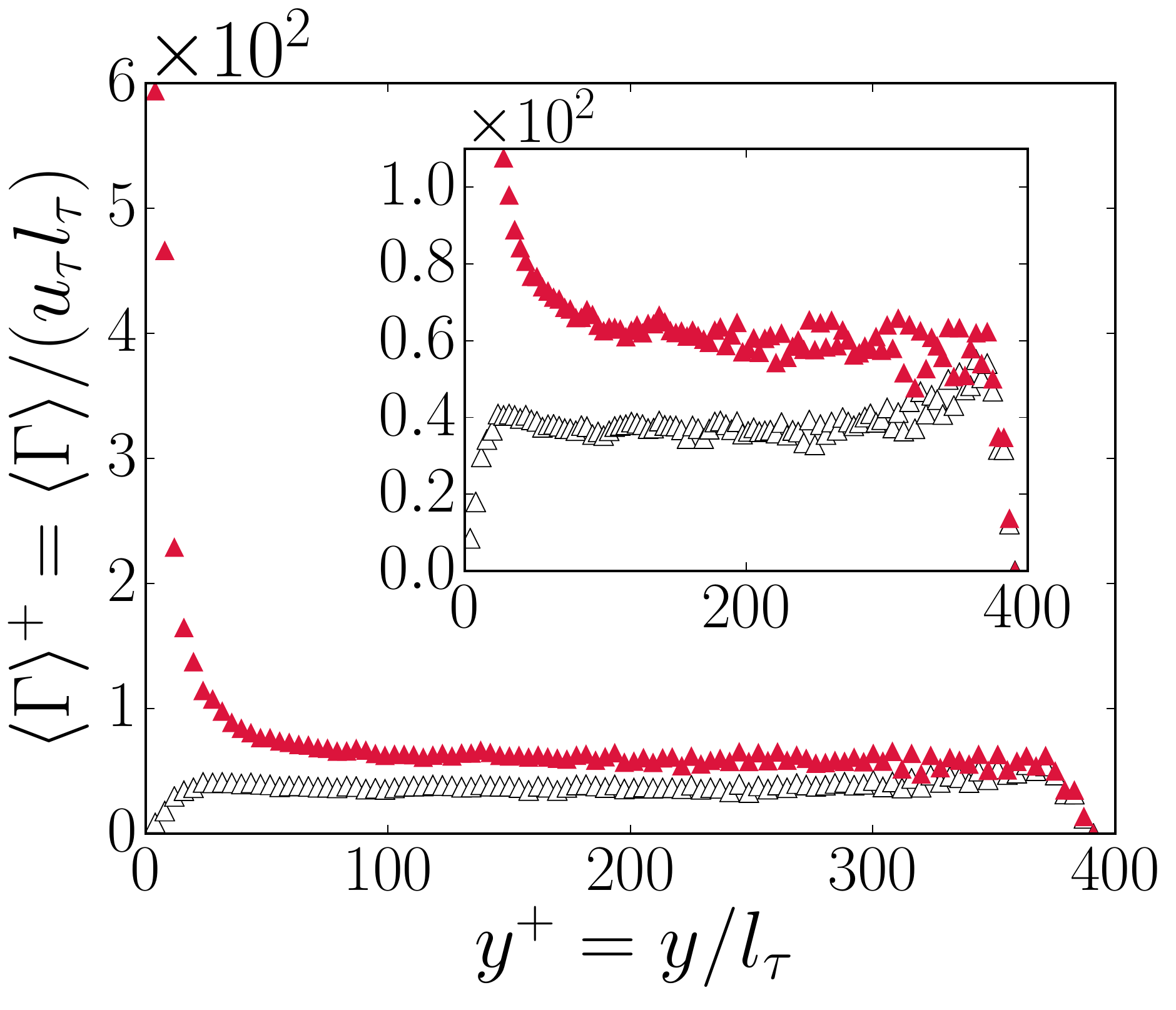}
            \put (-80,75){\makebox[0.05\linewidth][r]{(c)}}
         \end{minipage}	
						\begin{minipage}{0.48\linewidth}
           \includegraphics[width=\linewidth]{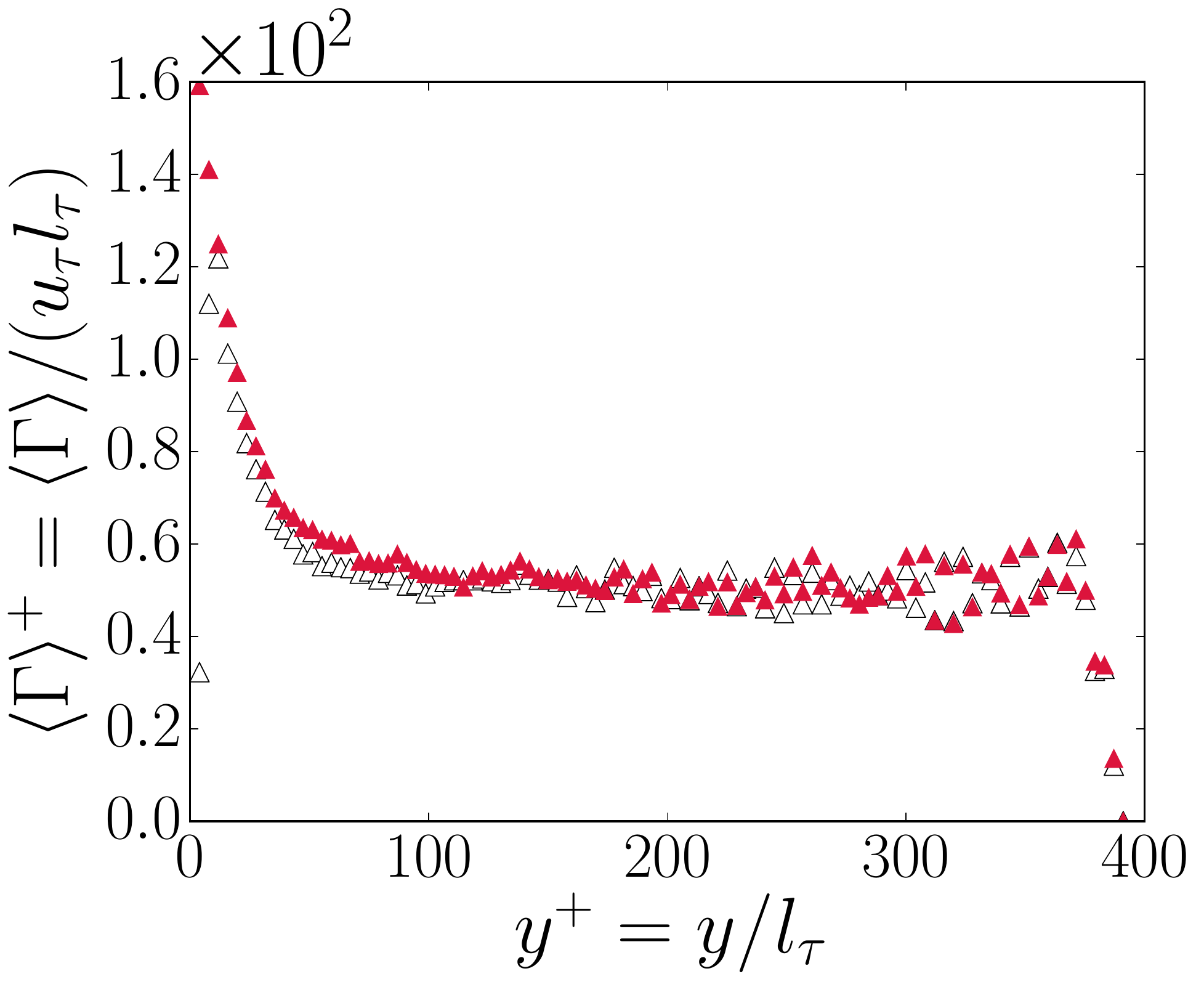}
            \put (-80,72){\makebox[0.05\linewidth][r]{(d)}}
         \end{minipage}          
      \caption{Absolute mean values of the circulation for retrograde (open symbols) and prograde (solid symbols) vortices, as a
               function of the distance to the wall. All the quantities are given in wall units (friction velocity $u_\tau
               \simeq  0.34$ and viscous length $l_\tau \simeq 2.5 \times 10^{-3})$. Plots (a) and (b) are associated to
               vortex identification by the $\lambda_{ci}$-criterion, while (c) and (d) are associated to the 
               $\lambda_\omega$-criterion. The background subtraction procedure has been applied only for the results depicted in 
               (b) and (d).}
      \label{fig:dns3}      
    \end{figure}

\begin{figure}[ht]
         \begin{minipage}{0.48\linewidth}
           \includegraphics[width=\linewidth]{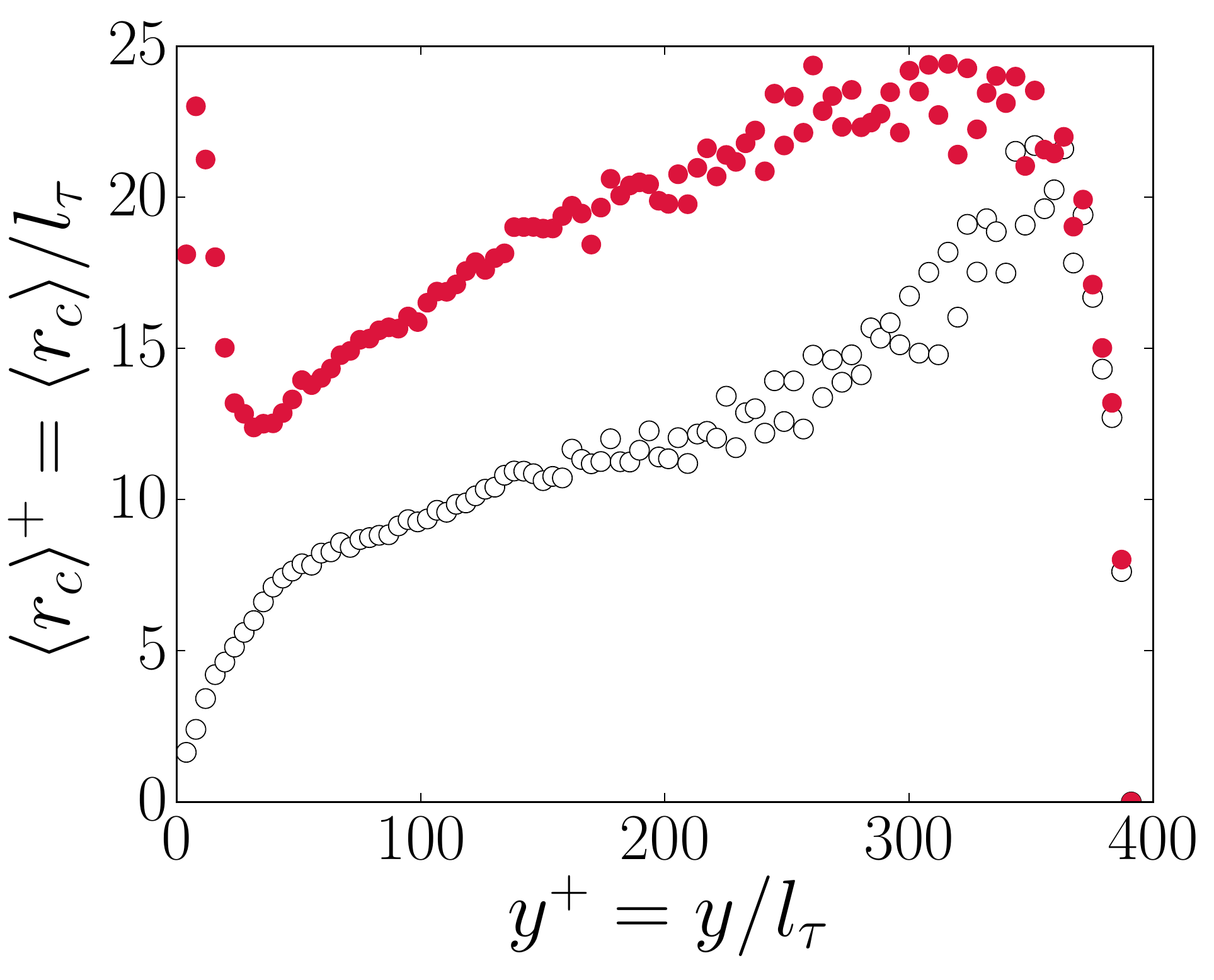}
           \put (-77,72){\makebox[0.05\linewidth][r]{(a)}}\\
           \includegraphics[width=\linewidth]{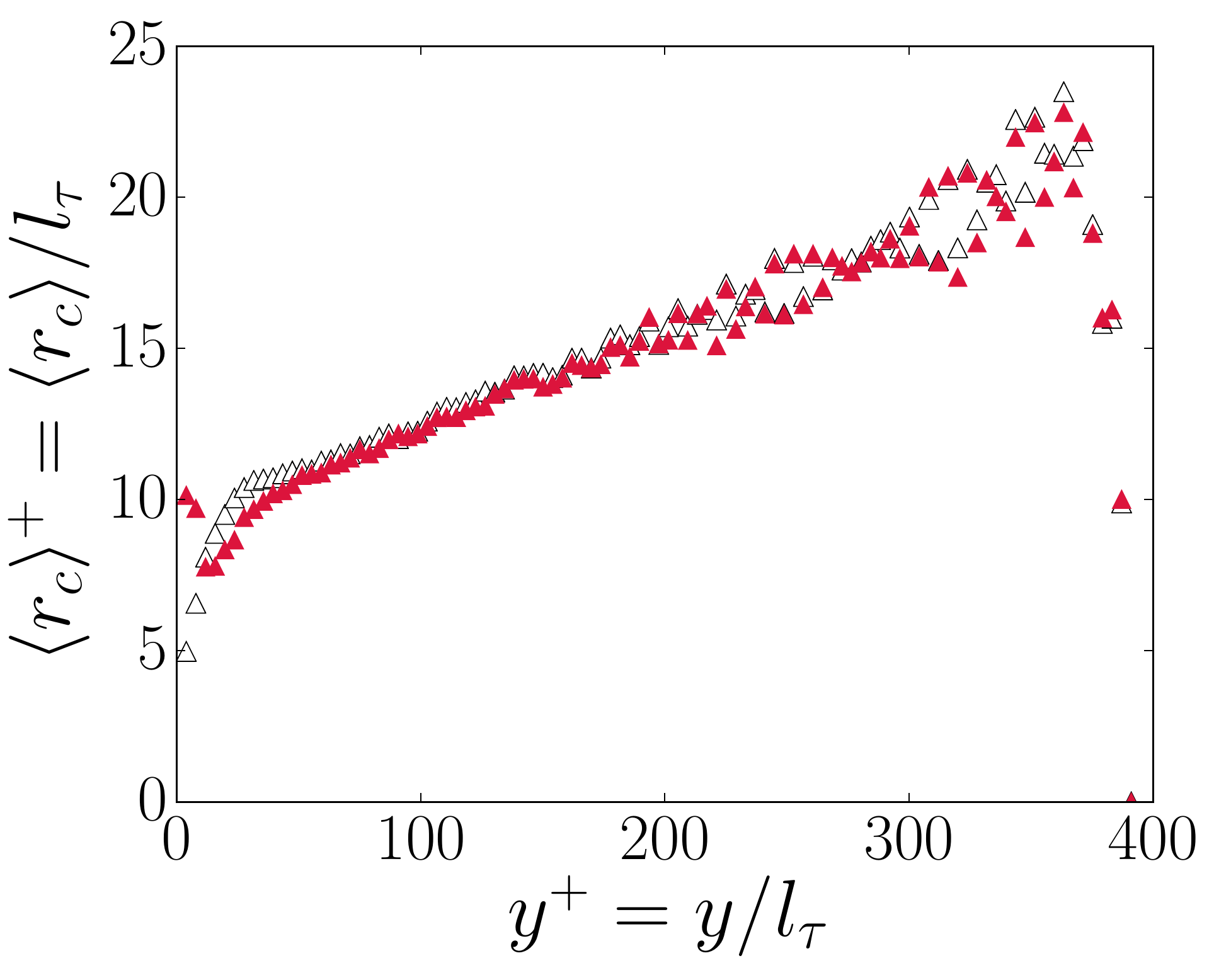}
           \put (-77,72){\makebox[0.05\linewidth][r]{(c)}}
         \end{minipage}
         \begin{minipage}{0.48\linewidth}
           \includegraphics[width=\linewidth]{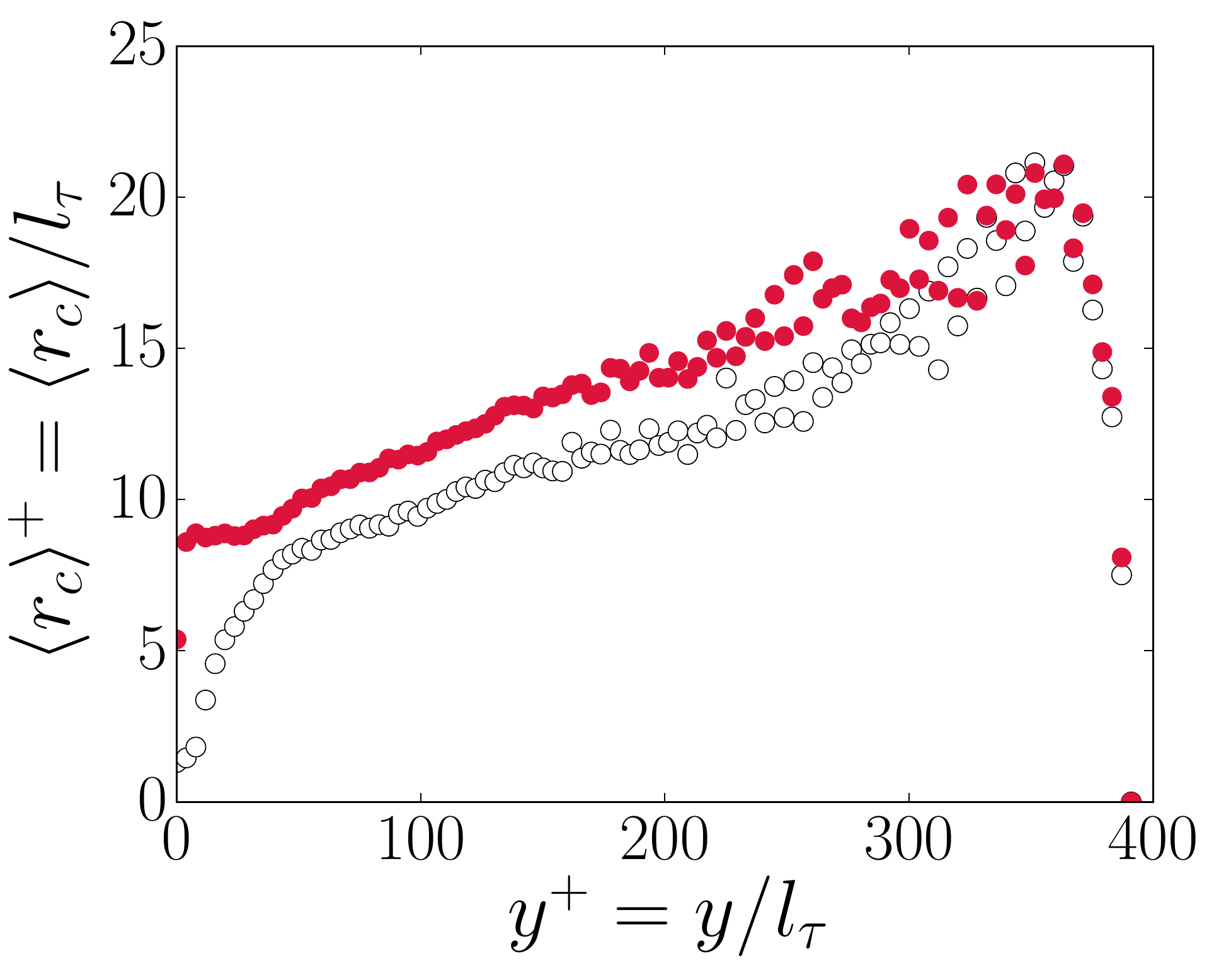}
           \put (-77,72){\makebox[0.05\linewidth][r]{(b)}}\\
           \includegraphics[width=\linewidth]{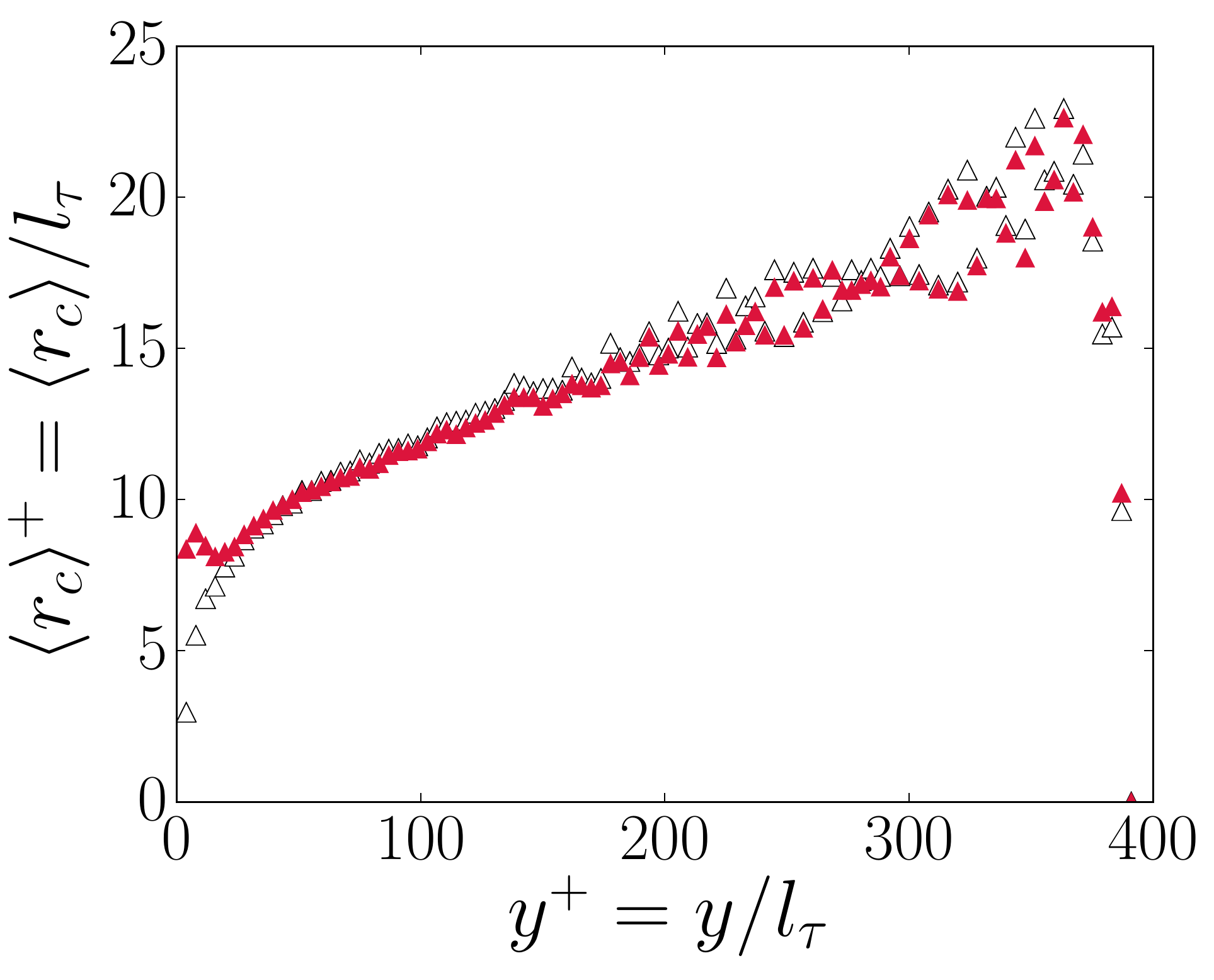}
           \put (-77,72){\makebox[0.05\linewidth][r]{(d)}}
         \end{minipage}          
        
      \caption{Mean radius values for retrograde (open symbols) and prograde (solid symbols) vortices, as a function of the distance 
               to the wall. All the rest as in the caption of Fig. \ref{fig:dns3}.}
      \label{fig:dns4}      
    \end{figure}

\begin{figure}[ht]
         \begin{minipage}{0.48\linewidth}
           \includegraphics[width=\linewidth]{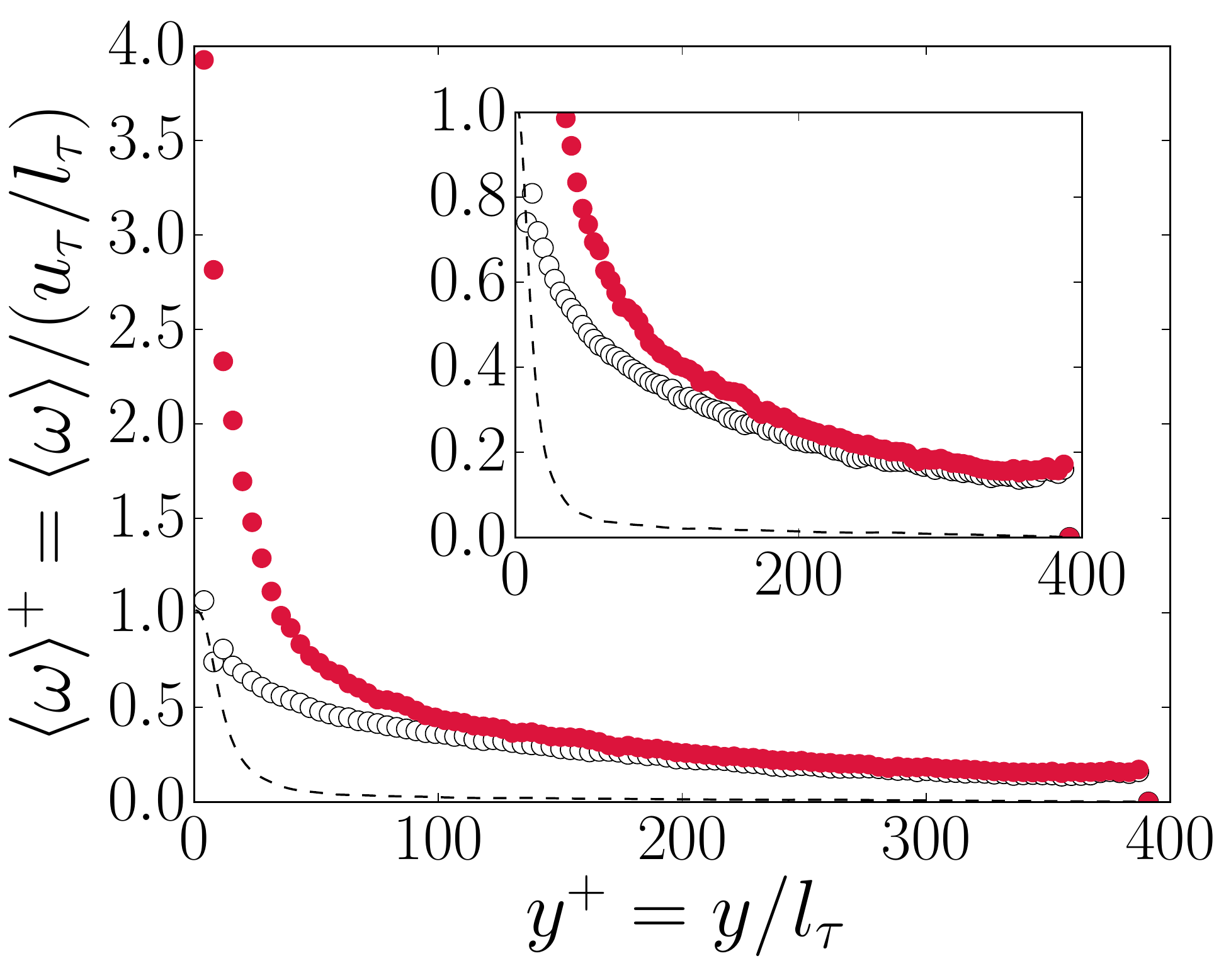}
           \put (-80,73){\makebox[0.05\linewidth][r]{(a)}}\\
           \includegraphics[width=\linewidth]{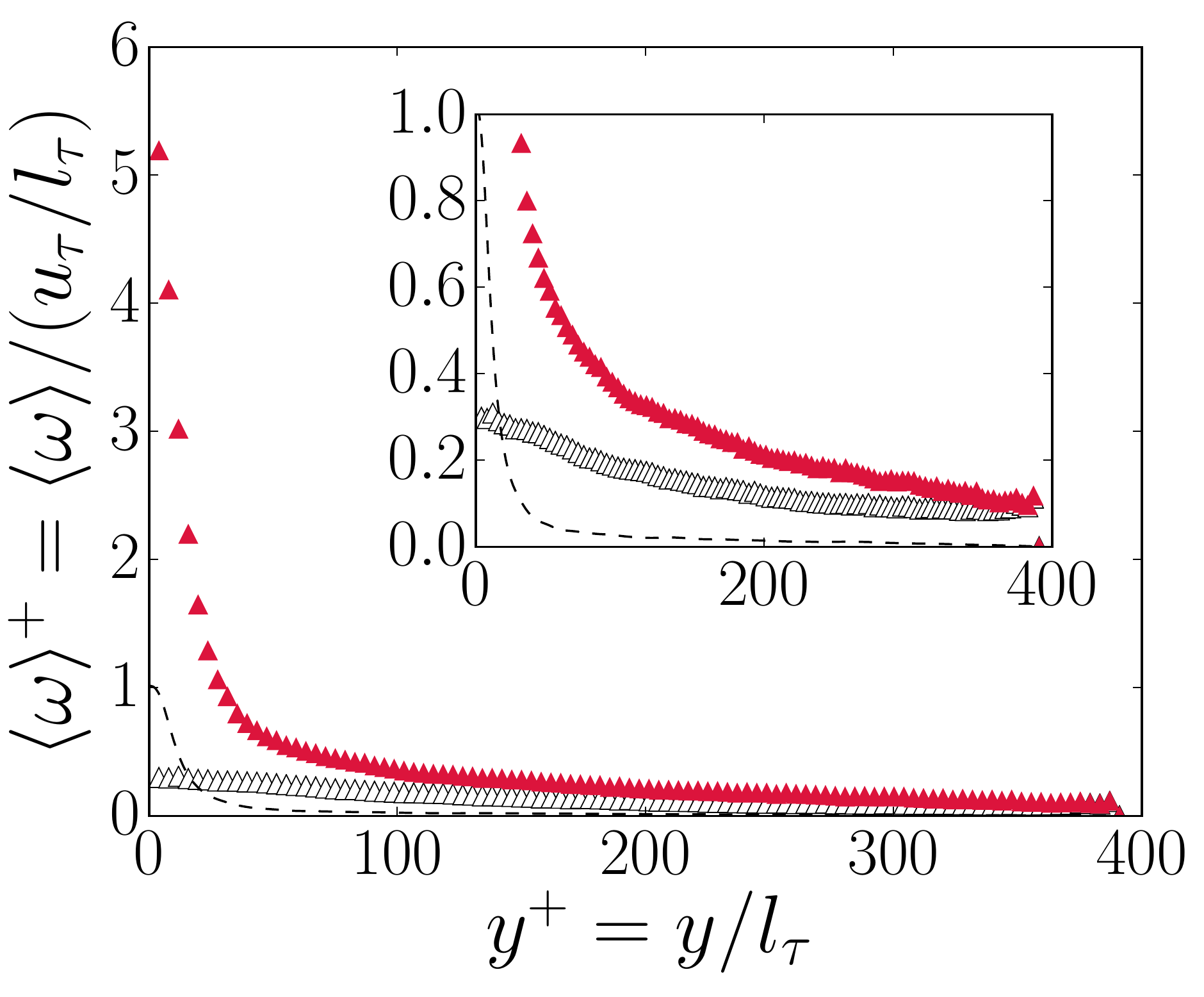}
           \put (-80,73){\makebox[0.05\linewidth][r]{(c)}}
         \end{minipage}
         \begin{minipage}{0.48\linewidth}
           \includegraphics[width=\linewidth]{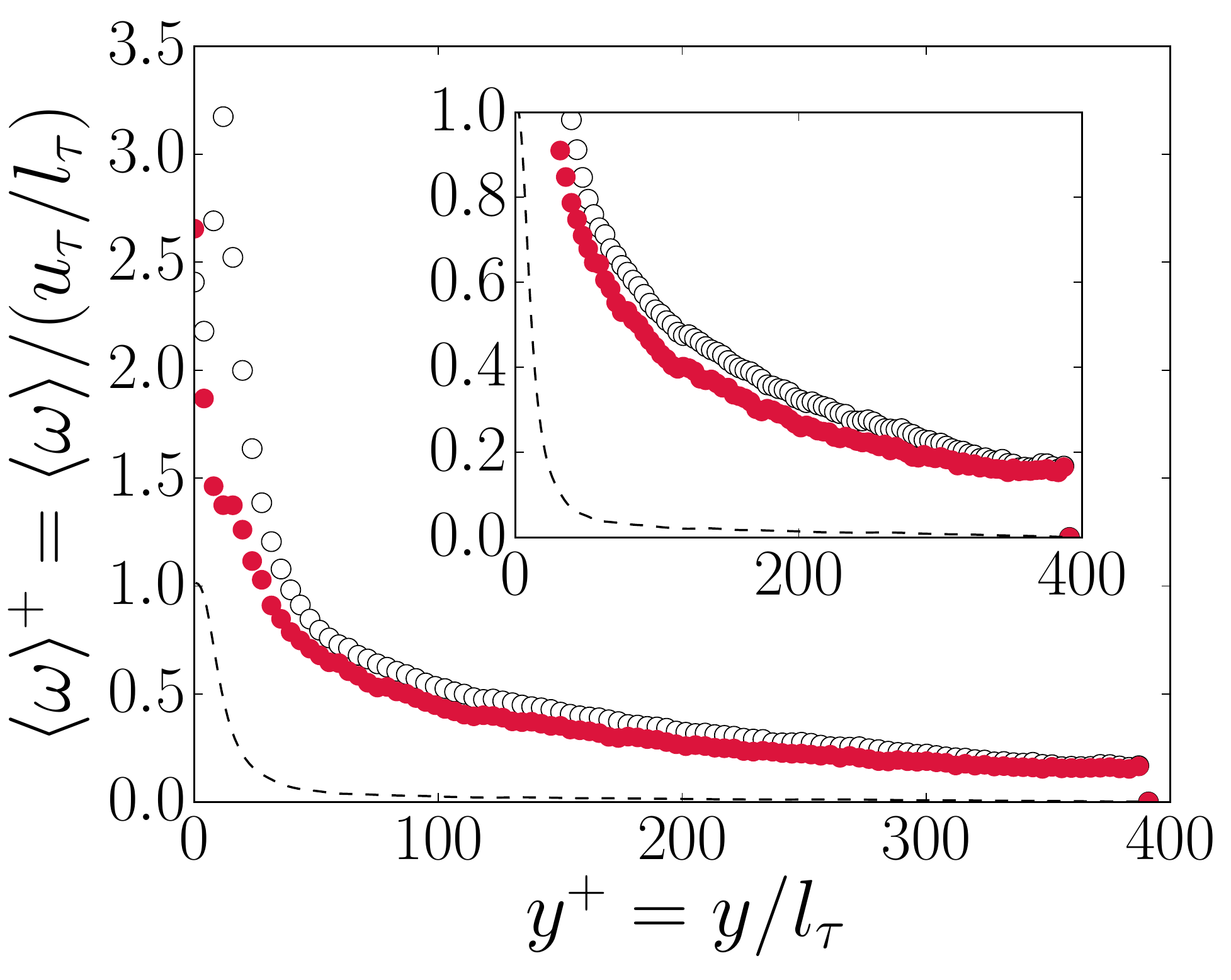}
           \put (-80,72){\makebox[0.05\linewidth][r]{(b)}}\\
           \includegraphics[width=\linewidth]{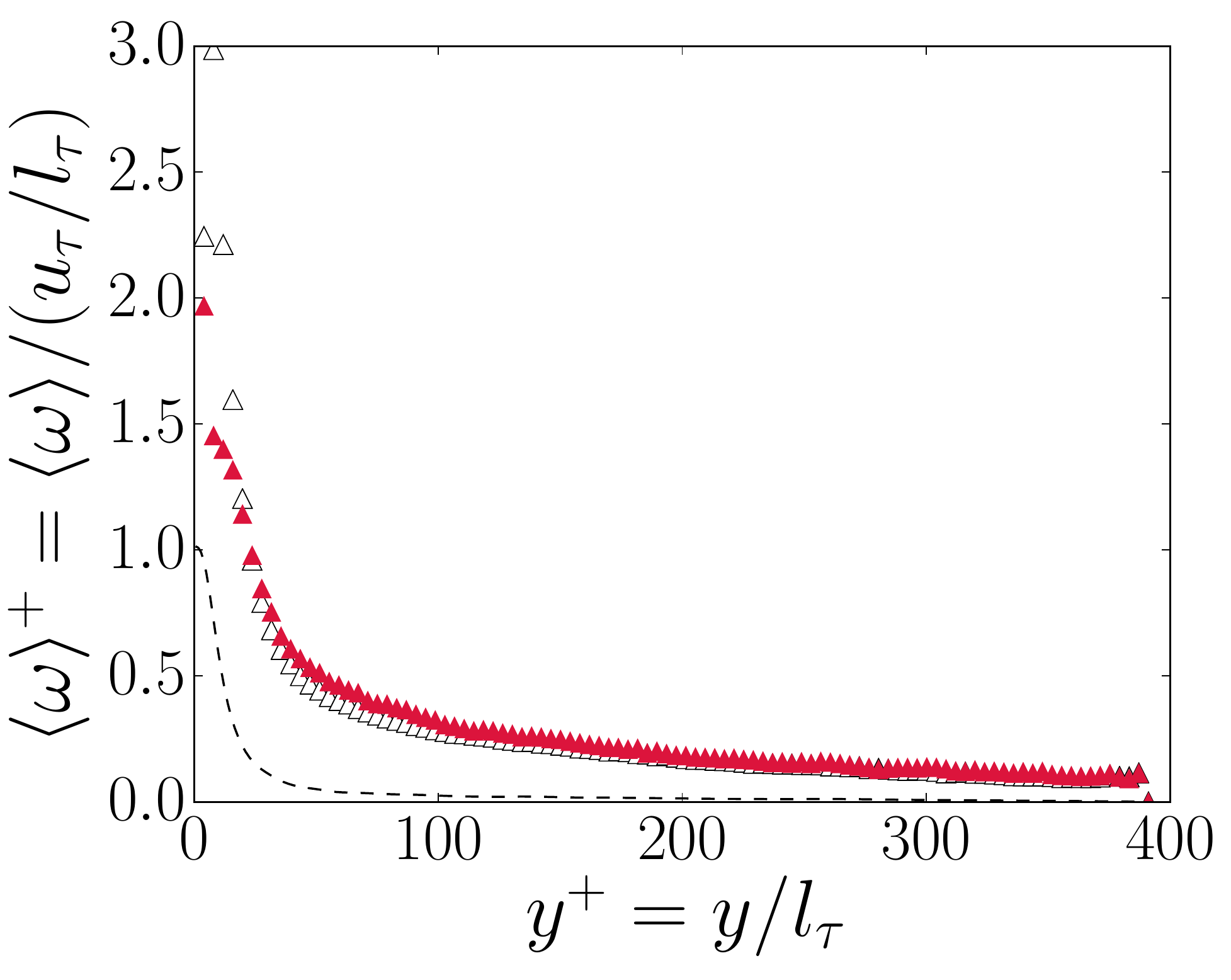}
           \put (-80,72){\makebox[0.05\linewidth][r]{(d)}}
         \end{minipage}          

      \caption{Absolute mean values of the peak vorticity, i.e., $|\langle \Gamma / \pi r_c^2 \rangle |$ for retrograde (open symbols) and prograde
               (solid symbols) vortices, as a function of the distance to the wall. The dashed line is the average vorticity of 
							 the turbulent channel (which closely agrees with the law of the wall). All the rest as in the caption of 
							 Fig. \ref{fig:dns3}.}
      \label{fig:dns5}      
    \end{figure}

\begin{figure}[ht]
      \begin{minipage}{0.48\linewidth}
           \includegraphics[width=\linewidth]{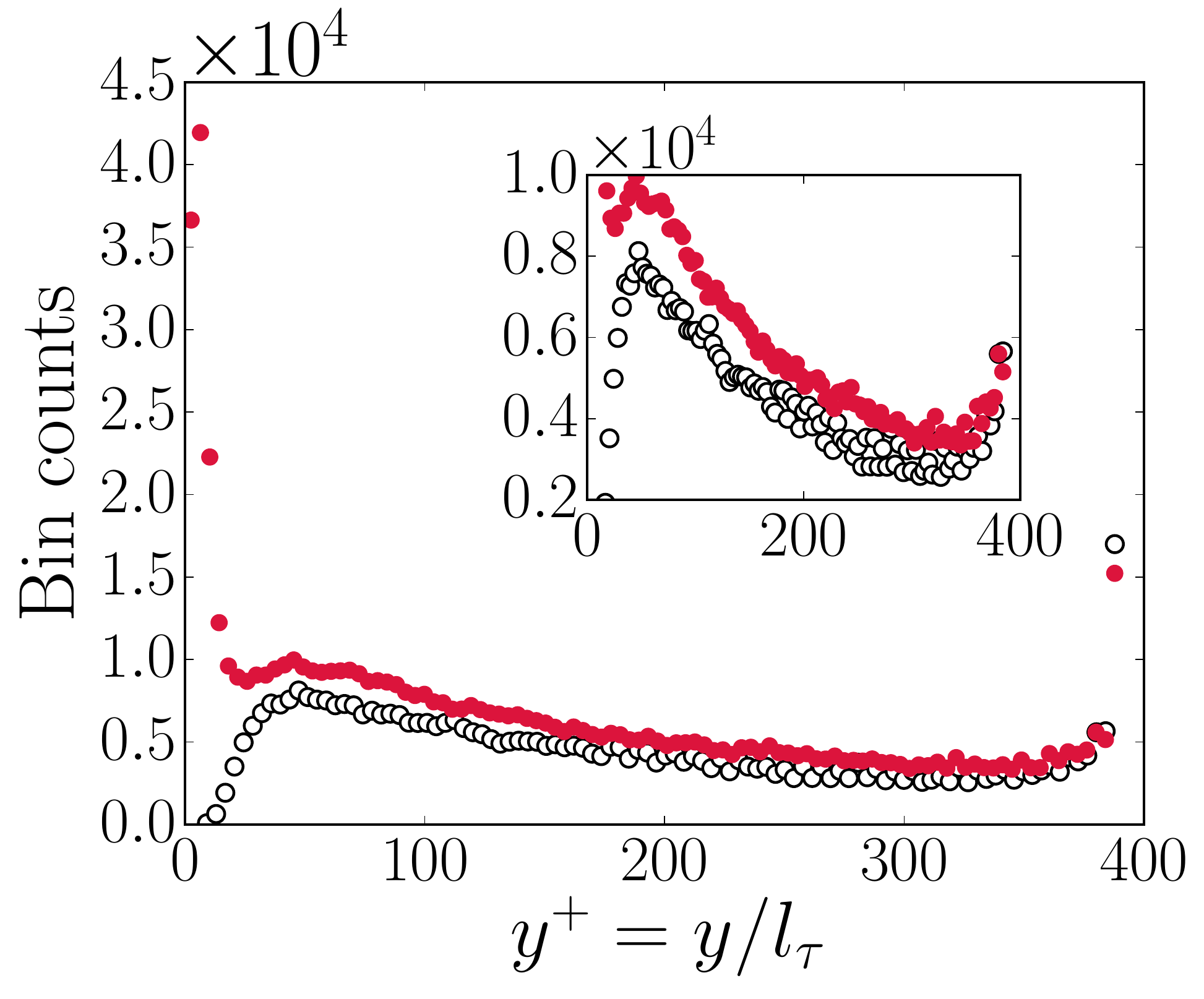}
           \put (-77,72){\makebox[0.05\linewidth][r]{(a)}}\\
           \includegraphics[width=\linewidth]{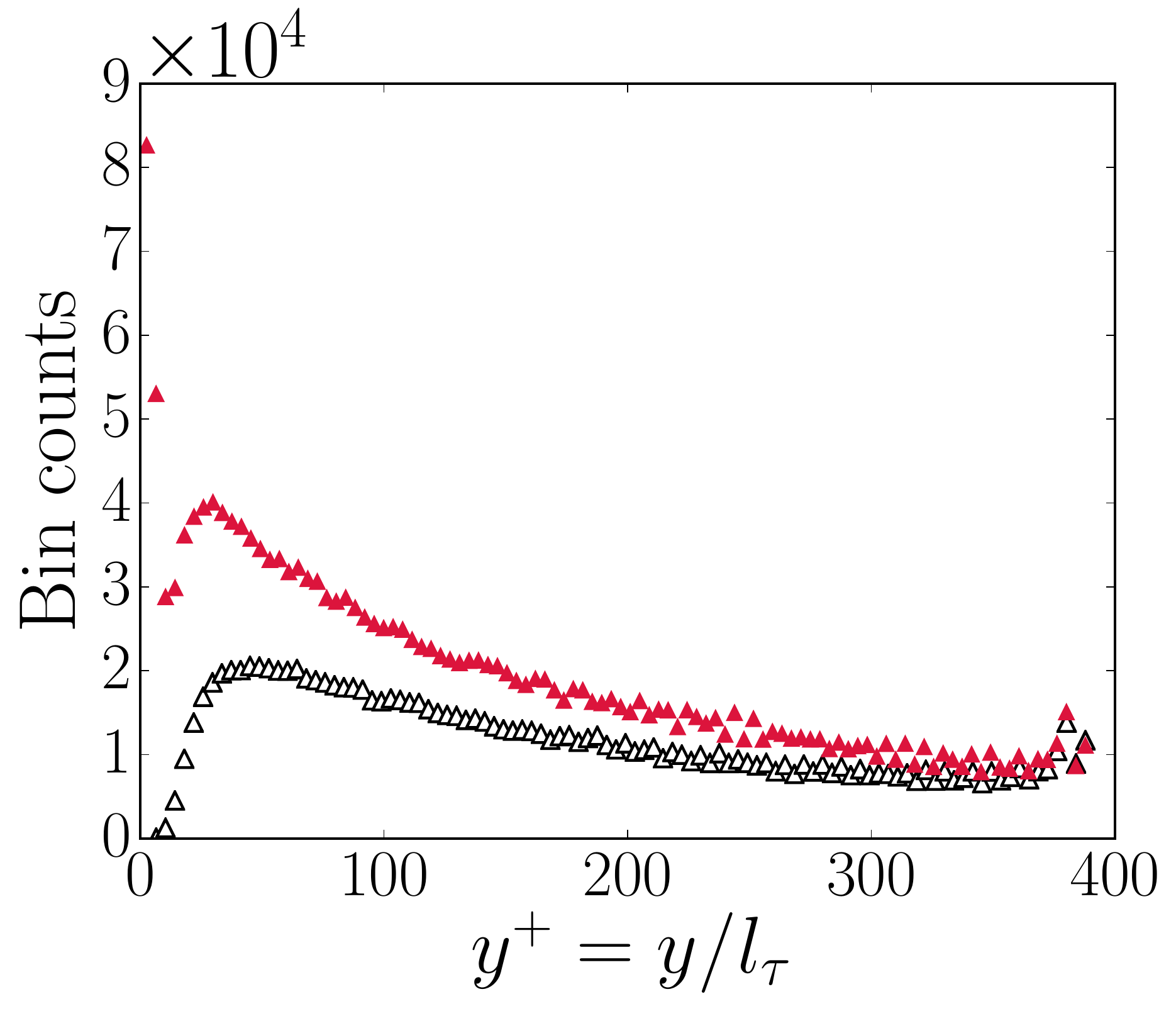}
           \put (-77,72){\makebox[0.05\linewidth][r]{(c)}}
         \end{minipage}
         \begin{minipage}{0.48\linewidth}
           \includegraphics[width=\linewidth]{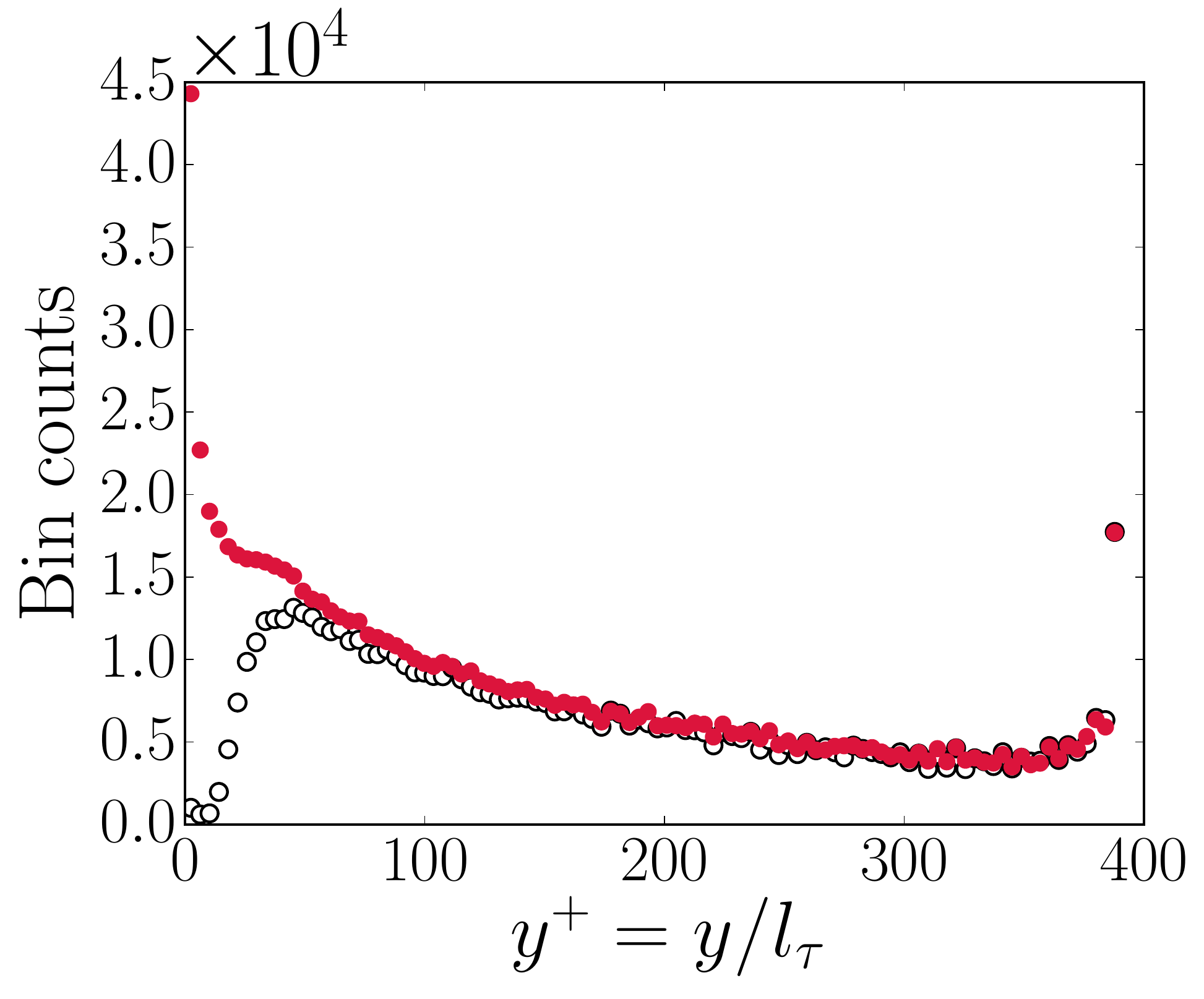}
           \put (-77,72){\makebox[0.05\linewidth][r]{(b)}}\\
           \includegraphics[width=\linewidth]{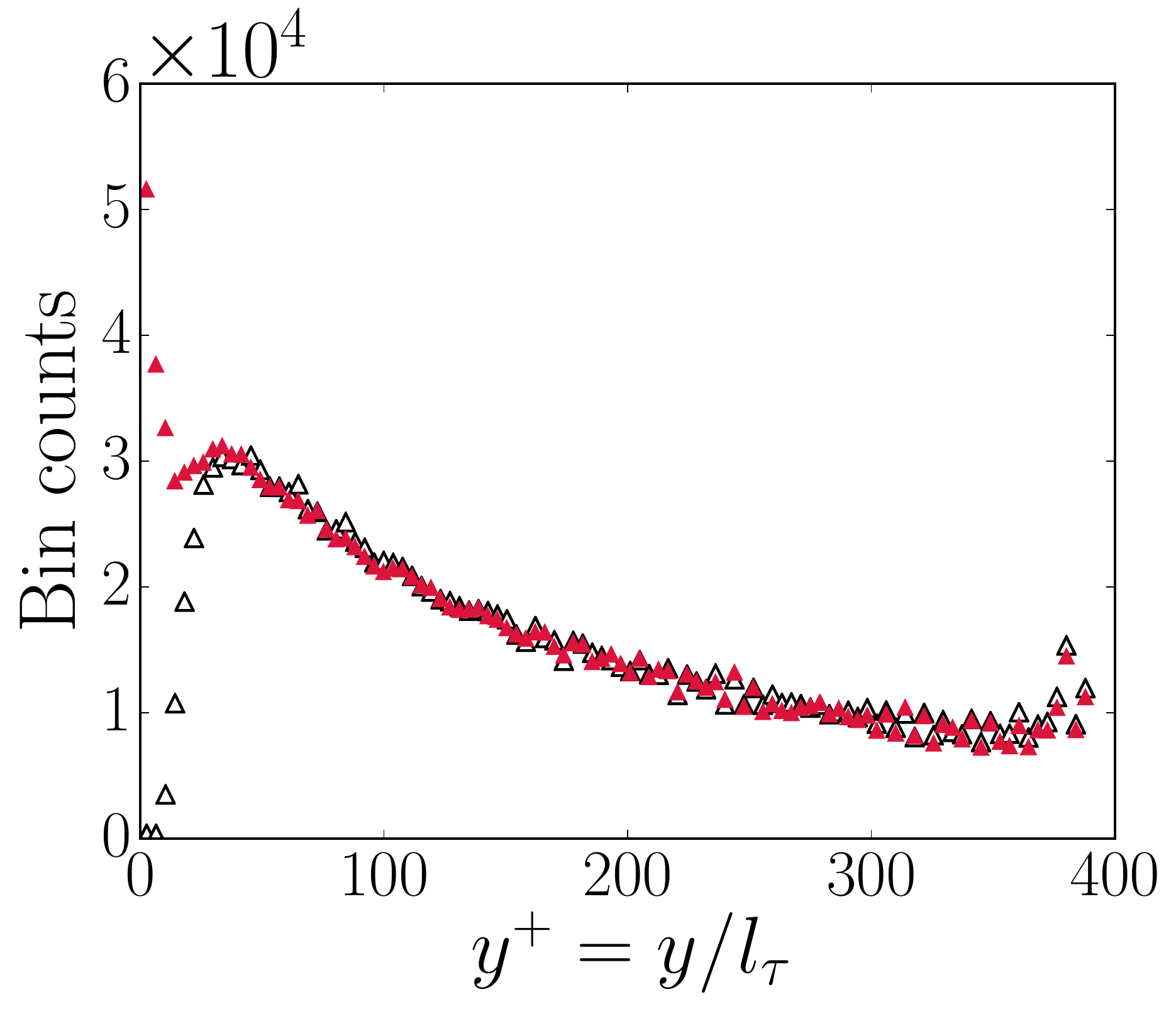}
           \put (-77,72){\makebox[0.05\linewidth][r]{(d)}}
         \end{minipage} 
         
         \caption{Vortex counting per stripe of width $\Delta y^+ = 4$, for retrograde (open symbols) and prograde (solid symbols)
                  vortices, as function of the distance to the wall. All the rest as in the caption of 
							    Fig. \ref{fig:dns3}.}
      \label{fig:dns1}      
    \end{figure}

    The streamwise/wall normal plane snapshots of the turbulent channel flow are partitioned in thin streamwise stripes which have
  vertical width (bin size) $\Delta y^+ \approx 4$. Through a computational strategy analogous to the one discussed in the
  previous section, we identify vortices for each one of the stripes, and determine their mean circulation, peak vorticity, 
	mean radius, and mean number as a function of the stripe distance to the wall. Results are reported in 
	Figs. \ref{fig:dns3}-\ref{fig:dns1}. We provide, for some of the pictures, insets which magnify their details,
	for the sake of better visual inspection.
  
    Similar evaluations of the mean vorticy and mean vortex radii as a function of the distance to the wall have been
  discussed in Ref. \cite{herpin_a,herpin}, where, however, vortex
  parameters are obtained from Levenberg-Marquardt fittings of the identified structures to the Lamb-Oseen vortex pattern. 
  Their results, derived from a large turbulent database are compatible with ours, in the context of the $\lambda_{ci}$-criterion.
  
    The application of the $\lambda_\omega$-criterion to the turbulent channel DNS data brings a phenomenologically
  interesting perspective on the statistical properties of the spanwise vortices. It is clear, from Figs. \ref{fig:dns3} -
  \ref{fig:dns5}, that even with the use of the background subtraction procedure, the $\lambda_{ci}$-criterion gives, for all
  the heights, distinct absolute values of the mean circulations, vorticities, and radii for the populations of positive
  (retrograde) and negative (prograde) vortices. The application of the background subtraction procedure in the
  $\lambda_\omega$-criterion yields, on the other hand, a fine collapse of these quantities for $y^+ > 50$, which extends all
  throughout the logarithm boundary layer, as it can be appreciated from Figs. \ref{fig:dns3}d, \ref{fig:dns4}d,
  and \ref{fig:dns5}d. If we now take a look at the populations of prograde and retrograde vortices in Figs. \ref{fig:dns1}b
  and \ref{fig:dns1}d, they are found to match each other in both criteria, but only after the background subtraction
  procedure is carried out.
  
      We know, from the law of the wall, that the mean vorticity background is, in the logarithm layer, $\langle \omega^+
  \rangle = 2.5 / y^+$. It is clear, thus, from the inspection of Fig. \ref{fig:dns5}, that the mean peak vorticity of the
  vortex structures is well above the vorticity background value for $y^+ > 50$, which tells us that the 
	there is in fact a weak background shear regime in the log-layer, following the convention put forward in the previous 
	section. However, as it is suggested from Fig. \ref{fig:dns5}d, the buffer layer is likely to be the region where
	shear effects can become relevant in the problem of vortex identification.
	
	{\black{From the above compilation of statistical results, we find that the detected vortical structures have their vorticities and circulations enhanced within the region $5< y^+ < 30$. This is likely to be related to the observation that near the bottom of the buffer layer, streamwise velocity fluctuations become more intermittent as the distance to the wall decreases, as quantified by a kurtosis analysis \cite{lorko}. A simple explanation of why individual vortices carry stronger vorticity as they get closer to the wall can be addressed from a combination of the no-slip boundary condition with the attached eddy hypothesis \cite{mori}. It is expected, of course, that fluctuations will disappear deep down in the viscous layer, $y^+ < 5$, which, unfortunately, is poorly resolved in our data.}}
	
	    The data collapse attained in Figs. \ref{fig:dns3}d, \ref{fig:dns4}d, \ref{fig:dns5}d, and \ref{fig:dns1}d is an important 
		point for the consolidation of the $\lambda_\omega$-criterion, once it supports the long-standing phenomenological assumption 
		of small scale turbulence isotropization in turbulent boundary layers \cite{corrsin,saddoughi,xu,tsinober}. The $\lambda_{ci}$-criterion 
		yields data collapse only for the vortex counting histogram, Fig. \ref{fig:dns1}b, failing to do so in the evaluations of vortex 
		circulation, peak vorticity and radius parameters, as it can be clearly seen from Figs. \ref{fig:dns3}b, \ref{fig:dns4}b, 
		and \ref{fig:dns5}b.
  
	The validity of the isotropic turbulence hypothesis in the turbulent
  boundary logarithm layer has been usually checked with the help of general theoretical relations that should
  hold for the expectation values of some local fluid dynamical observables \cite{corrsin,saddoughi,xu,tsinober}. This is a relevant aspect 
	of the turbulent boundary layer phenomenology that has lacked so far proper corroboration within structural analyses, a fact due,
  essentially, to the limitations of the standard $\lambda_{ci}$ vortex identification methodology.
  
  								  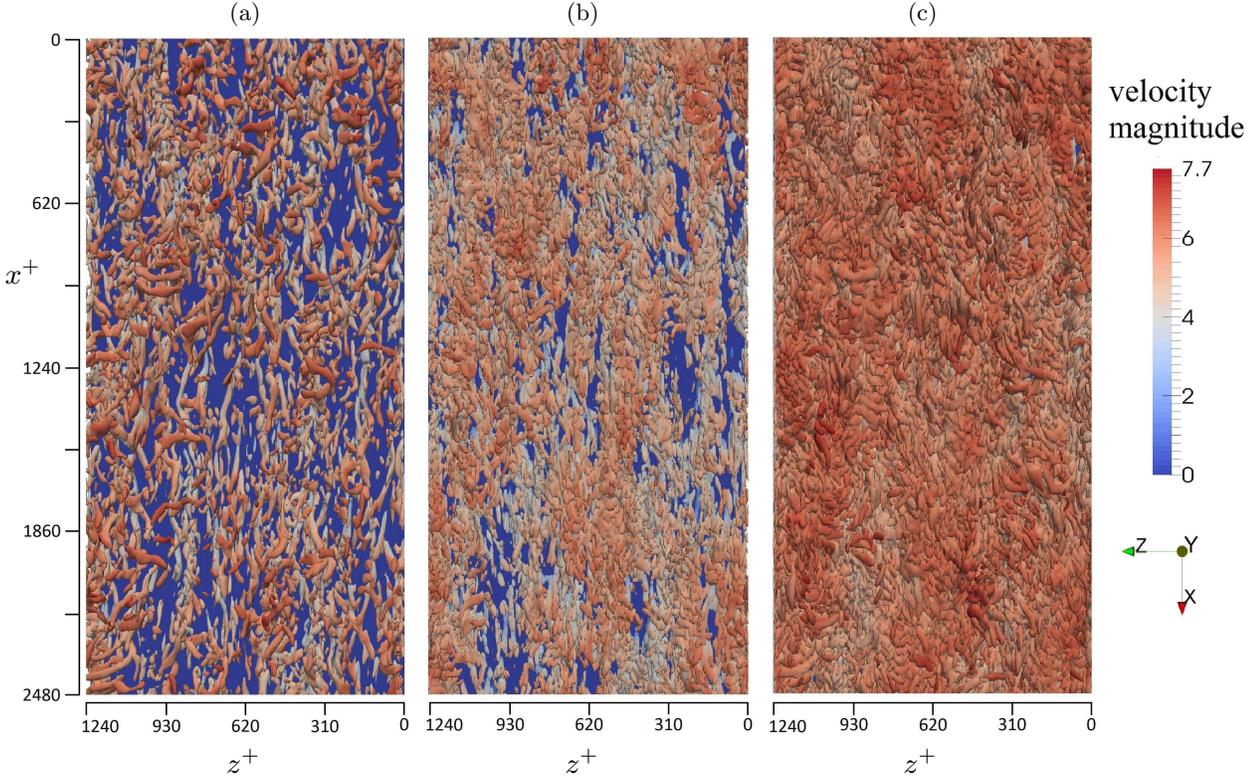
\begin{figure*}[ht]
          \includegraphics[width=\linewidth]{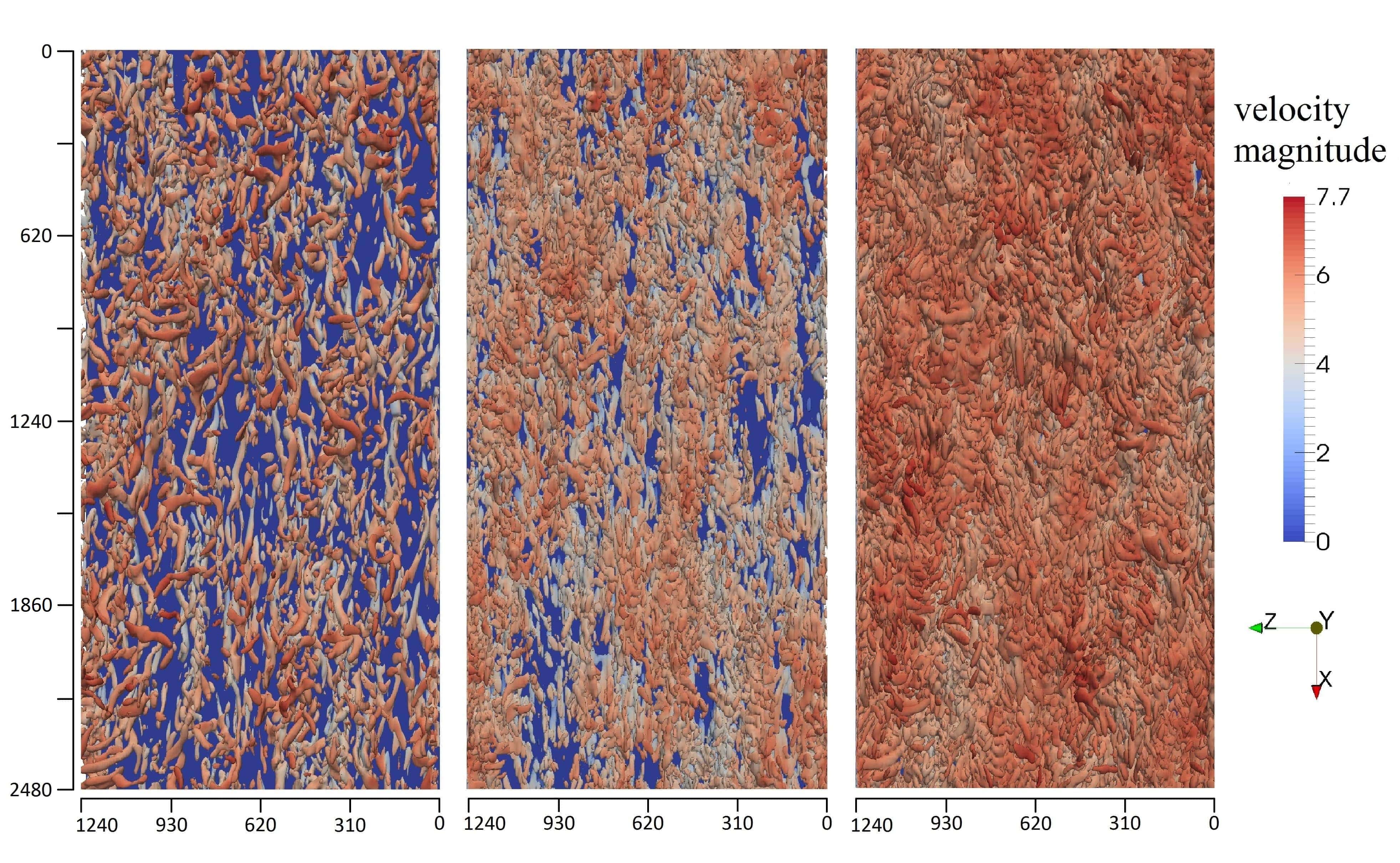}				
          \put (-400,280){\makebox[0.05\linewidth][r]{(a)}}
          \put (-272,280){\makebox[0.05\linewidth][r]{(b)}}
          \put (-144,280){\makebox[0.05\linewidth][r]{(c)}}
          \put (-480,180){\makebox[0.05\linewidth][r]{ \scalebox{1.2}{$x^+$} }}
          \put (-400,-5){\makebox[0.05\linewidth][r]{\scalebox{1.2}{$z^+$}}}
          \put (-272,-5){\makebox[0.05\linewidth][r]{\scalebox{1.2}{$z^+$}}}
          \put (-144,-5){\makebox[0.05\linewidth][r]{\scalebox{1.2}{$z^+$}}}
          \caption{Vortex identification, {\black{with the use of thresholds}}, as seen from the top of the turbulent channel flow, {\black{according to the}} DNS addressed in Sec. \ref{sec:dns}. (a) $Q$-criterion with no background subtraction, $Q>10^2$; (b) $Q_\omega$-criterion with background subtraction, $Q_\omega >1.1 \times 10^9$; (c) $Q_\omega$-criterion with background subtraction, $Q_\omega >1.2 \times 10^8$. {\black{The color scheme gives the magnitude of the velocity field on the coherent structures. The bottom of the channel is depicted as a uniform blue background.}}}
          \label{fig:qqtilda180}
          \end{figure*}
  
\section{Extension to Three-Dimensional Velocity Fields}\label{sec:3dvc}
    It is interesting to devise three-dimensional generalizations of the $\lambda_\omega$-criterion as a way to investigate the
  coherent structures that are behind their identified two-dimensional cross sections. There are several ways to do that,
  following two essential principles that all of the three-dimensional extensions have to satisfy. They have to
    \vspace{0.2cm}
		
      \noindent (i) be covariant under rotations 
			
    \vspace{0.2cm}
		
      \noindent and
			
   \vspace{0.2cm}
	
     \noindent (ii) reduce to the $\lambda_\omega$-criterion in two-dimensional slices of the flow. 
   \vspace{0.2cm}
	
    With the above constraints in mind, let $\vec \omega (\vec r)$ be the three-dimensional vorticity vector field, so that
   we can define, analogously to Eqs. (\ref{pseudov}) and (\ref{pseudo_omega}), the pseudo-velocity and the pseudo-vorticity
   vector field {\black{components}}, respectively, as
     \begin{linenomath*} \be 
       \tilde v_i (\vec r) = \epsilon_{ijk} \partial_j \omega_k (\vec r) \label{pseudov2}
     \ee \end{linenomath*}
     and
     \begin{linenomath*} \be
       \tilde \omega_i (\vec r)  = - \partial^2 \omega_i (\vec r) \ . \ \label{pseudo_omega2}
     \ee \end{linenomath*}
    We can then pick up any of the standard three-dimensional vortex identification methods, like the $Q$ or $\Delta$ criteria, to
  write down a straightforward generalization of the $\lambda_\omega$-criterion. Taking the {\black{extensively used}} $Q$-criterion \cite{okubo,weiss,hunt},
	as our specific example, recall that 
    \begin{linenomath*} \be
      Q(\partial_j v_i) = -\frac{1}{2} \partial_i v_j \partial_j v_i  \ . \
    \ee \end{linenomath*}
    Vortex regions are defined as the connected sets of points where $Q>0$. Resorting to the
  pseudo-velocity and pseudo-vorticity vector fields, the $Q_\omega$-criterion, which extends the $\lambda_\omega$-criterion to
  three dimensions, is defined from the scalar field
    \begin{linenomath*} \be
      Q_\omega(\vec r) = \Theta(\omega_i \tilde \omega_i) Q(\partial_j \tilde v_i)  \ . \
    \ee \end{linenomath*}
    {\black{The filtering function previously used in the two-dimensional context is re-written above in terms of the three-dimensional vorticity field. We cannot get rid of it in the definition of the $Q_\omega$-criterion, otherwise we would surely recover the vortex identification problems for the cases where the flow is quasi two-dimensional, where $Q_\omega(\vec r)$ becomes essentially equivalent to $\lambda_\omega(\vec r)$, Eq. (\ref{lomega}).}}
    
    In the same fashion as it is done with the $Q$-criterion, we look now for regions of the flow which have $Q_\omega > 0$ in order to find 
		vortices. 
		%It is clear that as the $Q$-criterion reduces to the $\lambda_{ci}$-criterion in two-dimensions, an analogous result %follows for the relation between the $Q_\omega$ and the $\lambda_\omega$-criterion. 
		The implementation of the background subtraction procedure can be readily done by the substitution of the velocity field by its fluctuation around the mean, exactly as given in the Reynolds decomposition prescription defined by Eq. (\ref{back_remov}).
	
	{\black{To constrast the role of locality in the definitions of the $Q$ and the $Q_\omega$ criteria, note that we may write, as it is well known, $Q = (\Omega_{ij}^2 - S_{ij}^2)/2$, where $\Omega_{ij}$ and $S_{ij}$ are the matrix components of the rotation and the rate of the strain tensors, respectively. Even though the rotation tensor content is identical to the one given by the set of vorticity field components, the $Q$-criterion is, in fact, not fundamentally dependent on local properties of the vorticity field (as it is the case for the $Q_\omega$-criterion). To understand it more clearly, just recall that the strain tensor contribution to $Q$ can be expressed as a non-local functional of the vorticity field, as a direct consequence of Eq. (\ref{biot-savart}).}}
  
  In Fig. \ref{fig:qqtilda180}, we show how the $Q$ and the $Q_\omega$ criteria perform for the simulation of the
  turbulent channel flow considered in Sec. \ref{sec:dns}. As expected, there are many more, and better resolved, structures
  obtained from the use of the $Q_\omega$-criterion. The color scale indicates the absolute value of the velocity field, which 
	turns out to be a bit more intense for general regions of the flow in the case where the background subtraction procedure has been 
	carried out.
	
	We show, in these pictures, regions which have $Q$ or $Q_\omega$ fields greater than prescribed thresholds, in order to obtain a clear 
	visualization of flow structures at different distances from the wall. Figs. \ref{fig:qqtilda180}a and \ref{fig:qqtilda180}b are maps of 
	coherent structures detected, approximately, for heights $y^+ < 50$, while Fig. \ref{fig:qqtilda180}c is related to structures found within 
	$y^+ < 100$. 
	
	The $Q_\omega$ images, at variance with the $Q$ ones, suggest long-range correlations between regions which have higher magnitudes of the velocity 
	field and the presence of vortex packets, a fact than can be related to the existence of the very large-scale motions (VLSMs) observed in boundary 
	layer flows \cite{kim,bailey}.
	
	Also, when we compare Figs. \ref{fig:qqtilda180}b and \ref{fig:qqtilda180}c, it is tempting to evoke here the conjecture that 
	low speed streaks are connected with the formation of aligned packets of hairpin vortices, as it has been put foward in Ref. \cite{adrian1}. 
	
	The $Q_\omega$-criterion seems, therefore, to be a promising tool to address the three-dimensional organization of vortex structures
	in boundary layers at high Reynolds numbers. However, since our aim in this section is just to give a first glimpse 
	on three-dimensional vortex identification, we left this and other interesting issues to further comprehensive studies.
	
  \section{Conclusions} \label{sec:conclusions}

    We have introduced in this work an alternative vortex identification method -- the $\lambda_\omega$-criterion (or
  ``vorticity curvature" criterion) --, which is fundamentally based on the local properties of the vorticity field. As the
  starting point of our approach, we have critically revisited the usual swirling strength, $\lambda_{ci}$-criterion, in 
  order to classify its main shortcomings in simple two-dimensional vortex configurations {\black{(in two dimensions, 
  most of the velocity gradient-based vortex identification methods become equivalent to the $\lambda_{ci}$-criterion, which,
  then, has a central status in the general problem of vortex recognition)}}. A careful and rigorous benchmarking
  analysis has then been carried out, through an extensive statistical Monte Carlo treatment of synthetic vortex systems, in
  order to compare the performances of the $\lambda_\omega$ and the $\lambda_{ci}$ criteria. We have been able to find, in
  this way, that the $\lambda_\omega$-criterion leads, in general, to a considerably better and accurate identification of 
  two-dimensional vortices, as well as of their parameters of circulation, size and position. We have also shown how to deal 
  with possible spoiling external shear effects, by means of a simple background subtraction procedure, which amounts in 
  the use of the local Reynolds decomposition of the velocity field.
	
	We note that some further, but not very expressive, refinement of the $\lambda_\omega$-criterion may be necessary for the 
	cases of moderate/strong background shear in anisotropic vortex distributions (i.e., sytems which have more negative than positive 
	vortices, for instance), which may be relevant in flow conditions like the turbulent boundary viscous sublayer. 
  
    There are two crucial points that explain the observed good performance of the $\lambda_\omega$-criterion: (i) the interesting
  local properties of the vorticity field, when compared to the ones of the streamfunction (which is a non-local function of
  vorticity), and (ii) the use of the filtering Heaviside function $\Theta(\omega \tilde \omega)$ in the definition
  of the $\lambda_\omega$-criterion, as given in Sec. \ref{sec:vcurv}. This filter removes most of the spurious vortices and 
	{\black{renders the vorticity curvature method essentially free from the need of subjective threshold parameters.}}
  
  We have provided evidence which supports the application of the $\lambda_\omega$-criterion to flow configurations
  obtained by direct numerical simulations, taking the paradigmatical turbulent channel flow as an example. It turns out that
  DNS velocity fields are smooth enough to allow the use of the $\lambda_\omega$-criterion, a third order derivative scheme.
  We have been able, in this way, to address issues of isotropization in the turbulent boundary layer, which have, so far, eluded 
	the structural approach. {\black{The application}} of the $\lambda_\omega$-criterion to the 
	turbulent channel flow problem has led, for the first time (up to the authors knowledge), to a clear indication of isotropization 
	in the turbulent boundary layer, within the structural point of view. More work is needed here, of course, in combination with 
	the investigation of the three-dimensional coherent structures.

    The $\lambda_\omega$-criterion is directly generalizable to three-dimensions in more than one way. We have explored the
  three-dimensional extension motivated by the definition of the standard $Q$-criterion, which we have denoted as the 
  ``$Q_\omega$-criterion". Preliminary visualizations based on the $Q_\omega$-criterion show a profusion of well-resolved vortex 
	structures, not revealed in any of the previous standard analyses based on the $Q$-criterion ({\black{likely to be affected}} by both vortex coalescence 
	and erasing due to thresholding), and may shed light on the nature of the VLSMs, once they suggest some correlation between 
	percolating stronger velocity fluctuations and the formation of vortex packets in the turbulent boundary layer. 
	
	The study of other important boundary layer aspects is in order, which can now be more accurately addressed. We mean, for instance, an investigation of coherent structures in the turbulent viscous layer, and their role in the production of viscous drag. {\black{In this respect, it is worthwhile mentioning that phenomenological elements like the VLSMs and quasi-streamwise vortices, which can be identified with improved resolution through the $Q_\omega$-criterion, have been, actually, the subject of previous works focused on wall shear-stress fluctuations \cite{kravechenko,abe}.}}
	
		{\black{An interesting discussion, which we touch in passing, leaving a detailed account for a future study, is related to the description of coherent structures in terms of Komolgorov scales as developed in Refs. \cite{herpin_a,tanashi}. Consistently with the results of these works, we have found, through an application of the $\lambda_{ci}$-criterion to the streamwise/wall normal planes of our turbulent channel DNS samples, that the Kolmogorov-rescaled vortex radii, mean circulations and mean vorticities become very approximately constant for $y^+ > 50$. This, again, is a strong indication that the local Reynolds number (a function of $y/\eta$ where $\eta$ is the Kolmogorov dissipation length scale) is stable within the large regions of the flow where turbulence can be considered to be effectively isotropic.}}	
	
	  {\black{To put the bulk of our findings into a proper context, it is is important to stress that the $\lambda_{ci}$-criterion (or the $Q$-criterion as well) still offers a reasonably good computational cost-benefit ratio for the investigation of high Reynolds number flows, both in experimental and numerical studies. As it can be clearly seen from the turbulent channel analysis put forward in Sec. V, results found from the use of the $\lambda_{ci}$-criterion can be seen as a first approximation to the more accurate ones related to the application of the $\lambda_\omega$-criterion, as far as coherent structure resolution and background shear effects are not points of concern. In such cases, the $\lambda_{ci}$-criterion can be loosely interpreted as a low-pass filtered version of the $\lambda_\omega$-criterion.
	  
	  While the $\lambda_{ci}$-criterion relies on the set of first spatial derivatives of the velocity field, and its application to DNS or Particle Image Velocimetry (PIV) data is, therefore, comparatively less affected by numerical/measurement errors, some special care is necessary when the $\lambda_\omega$-criterion comes into play, once it is a higher-order derivative method.
	  
	   In order to deal with PIV or DNS data at higher Reynolds numbers, we point out here the main points related to the accuracy of the $\lambda_\omega$-criterion. On practical grounds, it is necessary to comply with two basic conditions, namely, (i) the data must be smooth enough to support accurate velocity derivatives up to third order and (ii) the grid resolution has to be fine enough to resolve both the boundaries and interior of the vortex regions. {\black{In a general DNS, one can assure that computations of velocity fields and their second derivatives are accurate if $k^4E(k)$, where $E(k)$ is the energy spectrum, is smooth and peaked at inertial range scales \cite{abe_etal, vreman-kuerten}}}. However, the condition (i) can only be achieved if the resolution is high enough so that the tail of the energy spectrum is steeper than $k^{-7}$, which can be sometimes a stringent requirement. Of course, smooth velocity fields can be artificially attained through low-pass filtering, as long as some resolution lost is still acceptable. On the other hand, while condition (ii) is not very problematic in applications of the swirling strength criterion, which usually produce well resolved large vortex regions, it can be a matter of concern for the vorticity curvature criterion. If the data is already smooth enough, it may be necessary to use a high order interpolation scheme to reach the grid resolution that would resolve vortex domains. In this way, not only PIV, but also DNS data may require careful post-processing for use along the lines of the vorticity curvature criterion.
    
       The application of the $\lambda_\omega$-criterion to conventional PIV data can be pursued without much worry when the goal is to study large scale vortices in the turbulent boundary layer (length scales within and above the logarithmic layer) after the procedure of velocity field smoothing is carried out. Hopefully, smaller structures, within viscous layer dimensions, could be also identified with the help of high resolution PIV data, a subject we deserve for future research.}}

  \acknowledgments
    We would like to thank D.J.C. Dennis, R.M. Pereira, and J.M. Wallace for the many fruitful discussions during 
		the course of this work. H. Anbarloei is specially acknowledged for his help with the numerical simulations of the 
		turbulent channel flow discussed in this paper. The scientific atmosphere of the N\'ucleo Interdisciplinar de Din\^amica de 
		Fluidos (NIDF) where part of this work has been developed has been a major source of motivation.
    
		This work has been partially supported by CNPq (CT-CNPq/402059/2013-1) and FAPERJ. 
   \vspace{-0.2cm}
   
%\end{linenumbers}

\end{document}